%% file: QMfromPuriPRA-resub.tex
\documentclass[twocolumn,aps,pra,showpacs,groupedaddress]{revtex4-1}
\usepackage{epsfig,amssymb,amsmath,mathrsfs,color} 
\usepackage[hypertex,linkcolor=red]{hyperref}
\usepackage[matrix,frame,arrow]{xypic}\input{myQcircuit}
\newtheorem{lemma}{Lemma} \newtheorem{proposition}{Proposition}

\newtheorem{corollary}{Corollary} \newtheorem{theorem}{Theorem}
\newtheorem{axiom}{Axiom}\newtheorem{postulate}{Postulate}
 \newtheorem{definition}{Definition}
 
\def\spc #1{{\mathscr #1}}
\def\Proof{\medskip\par\noindent{\bf Proof. }}
\def\qed{$\,\blacksquare$\par} 
\def\Cmplx{\mathbb C}
\def\>{\rangle}
\def\<{\langle} \def\trnsfrm#1{\mathscr
  #1} \def\rA{{\rm A}}\def\rB{{\rm
    B}}\def\rC{{\rm C}}\def\rD{{\rm D}}    \def\rI{{\rm I}}   \def\rX{{\rm X}} \def\rY{{\rm Y}} \def\rZ{{\rm Z}}
\def\tA{\trnsfrm A}\def\tB{\trnsfrm B}\def\tC{\trnsfrm
  C}\def\tD{\trnsfrm D} \def\tN{\trnsfrm N} \def\tI{\trnsfrm
  I}\def\tU{\trnsfrm U} \def\tV{\trnsfrm V}
\def\tW{\trnsfrm W}
\def\tE{\trnsfrm E} \def\tR{\trnsfrm R}

\def\Cntset{{\mathsf{Eff}}} 
\def\Stset{{\mathsf{St}}} 
\def\Trnset{{\mathsf{Transf}}}

\def\grp#1{{\mathbf{#1}}} \def\Span{\mathsf{Span}}
\def\K#1{\left|#1\right)}  \def\B#1{\left(#1\right|}
\def\SC#1#2{\left(#1\right|\left.\!#2\right)}  \def\Tr{{\rm Tr}}

\def\Reals{{\mathbb R}}

\begin{document}

\title{Informational derivation of Quantum Theory}

\author{Giulio Chiribella}\email{gchiribella@perimeterinstitute.ca} 
\affiliation{Perimeter Institute for Theoretical Physics, 31 Caroline Street North,  Ontario, Canada N2L 2Y5.}
\homepage{http://www.perimeterinstitute.ca}
\author{Giacomo Mauro D'Ariano}\email{dariano@unipv.it}
\affiliation{{\em QUIT Group}, Dipartimento di Fisica  ``A. Volta'' and  INFN Sezione di Pavia, via Bassi 6, 27100 Pavia, Italy}
\homepage{http://www.qubit.it}
\author{Paolo Perinotti}\email{paolo.perinotti@unipv.it} 
\affiliation{{\em QUIT Group}, Dipartimento di Fisica  ``A. Volta'' and  INFN Sezione di Pavia, via Bassi 6, 27100 Pavia, Italy}
\homepage{http://www.qubit.it}
\date{\today}

\begin{abstract} We derive Quantum Theory from purely informational principles.  Five elementary
  axioms---causality, perfect distinguishability, ideal compression, local distinguishability, and
  pure conditioning---define a broad class of theories of information-processing that can be
  regarded as standard. One postulate---purification---singles out quantum theory within this class.
\end{abstract}
\maketitle

\tableofcontents
\section{Introduction}

More than eighty years after its formulation, quantum theory is still mysterious.  The theory has a
solid mathematical foundation, addressed by Hilbert, von Neumann and Nordheim in 1928  \cite{HNN28}
and brought to completion in the monumental work by von Neumann \cite{von32}.  However, this
formulation is based on the abstract framework of Hilbert spaces and self-adjoint operators,  which,
to say the least, are far from having an intuitive physical meaning.   For example, the postulate stating that the pure states of a physical system are represented by unit vectors in a suitable Hilbert space appears as rather artificial:  which are the physical laws that lead to this very specific choice of mathematical representation?     
The problem with the standard textbook formulations of quantum theory is that the postulates therein impose particular mathematical structures without providing any fundamental reason for this choice:  the mathematics of Hilbert spaces is adopted without further questioning as a prescription that ``works well"  when used as a black box to produce experimental predictions. 
In a satisfactory axiomatization of Quantum Theory, instead, the mathematical structures of Hilbert spaces (or C*-algebras) should emerge as consequences of physically meaningful postulates, that is, postulates formulated exclusively in the language of physics: this language refers to notions like physical system, experiment, or physical process and not to notions like Hilbert space,  self-adjoint operator, or unitary operator.   Note that any serious axiomatization has to be based on postulates that can be precisely translated in mathematical terms. However, the point with the present status of quantum theory is that there are postulates that have a precise mathematical statement, but cannot  be translated back into language of physics. Those are the postulates that one would like to avoid.

The need for a deeper understanding of quantum theory in terms of  fundamental principles was clear
since the very beginning.  Von Neumann himself expressed his dissatisfaction with his mathematical formulation of Quantum Theory with the surprising words ``I don't believe in Hilbert space anymore", reported by Birkhoff in  \cite{Birk61}.  Realizing
the physical relevance of the axiomatization problem, Birkhoff and von Neumann made an attempt to understand quantum theory as a new form of logic \cite{BirkVN36}: the
key idea was that propositions about the physical world must be treated in a suitable logical framework, different from classical logics, where the operations AND and OR are no longer distributive.    This work inaugurated
the tradition of quantum logics, which led to several attempts to axiomatize quantum theory, notably
by Mackey \cite{Mack63} and Jauch and Piron \cite{JauPir63} (see Ref.  \cite{coeckerev} for a
review on the more recent progresses of quantum logics).  In general,  a certain degree of
technicality, mainly related to the emphasis on infinite-dimensional systems, makes these results far from providing a clear-cut description of quantum theory in terms of fundamental principles. Later Ludwig initiated an axiomatization
program \cite{Lud83} adopting an operational approach, where the basic notions are those of preparation devices and measuring devices and the postulates specify how preparations and measurements combine to give the probabilities of experimental outcomes.   
However, despite the original intent,  Ludwig's axiomatization did not succeed in deriving Hilbert spaces from purely operational notions, as some of the postulates still contained mathematical notions with no operational interpretation.

More recently, the rise of quantum information science moved the emphasis from logics to
information processing. The new field clearly showed that  the mathematical principles of  quantum theory
imply an enormous amount of information-theoretic consequences,  such as the no-cloning theorem
\cite{wootterszurek,dieks}, the possibility of teleportation \cite{tele}, secure key distribution
\cite{wiesner,bb84,e91}, or of factoring numbers in polynomial time \cite{shor}.   The natural
question is whether the implication can be reversed:  is it possible to retrieve quantum theory
from a set  of purely informational principles? Another  contribution of quantum information has
been to shift the emphasis to finite dimensional systems, which allow for a simpler treatment but
still possess all the remarkable quantum features.    In a sense, the study of finite
dimensional systems allows one to decouple the conceptual difficulties in our understanding of quantum
theory from the technical difficulties of infinite dimensional systems.  

In this scenario, Hardy's 2001 work \cite{Har01} re-opened the debate about the axiomatizations of
quantum theory with fresh ideas. Hardy's proposal was based on five main assumptions about the relation between
dimension of the state space and the number of perfectly distinguishable states of a given system, about 
the structure of composite systems, and about the possibility of connecting any two pure states of a
physical system through a continuous path of reversible transformations.  However, some of these assumptions
directly refer to the mathematical properties of the state space (in particular, the ``Simplicity Axiom" 2, which is an abstract statement about the functional dependence of the state space dimension on the number of perfectly distinguishable states).    Very recently, building on Hardy's work
there have been two new attempts of axiomatization by Dakic and Brukner \cite{DakBru09} and Masanes
and M\"uller \cite{Mas10}. Although these works succeeded in removing the ``Simplicity Axiom", they
still contain mathematical assumptions that cannot be understood in elementary physical terms (see e.g.
requirement 5 of Ref. \cite{Mas10}, which assumes that ``all mathematically well-defined
measurements are allowed by the theory").

Another approach to the axiomatization of quantum theory was pursued by one of the authors in a
series of works \cite{maurofirst} culminated in Ref. \cite{maurolast}.  These works tackled the
problem using operational principles related to tomography and calibration of physical devices,
experimental complexity, and to the composition of elementary transformations.  In particular this
research introduced the concept of dynamically faithful states, namely states that can be used for
the complete tomography of physical processes.  Although this approach went very close to deriving
quantum theory also in this case one mathematical assumption without operational interpretation was needed (see the CJ postulate of
Ref. \cite{maurolast}).

In this paper we provide a complete derivation of finite dimensional quantum theory based of purely operational
principles. Our principles do not refer to abstract properties of the mathematical structures that we use to represent
states, transformations or measurements, but only to the way in which states, transformations and measurements combine with each other.  More specifically, our principles are of \emph{informational} nature:  they assert basic properties of information-processing, such as the possibility or impossibility to carry
out certain  tasks by manipulating physical systems. In this approach the rules by which information can be processed determine the physical theory,  in  accordance with Wheeler's program ``it from bit", for which he argued that   ``all things physical are information-theoretic in origin" \cite{wheeler}.   Note that, however, our axiomatization of quantum theory is relevant, as a rigorous result,  also for those who do not share Wheeler's ideas on the informational origin of physics. In particular, in the process of deriving quantum theory we provide alternative proofs for many key features of the Hilbert space formalism, such as the spectral decomposition of  self-adjoint operators or the existence of projections.  The interesting feature of these proofs is that they are obtained by manipulation of the principles, without assuming Hilbert spaces form the start.

The main message of our work is simple: within a standard class of theories of information processing,
quantum theory is uniquely identified by a single postulate: \emph{purification}.  The purification
postulate, introduced in Ref. \cite{purification}, expresses a distinctive feature of quantum
theory, namely that the ignorance about a part is always compatible with the maximal knowledge of the
whole. The key role of this feature was noticed already in 1935 by Schr\"odinger in his discussion
about entanglement \cite{Schr35}, of which he famously wrote ``I would not call that \emph{one} but
rather {\em the} characteristic trait of quantum mechanics, the one that enforces its entire
departure from classical lines of thought".  In a sense, our work can be viewed as the concrete
realization of Schr\"odinger's claim: the fact that every physical state can be viewed as the
marginal of some pure state of a compound system is indeed the key to single out quantum theory within a standard set of possible theories.  It
is worth stressing, however, that the purification principle assumed in this paper includes a requirement that was not explicitly mentioned in Schr\"odinger's discussion: if two pure states of a composite system $\rA\rB$ have the same marginal on system $\rA$, then they are connected by some
reversible transformation on system $\rB$.  In other words, we assume that all purifications of a given mixed state are equivalent under local reversible operations
\cite{anchese...}.    

The purification principle expresses a law of \emph{conservation of information}, stating that at
least in principle, irreversibility can always be reduced to the lack of control over an
environment. More precisely, the purification principle is equivalent to the statement that every
irreversible process can be simulated in an essentially unique way by a reversible interaction of
the system with an environment, which is initially in a pure state \cite{purification}.  This
statement can also be extended to include the case of measurement processes, and in that case it
implies the possibility of arbitrarily shifting the cut between the observer and the observed system
\cite{purification}.  The possibility of such a shift was considered by von Neumann as a
``fundamental requirement of the scientific viewpoint" (see p. 418 of \cite{von32}) and his
discussion of the measurement process was exactly aimed to show that quantum theory fulfils this
requirement.

Besides Schr\"odinger's discussion on entanglement and von Neumann's discussion of the measurement process, the purification principle is deeply rooted in
the structure of quantum theory.  At the purely mathematical level, it plays a crucial role in the
theory of C*-algebras of operators on separable Hilbert spaces, where the purification principle is
equivalent to the Gelfand-Naimark-Segal (GNS) construction \cite{arveson} and implies the celebrated
Stinespring's theorem \cite{stine}.  On the other hand, purification is a cornerstone of quantum information, lying at the origin of most
quantum protocols.  As it was shown in Ref. \cite{purification}, the purification 
principle directly implies crucial features like no-cloning, teleportation, no-information without
disturbance, error correction, the impossibility of bit commitment, and the ``no-programming"
theorem of Ref. \cite{no-prog}.

In addition to the purification postulate, our derivation of quantum theory is based on five
informational axioms.  The reason why we call them ``axioms", as opposed to the the purification
``postulate", is that they are not at all specific of quantum theory.  These axioms represent 
standard features of information-processing that everyone would,
more or less implicitly, assume. They define a class of theories of information-processing 
that includes, for example, classical information theory, quantum information theory, and quantum theory with superselection rules.  The question 
whether there are other theories satisfying our five axioms  and, in case of a positive answer, the full classification of these theories is currently an open problem. 

Here we informally illustrate the five axioms, 
leaving the more detailed description to the remaining part of the paper: \begin{enumerate}
\item \emph{Causality}:  the probability of a measurement outcome at a certain time does not depend
  on the choice of measurements that will be performed later.    
\item \emph{Perfect distinguishability}:   if a state is not completely mixed (i.e. if it cannot be
  obtained as a mixture from any other state), then there exists at least one state that can be
  perfectly distinguished from it.  
\item \emph{Ideal compression:} every source of information can be encoded in a suitable physical
  system in a lossless and maximally efficient fashion. Here \emph{lossless} means that the information
  can be decoded without errors and \emph{maximally efficient} means that every state of the encoding
  system represents a state in the information source.
\item \emph{Local distinguishability}: if two states of a composite system are different, then we
  can distinguish between them from the statistics of local measurements on the component systems.
\item \emph{Pure conditioning:} if a pure state of system $\rA\rB$ undergoes an atomic measurement
  on system $\rA$, then each outcome of the measurement induces a pure state on system $\rB$.  (Here
  \emph{atomic measurement} means a measurement that cannot be obtained as a coarse-graining of  another measurement).
\end{enumerate}  
All these axioms are satisfied by classical information theory.     Axiom 5 is even trivial for classical theory, because the only pure
states of a composite system $\rA\rB$ are the product of pure states
of the component systems $\rA$ and $\rB$, and hence the state of
system $\rB$ will be pure irrespectively of what we do on system
$\rA$.

A stronger version of axiom 5, introduced in Ref.
\cite{maurolast}, is the following:
\begin{enumerate} \item[5'] \emph{Atomicity of composition}: the
  sequential composition of two atomic operations is atomic.  (Here
  \emph{atomic transformation} means a transformation that cannot
  be obtained from coarse-graining).
\end{enumerate}
However, it turns out that Axiom 5 is enough for our derivation:
thanks to the purification postulate we will be able to show the
non-trivial implication: Axiom 5 $\Rightarrow $ Axiom 5' (see lemma
\ref{lem:atomicity}).

The paper is organized as follows. In Sec. \ref{sec:framework} we
review the framework of {\em operational-probabilistic theories} 
introduced in Ref. \cite{purification}. This framework will provide the basic
notions needed for the formulation of our principles. In Sec.
\ref{sec:principles} we introduce the principles from which we will
derive Quantum Theory. In Sec.  \ref{sec:consequences} we prove some direct consequences of the principles that will be used later in the paper. In Sec. \ref{sec:distinguishable} we
discuss the properties of perfectly distinguishable states,  while in Sec. \ref{sec:duality} we prove the
existence of a duality between pure states and
atomic effects.

The results about distinguishability and duality of pure states and atomic effects allow us to show
in Sec. \ref{sec:dimension} that every system has a well defined \emph{informational
  dimension}---the operational counterpart of the Hilbert space dimension. Sec.
\ref{sec:decomposition} contains the proof that every state can be decomposed as a convex
combination of perfectly distinguishable pure states. Similarly, any element of the vector space
spanned by the states can be written as a linear combination of perfectly distinguishable states.
This result corresponds to the spectral theorem for self-adjoint operators on complex Hilbert
spaces. In Sec.  \ref{sec:teleportation} we prove some results about the maximum teleportation
probability, which allow us to derive a functional relation between the dimension of the state space
and the number of perfectly distinguishable states of the system. The mathematical representation of
systems with two perfectly distinguishable states is derived in Sec.  \ref{sec:qubits}, where we
prove that such systems are indeed two-dimensional quantum systems---a.k.a. qubits. In Sec.
\ref{sec:projections} we construct projections on the faces of the state space of any system and
prove their main properties. These results lead to the derivation of the operational analogue of the
superposition principle in Sec. \ref{sec:superposition} which allows to prove that systems with the
same number of perfectly distinguishable states are operationally equivalent (Subsec.
\ref{subsec:equivalence}).  The properties of the projections and the superposition principle are
then exploited in Sec.  \ref{sec:matrix}---where we extend the density matrix representation from
qubits to higher-dimensional systems, thus proving that a system with $d$ perfectly distinguishable
states is indeed a quantum system with $d$-dimensional Hilbert space. We conclude the paper with
Sec.  \ref{sec:conclusion}, where we review our results, 
discussing future directions for this research.

\section{The framework}\label{sec:framework}

This Section provides a brief summary of the framework of \emph{operational-probabilistic theories},
which was formulated in Ref. \cite{purification}. We refer to Ref. \cite{purification} for an exhaustive presentation of the details of the framework and of the ideas behind it.   The operational-probabilistic framework combines the operational language of circuits with the toolbox of probability theory: on
the one hand, experiments are described by circuits resulting from the connection of physical
devices, on the other hand each device in the circuit can have classical outcomes and the theory
provides the probability distribution of outcomes when the devices are connected to form closed
circuits (that is, circuits that start with a preparation and end with a measurement).

The notions discussed in this section will allow us to draw a precise distinction between  principles  with an operational content and exclusively mathematical principles: with the expression ''operational principle" we will mean a principle that can be expressed using only the basic notions of the the operational-probabilistic framework.

\subsection{Circuits with outcomes}  

A \emph{test} represents one use of a physical device, like a Stern-Gerlach magnet, a beamsplitter,
or a photon counter.  The device will have an \emph{input system} and an \emph{output system},
labelled by capital letters. The corresponding test can have different classical outcomes,
represented by different values of an index $i \in \rX$:
\begin{equation*}
 \Qcircuit @C=1em @R=.7em @! R {& \qw \poloFantasmaCn \rA & \gate {\{\tC_i\}_{i \in \rX}} & \qw \poloFantasmaCn \rB &\qw}
\end{equation*}
Each outcome $i \in \rX$ corresponds to a possible \emph{event}, represented as 
\begin{equation*}
  \Qcircuit @C=1em @R=.7em @! R {& \qw \poloFantasmaCn \rA & \gate {\tC_i} & \qw \poloFantasmaCn \rB &\qw}
\end{equation*}
We denote by $\Trnset (\rA, \rB)$ the set of all events from $\rA$ to $\rB$.   The reason for this notation is that in the next subsection the elements of $\Trnset (\rA,\rB)$  will be interpreted as  \emph{transformations}   with input system $\rA$ and output system $\rB$.    If $\rA = \rB$ we
simply write $\Trnset (\rA)$ in place of $\Trnset (\rA, \rA)$.

A test with a single outcome will be
called \emph{deterministic}.  This name is justified by the fact that, if there is a single possible
outcome, then this outcome will occur with certainty (cf. the probabilistic structure introduced in the next subsection).

Two devices can be composed in a sequence, as long as the input system of the second device is equal
to the output system of the first. The events in the composite test are represented as
\begin{equation*}
  \Qcircuit @C=1em @R=.7em @! R {& \qw \poloFantasmaCn \rA & \gate {\tC_i} & \qw \poloFantasmaCn \rB &\gate{\tD_j} & \qw\poloFantasmaCn\rC &\qw} 
\end{equation*}
and are written in formulas as $\tD_j \tC_i$. 

For every system $\rA$ one can  perform the \emph{identity-test} (or simply, the \emph{identity}), that is,  a test  $\{\tI_\rA\}$ with a single outcome, with the property 
\begin{equation*}
\begin{split}
  \Qcircuit @C=1em @R=.7em @! R {& \qw \poloFantasmaCn \rA & \gate {\tI_\rA} & \qw \poloFantasmaCn \rA &\gate{\tC_i} & \qw\poloFantasmaCn\rB &\qw} &=~   \Qcircuit @C=1em @R=.7em @! R {& \qw \poloFantasmaCn \rA &\gate{\tC_i} & \qw\poloFantasmaCn\rB &\qw} ~\forall \tC_i \in \Trnset (\rA, \rB) \\
&\\
  \Qcircuit @C=1em @R=.7em @! R {&\qw \poloFantasmaCn \rB &\gate{\tD_j} & \qw\poloFantasmaCn\rA  & \gate {\tI_\rA} &  \qw \poloFantasmaCn \rA &\qw} &=~   \Qcircuit @C=1em @R=.7em @! R {& \qw \poloFantasmaCn \rB &\gate{\tD_j} & \qw\poloFantasmaCn\rA &\qw} ~ \forall \tD_j \in \Trnset (\rB, \rA)
\end{split} 
\end{equation*}
The subindex $\rA$ will be dropped from  $\tI_\rA$ where there is no ambiguity. 

The letter $\rI$ will be reserved for the \emph{trivial system}, which simply means ``nothing''
\cite{nothingquantum}.  A device with input (resp. output) system $\rI$ is a device with no input
(resp. no output). The corresponding tests will be called \emph{preparation-tests} (resp.
\emph{observation-tests}).  In this case we replace the input (resp. output) wire with a round
portion:
\begin{equation}
 \Qcircuit @C=1em @R=.7em @! R {\prepareC{\rho_i}& \qw \poloFantasmaCn \rB &\qw}  \qquad  \left({\rm resp.}~  \Qcircuit @C=1em @R=.7em @! R {& \qw \poloFantasmaCn \rA &\measureD {a_j}} \right). 
\end{equation}
In formulas we will write $\K {\rho_i}_\rB$ (resp.  $\B {a_j}_\rA$). The sets $\Trnset (\rI, \rA)$
and $\Trnset (\rA, \rI)$ will be denoted as $\Stset(\rA)$ and $\Cntset (\rA)$, respectively.  The
reason for this special notation is that in the next subsection the elements of $\Stset (\rA)$
(resp. $\Cntset (\rA)$) will be interpreted as the \emph{states} (resp. \emph{effects}) of system $\rA$.

From every pair of systems $\rA$ and $\rB$ one can form a \emph{composite system}, denoted by
$\rA\rB$.  Clearly, composing system $\rA$ with nothing still gives system $\rA$, in formula $\rA\rI
= \rI \rA = \rA$.  Two devices can be composed in parallel, thus obtaining a new device with
composite input and composite output systems. The events in composite test are represented as
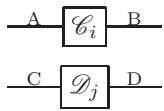
\begin{equation*}
\begin{aligned}
  \Qcircuit @C=1em @R=.7em @! R {& \qw \poloFantasmaCn \rA & \gate {\tC_i} & \qw \poloFantasmaCn \rB &\qw\\
& \qw \poloFantasmaCn \rC & \gate {\tD_j} & \qw \poloFantasmaCn \rD &\qw}
\end{aligned}
\end{equation*}
and are written in formulas as $\tC_i \otimes \tD_j$.  In the special case
of states we will often write $\K{\rho_i} \K{\sigma_j}$ in place of $\rho_i
\otimes \sigma_j$. Similarly, for effects we will write $\B {a_i}
\B{b_j}$ in place of $a_i \otimes b_j$.

Sequential and parallel composition commute: one has $(\tA_i \otimes \tB_j) (\tC_k\otimes \tD_l) =
\tA_i\tC_k \otimes \tB_j\tD_l$ for every $\tA_i,\tB_j,\tC_k,\tD_l$ such that the output of $\tA_i$
(resp. $\tB_j$) coincides with the input of $\tC_k$ (resp. $\tD_l$).

When one of the two tests is the identity, we will omit the box and draw only a straight line, as in 
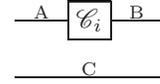
\begin{equation*}
\begin{aligned}
  \Qcircuit @C=1em @R=.7em @! R {& \qw \poloFantasmaCn \rA & \gate {\tC_i} & \qw \poloFantasmaCn \rB &\qw\\
& \qw  & \qw \poloFantasmaCn \rC &\qw&\qw}
\end{aligned}
\end{equation*} 

The rules summarized in this section define the operational language of circuits, which has been discussed in detail in a series of inspiring works by Coecke  (see in particular Refs. \cite{kinder, picturalism}).  
The language of circuits
allows one to represent the schematic of an experiment, like e.g.
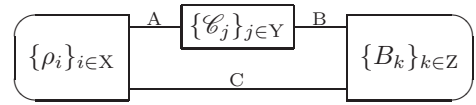
\begin{equation*}
\begin{aligned}
  \Qcircuit @C=1em @R=.7em @! R {  \multiprepareC{1} {\{\rho_i\}_{i \in \rX}} & \qw \poloFantasmaCn \rA & \gate {\{\tC_j\}_{j \in \rY}} & \qw \poloFantasmaCn \rB &\multimeasureD{1} {\{B_k\}_{k \in \rZ}}\\
\pureghost{\{\rho_i\}_{i \in \rX}}& \qw  & \qw \poloFantasmaCn \rC &\qw&\ghost{\{B_k\}_{k \in \rZ}}}
\end{aligned}
\end{equation*} 
and also to represent a particular outcome of the experiment
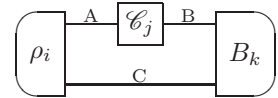
\begin{equation*}
\begin{aligned}
  \Qcircuit @C=1em @R=.7em @! R {  \multiprepareC{1} {\rho_i} & \qw \poloFantasmaCn \rA & \gate {\tC_j} & \qw \poloFantasmaCn \rB &\multimeasureD{1} {B_k}\\
\pureghost{\rho_i}& \qw  & \qw \poloFantasmaCn \rC &\qw&\ghost{B_k}}
\end{aligned}
\end{equation*} 
In formula, the above circuit is given by 
\begin{align*}
\B{ B_k}_{\rB\rC}  (\tC_j \otimes \tI_\rC)  \K{\rho_i}_{\rA\rC}.
\end{align*}

\subsection{Probabilistic structure:  states, effects and transformations}
On top of the language of circuits, we put a probabilistic structure \cite{purification}: we declare that the
composition of a preparation-test $\{\rho_i\}_{i \in \rX}$ with an observation-test $\{a_j\}_{j \in \rY}$ gives rise to a joint probability
distribution:
\begin{align}\label{probabilities}
\begin{aligned}
  \Qcircuit @C=1em @R=.7em @! R {  \prepareC{\rho_i} & \qw \poloFantasmaCn \rA & \measureD{a_j}}
\end{aligned}
= p(i,j),
\end{align}  
with 
$p(i,j)   \ge 0$ and  
$\sum_{i \in \rX}  \sum_{j \in \rY}   p(i,j)  = 1$.
In formula we write $p(i,j)  = \SC {a_j}  {\rho_i}$.
Moreover, if two experiments are run in parallel,  we assume that the joint probability distribution is given by the product:  
\begin{equation}
\begin{aligned}\label{productprobabilities}
  \Qcircuit @C=1em @R=.7em @! R {  \prepareC{\rho_i} & \qw \poloFantasmaCn \rA & \measureD{a_k} \\
  \prepareC {\sigma_j}   & \qw \poloFantasmaCn \rB  & \measureD{b_l}}
\end{aligned}  =  p(i,k)  q(j,l)  
\end{equation} 
where  $p(i,k): = \SC {a_k}{ \rho_i} , q(j,l) := \SC  {b_l} {\sigma_j}$. 

The probabilistic structure defined by Eq. (\ref{probabilities}) turns
every event $\rho_i \in \Stset(\rA)$ into a function $\hat \rho_i :
\Cntset (\rA) \to \Reals$, given by $\hat \rho_i (a_j) := \SC {a_j}
{\rho_i}$.  If two events $\rho_i, \rho_i' \in \Stset (\rA)$ induce 
the same function, then it is impossible to distinguish between them
from the statistics of the experiments allowed by our theory.  This
means that for our purposes the two events are the same: accordingly,
we will take equivalence classes with respect to the relation $\rho_i
\simeq \rho_i'$ if $\hat \rho_i = \hat \rho_i'$. To avoid introducing
new notation, from now on we will assume that the equivalence classes
have been taken since the start. We will identify the event
$\rho_i\in\Stset (\rA)$ with the corresponding function $\hat \rho_i$
and will call it \emph{state}.  Accordingly, we will refer to
preparation-tests as collections of states $\{\rho_i\}_{i \in \rX}$.
Note that, since one can take linear combinations of functions, the
states in $\Stset (\rA)$ generate a real vector space, denoted by
$\Stset_\Reals (\rA)$.

The same construction holds for observation-tests: every event $a_j \in \Cntset(\rA)$ induces a
function $\hat a_j : \Stset (\rA) \to \Reals$, given by $\hat a_j (\rho_i) := \SC {a_j} {\rho_i}$.
If two events $a_j, a_j' \in \Cntset (\rA)$ induce to the same function, then it is impossible to
distinguish between them from the statistics of the experiments allowed in our theory.  This means
that for our purposes the two events are the same: accordingly, we will take equivalence classes
with respect to the relation $a_j \simeq a_j'$ if $\hat a_j = \hat a_j'$. To avoid introducing new
notation, from now on we will identify the event $a_j \in \Cntset (\rA)$ with the corresponding
function $\hat a_j$ and we will call it \emph{effect}.  Accordingly, we will refer to
observation-tests as collection of effects $\{a_j\}_{i \in \rY}$.  The effects in $\Cntset (\rA)$
generate a real vector space, denoted by $\Cntset_\Reals (\rA)$.

A vector in $\Stset_\Reals (\rA)$ (resp.  $\Cntset_\Reals (\rA)$) can be extended to a linear
function on $\Cntset_\Reals (\rA)$ (resp. $\Stset_\Reals (\rA)$).  In this way, states and effects
can be thought as elements of two real vector spaces, one dual to the other.  In this paper we will
restrict our attention to \emph{finite dimensional} vector spaces: operationally, this means that the state of a given physical system is completely determined by the statistics of a finite number of finite-outcome measurements.  The dimension of the vector space
$\Stset_\Reals (\rA)$, which by construction is equal to the dimension of its dual
$\Cntset_\Reals(\rA)$, will be denoted by $D_\rA$.  We will refer to $D_\rA$ as the \emph{size} of
system $\rA$.

Finally, the vector spaces $\Stset_\Reals (\rA)$ and $\Cntset_\Reals (\rA)$ can be equipped with
suitable norms, which have an operational meaning related to optimal discrimination schemes
\cite{purification}.  The norm of an element $\delta \in \Stset_\Reals (\rA)$ is given by
\cite{purification}
\begin{equation*}
|\!| \delta   |\!|  = \sup_{a_0 \in \Cntset(\rA)} \SC {a_0} \delta  - \inf_{a_1 \in \Cntset (\rA)}  \SC {a_1} \delta,
\end{equation*} 
while the norm of an element $\xi \in \Cntset_\Reals (\rA)$ is given by 
\begin{equation*}
|\!| \xi   |\!|  = \sup_{\rho \in \Stset(\rA)}  |  \SC \xi \rho  | .
\end{equation*} 

We will always take the set of states $\Stset(\rA)$ to be closed in
the operational norm. The convenience of this choice is the convenience of using real numbers
instead of rational ones: dealing with a single real number is much easier than dealing with
a Cauchy sequence of rational numbers. 
Operationally, taking $\Stset(\rA)$ to be closed is very natural: the fact that there is a sequence of states $\{\rho_n\}_{n=1}^\infty$
that converges to $\rho\in\Stset_\Reals (\rA)$ means that there is a procedure to prepare $\rho$
with arbitrary precision and hence that $\rho$ deserves the name of ``state".

We conclude this Subsection by noting that every event $\tC_k$ from $\rA$ to
$\rB$ induces a linear map $\hat {\tC_k}$ from $\Stset_\Reals (\rA)$ to
$\Stset_\Reals (\rB)$, uniquely defined by  
\begin{equation*}
\hat{\tC_k}: \K{\rho} \in \Stset (\rA)
\mapsto \tC_k \K{ \rho} \in \Stset (\rB).
\end{equation*}

Likewise, for every system $\rC$ the event $\tC_k \otimes \tI_\rC$ induces a linear map
$\widehat{\tC_k \otimes \tI_\rC }$ from $\Stset_\Reals (\rA\rC)$ to $\Stset_\Reals (\rB\rC)$. If two
events $\tC_k$ and $\tC_k'$ induce the same maps for every possible system $\rC$, then there is no
experiment in the theory that is able to distinguish between them.  This means that for our purposes the two
events are the same: accordingly, we will take equivalence classes with respect to the relation
$\tC_k \simeq \tC_k'$ if $\widehat { \tC_k \otimes \tI_\rC} = \widehat{ \tC_k' \otimes \tI_\rC}$ for
every system $\rC$.  In this case, we will say that two events represent the same
\emph{transformation}. Accordingly, we will refer to tests $\{\tC_i\}_{i \in\rX}$ as collections of
transformations.  The deterministic transformations (corresponding to single-outcome tests) will be
called \emph{channels}.

\subsection{Basic definitions in the operational-probabilistic framework }  
Here we summarize few elementary definitions that will be used later in the paper. The meaning of
the definitions in the case of quantum theory is also discussed.

\subsubsection{Coarse-graining, refinement, atomic transformations, pure, mixed and completely mixed states}
First, we start from the notions of \emph{coarse-graining} and \emph{refinement}.  Coarse-graining
arises when we join together some outcomes of a test: we say that the test $\{\tD_j\}_{j \in \rY}$
is a coarse-graining of the test $\{\tC_i\}_{i \in \rX}$ if there is a disjoint partition
$\{\rX_j\}_{j\in \rY} $ of $\rX$ such that
\begin{equation*}
  \tD_j  = \sum_{i \in \rX_j} \tC_i.
\end{equation*}
Conversely, if $\{\tD_j\}_{j \in \rY}$ is a coarse-graining of $\{\tC_i\}_{i \in \rX}$, we say that
$\{\tC_i\}_{i \in \rX}$ is a refinement of $\{\tD_j\}_{j \in \rY}$.  Intuitively, a test that
refines another is a test that extracts information in a more precise way:  it is a test with  better ``resolving power".

The notion of refinement also applies to a single transformation: a refinement of the transformation
$\tC$ is given by a test $\{\tC_i\}_{i \in \rX}$ and a subset $\rX_0$ such that 
\begin{equation*}
\tC = \sum_{i \in
  \rX_0} \tC_i.
  \end{equation*}  
Accordingly, we say that each transformation $\tC_i, i \in \rX_0$ is a \emph{refinement} of $\tC$.
A transformation $\tC$ is \emph{atomic} if it has only trivial refinements: if $\tC_i$ refines
$\tC$, then $\tC_i =p \tC$ for some probability $p \ge 0 $.
A test that consists of atomic transformations is a test whose ``resolving power" cannot be further improved.

When discussing states (i.e. transformations with trivial input) we will use the word \emph{pure} as a
synonym of atomic.  A pure state describes a situation of maximal knowledge about the system's preparation, a knowledge that cannot be further refined.

As usual, a state that is not pure will be called \emph{mixed}. An important notion is that of \emph{completely mixed state}:  
\begin{definition}[Completely mixed state]\label{def:complemix}
 A state is \emph{completely mixed} if any other state can refine it:
 precisely, $\omega\in\Stset(\rA)$ is completely mixed if for every $\rho \in\Stset(\rA)$ there is
 a non-zero probability $p>0$ such that $p \rho$ is a refinement of $\omega$.  
 \end{definition}
Intuitively, a completely mixed state describes a situation of complete ignorance about the system's preparation:   if a system is described by a completely mixed state, then it  means that we know so little about its preparation that, in fact, every preparation is possible.

We conclude this paragraph with a couple of definitions that will be used throughout the paper: 
\begin{definition}[Reversible transformation]
A transformation $\tU \in \Trnset (\rA, \rB)$ is \emph{reversible} if there exists another
transformation $\tU^{-1}\in\Trnset(\rB,\rA)$ such that $\tU^{-1}\tU = \tI_\rA $ and $\tU \tU^{-1} =
\tI_\rB$.  When $\rA=\rB$ the reversible transformations form a group, indicated as $\grp G_\rA$.
\end{definition}
\begin{definition}[Operationally equivalent systems]\label{def:opeq}
Two systems $\rA$ and $\rB$ are \emph{operationally equivalent} if there exists a reversible
transformation $\tU$ from $\rA$ to $\rB$.  
\end{definition}
When two systems are operationally equivalent one can
convert one into the other in a reversible fashion.

\subsubsection{Examples in in quantum theory }  

Consider a quantum system with Hilbert space $\spc H = \Cmplx^d, d <
\infty$.  In this case a preparation-test is a collection of unnormalized density matrices $\{\rho_i\}_{i \in \rX}$ (i.e. of non-negative $d \times d$  complex matrices with trace bounded by 1)  such that 
\begin{equation*}
\sum_{i \in \rX} \Tr[ \rho_i]  =1  .  
\end{equation*} 
Preparation-tests are often called \emph{quantum information sources}  in quantum information theory.  
A generic state $\rho$ is an unnormalized density matrix. A deterministic state, corresponding to a single-outcome preparation-test is a normalized density matrix $\rho$, with $\Tr[\rho]=1$.

Diagonalizing $\rho = \sum_i \alpha_i |\psi_i\>\<\psi_i|$ we then
obtain that each matrix $\alpha_i |\psi_i\>\<\psi_i|$ is a refinement
of $\rho$. More generally, every matrix $\sigma$ such that $\sigma \le
\rho$ is a refinement of $\rho$.  Up to a positive rescaling, all
matrices with support contained in the support of $\rho$ are
refinements of $\rho$.  A quantum state $\rho$ is atomic (pure) if and
only if it is proportional to a rank-one projection.  A quantum state
is completely mixed if and only if its density matrix has full rank. Note that the quantum state $\chi  = \frac {I_d} d$, where ${I_d}$ is the identity $d\times d$ matrix, is a particular example of completely mixed state, but not the only example.  Precisely, $\chi = \frac {I_d} d $ is the unique unitarily invariant state in dimension $d$.   

Let us now consider the case of observation-tests:  in quantum theory an observation-test is given by a \emph{POVM (positive operator-valued measure)}, namely by a collection $\{P_j\}_{j \in \rY}$ of non-negative $d\times d$ matrices such that  
\begin{equation*}
\sum_{j \in \rY}  P_j  =  I_d.
\end{equation*}
An effect is then a non-negative matrix $P\ge 0 $ upper bounded by the identity.  In quantum theory there is only one deterministic effect, corresponding to a single-outcome observation test:  the unique deterministic effect given by the identity matrix.   As we will see in the following section, the fact that the deterministic effect is unique is equivalent to the fact that quantum theory is a \emph{causal} theory.   

An effect $P$ is atomic if and only if $P$ is
proportional to a rank-one projector.  An observation-test is atomic if it is a POVM with rank-one elements.   

Finally, a general test from an input system with Hilbert space $\spc H_1= \Cmplx^{d_1}$
to an output system with Hilbert space $\spc H_2 = \Cmplx^{d_2} $ is given by a \emph{quantum instrument}, namely by a collection $\{\tC_k\}_{k \in \rZ}$ of  completely positive trace non-increasing maps sending linear operators on
$\spc H_1$ to linear operators on $\spc H_2$, with the property that 
\begin{equation*}
\tC_\rZ  : = \sum_{k \in \rZ}  \tC_k  
\end{equation*}   
is trace-preserving.  A general transformation is then given by a  trace non-increasing map, called \emph{quantum operation}, whereas a deterministic transformation, corresponding to a single-outcome test, is given by a trace-preserving map, called \emph{quantum channel}.    

Any quantum operation $\tC$ can be written in the Kraus form $\tC
(\rho) = \sum_i C_i \rho C_i^\dag$, where $C_i : \spc H_1 \to \spc H_2$ are the Kraus operators.  Up
to a positive scaling, every quantum operation $\tD $ such that the Kraus operators of $\tD$ belong
to the linear span of the Kraus operators of $\tC$ is a refinement of $\tC$.  A map $\tC$ is atomic
if and only if there is only one Kraus operator in its Kraus form.  A reversible transformation in
quantum theory is a unitary map $\tU (\rho ) = U \rho U^\dag$, where $U: \spc H_1 \to \spc H_2$ is a
unitary operator, that is $U^\dag U = I_1$ and $U U^\dag = I_2$ where $I_1$ $(I_2)$ is the identity
operator on $\spc H_1$ ($\spc H_2$).  Two quantum systems are operationally equivalent if and only
if the corresponding Hilbert spaces have the same dimension.

\subsection{Operational principles}

We are now in position to make precise the usage of the expression ``operational principle'' in the
context of this paper.  By ``operational principle'' we mean here a principle that can be stated
using only the operational-probablistic language, i.e. using only 
\begin{itemize}
\item the notions of system, test, outcome,
probability, state, effect, transformation
\item their specifications: atomic, pure, mixed, completely mixed
\item more complex notions constructed from the above terms
(e.g. the notion of ``reversible transformation'').  
\end{itemize}

The distinction between operational principles and principles referring to abstract mathematical
properties, mentioned in the introduction, should now be clear: for example, a statement like ``the
pure states of a system cannot be cloned" is a valid operational principle, because it can be
analyzed in basic operational-probabilistic terms as ``for every system $\rA$ there exists no transformation $\tC $ with input
system $\rA$ and output system $\rA\rA$ such that $\tC \K \varphi = \K \varphi \K\varphi$ for every
pure state $\varphi$ of $\rA$ ".  On the contrary, a statement like ``the state space of a system
with two perfectly distinguishable states is a three-dimensional sphere" is not a valid operational principle, because
there is no way to express what it means for a state space to be a three-dimensional sphere in terms of basic operational notions.  The fact that a state spate is a sphere may be eventually derived from operational principles, but cannot be assumed as a starting point.

\section{The principles}\label{sec:principles}
We now state the principles used in our derivation.  The first five
principles express generic features that are shared by both classical
and quantum theory. They could be even included in the definition of
the background framework: they define the simple model of information
processing in which we try to single out quantum theory.  For this
reason we will call them \emph{axioms}.  The sixth principle in our
derivation has a different status: it expresses the genuinely quantum
features.  
A major message of our work is that, within a broad class of theories of information processing, quantum theory is completely described by the purification principle. 
To emphasize the special role of the sixth principle we will call it \emph{postulate}, in analogy with
the parallel postulate of Euclidean geometry.

\subsection{Axioms}

\subsubsection{Causality}

The first axiom of our list, \emph{causality} \cite{purification}, is so basic that could
be considered as part of the background framework. We decided to explicitly present it as an axiom for two reasons: The first reason is that the
framework of operational-probabilistic theories can be developed even
without this requirement (see Ref.\cite{purification} for the general
framework and Refs.  \cite{supermaps, comblong}  for two explicit
examples of non-causal theories).  The second reason is that we want
to stress that causality is an essential ingredient in our derivation.
This observation is important in view of possible extensions of
quantum theory to quantum gravity scenarios where the causal structure
is not defined from the start (see e.g. Hardy in Ref. \cite{causaloid}).

\begin{axiom}[Causality]\label{ax:caus}  
The probability of  preparations is independent  of the choice of observations.    
\end{axiom}  
In technical terms:  if  $\{\rho_i\}_{i \in \rX} \subset \Stset(\rA)$ is a preparation-test, then the conditional probability of the preparation $\rho_i$  given the choice of the 
  observation-test $\{a_j\}_{j \in \rY}$    is the marginal
  \begin{equation*} 
  p \left(i |  \{a_j\}  \right) := \sum_{j\in\rY} \SC {a_j} {\rho_i} 
  \end{equation*}
  The axiom states that the marginal probability $ p \left(i |  \{a_j\}  \right) $ is independent of the choice of the observation-test $\{a_j\}$: if $\{a_j\}_{j \in \rY}$ and
  $\{b_k\}_{k \in \rZ}$ are two different observation-tests, then one has $p(i| \{a_j\}) = p(i|
  \{b_k\})$. Loosely speaking, one may refer to causality as a requirement of \emph{no-signalling
    from the future}: indeed, causality is equivalent to the fact that the probability of an outcome at a certain time
  does not depend on the choice of operations that will be done at later times \cite{maurolast}.

  An operational-probabilistic theory that satisfies the causality axiom \ref{ax:caus} will be
  called \emph{causal}.  As we already mentioned, causality is a very basic requirement and could be
  considered as part of the framework: it provides the notions used to state the other axioms and it
  implies several facts that will be used frequently in the paper.  In fact, in our derivation we do
  not use the causality axiom directly, but only through its consequences.  In the following we briefly summarize the facts and the
  notations that characterize the framework of \emph{causal operational-probabilistic theories},  introduced and discussed in detail in Ref.
\cite{purification}.   Similar structures have been subsequently considered in Refs. \cite{duotenz,caucat} within a formal description of  circuits in foliable spacetime regions. 

  First, causality is equivalent to the existence of an effect $e_\rA$ such that $ e_\rA = \sum_{j
    \in \rX} a_j$ for every observation-test $\{a_j\}_{j \in \rY}$. We call the effect $e_\rA$ the
  \emph{deterministic effect} for system $\rA$.  By definition, the effect $e_\rA$ is unique.  The
  subindex $\rA$ in $e_\rA$ will be dropped when no confusion can arise.

  In a causal theory every test $\{\tC_i\}_{i \in \rX} \subset \Trnset(\rA,\rB)$ satisfies the
  condition
\begin{equation*}
\sum_{i \in \rX} \B {e_\rB}  \tC_i  = \B {e_\rA}.
\end{equation*}  
As a consequence, a transformation $\tC \in \Trnset (\rA, \rB)$ satisfies the condition 
\begin{equation}\label{channel normalization}
\B {e_\rB}  \tC  \le \B {e_\rA},
\end{equation}  
with the equality if and only if $\tC$ is a channel (i.e. a deterministic transformation,
corresponding to a single-outcome test). In Eq. (\ref{channel normalization}) we used the notation
$\B a \le \B {a'}$ to mean $\SC a \rho \le \SC {a'} \rho$ for every $\rho \in \Stset(\rA)$.

In a causal theory the norm of a state $\rho_i \in\Stset (\rA)$ is given by $|\!| \rho_i |\!| = \SC
e {\rho_i}$.  Accordingly, one can define the normalized state
\begin{equation*}\bar \rho_i  :=  \frac{\rho_i} {\SC e {\rho_i}}.
\end{equation*}     
In a causal theory one can always allow for \emph{rescaled preparations}: conditionally to the
outcome $i \in \rX$ in the preparation-test $\{\rho_i\}_{i \in \rX}$ we can say that we prepared the
normalized state $\bar \rho_i$. For this reason, every state in a causal theory is proportional to a
normalized state.

The set of normalized states will be denoted by $\Stset_1 (\rA)$. Since the set of all states
$\Stset(\rA)$ is closed in the operational norm, also the set of normalized states $\Stset_1 (\rA)$
is closed.  Moreover, the set $\Stset_1(\rA)$ is convex \cite{purification}: this means that for
every pair of normalized states $\rho_1, \rho_2 \in\Stset_1 (\rA)$ and for every probability $p \in
[0,1]$ the convex combination $\rho_p = p \rho_1 + (1-p) \rho_2$ is a normalized state.
Operationally, the state $\rho_p$ is obtained by
 \begin{enumerate}
 \item performing a binary test with outcomes $\{1,2\}$ and outcome
   probabilities $p_1= p$ and $p_2 =1-p$
 \item for outcome $i$ preparing $\rho_i$, thus realizing the preparation-test  $\{ p_i \rho_i \}_{i = 1,2}$
 \item coarse-graining over the outcomes, thus obtaining $\rho_p = p \rho_1 + (1-p)  \rho_2$.  
\end{enumerate}
The step 2 (preparation of a state conditionally on the outcome of a previous test) is possible
because the theory is causal \cite{purification}.

The pure normalized states are the extreme points of the convex set $\Stset_1 (\rA)$.  
For a normalized state $\rho \in\Stset_1 (\rA)$ we define the \emph{face identified by $\rho$} as follows:  
\begin{definition}[Face identified by a state]
The \emph{face identified by}  $\rho \in\Stset_1 (\rA)$  is the set $F_\rho$ of all normalized states $\sigma \in \Stset_1 (\rA)$ such that $\rho = p \sigma +
(1-p) \tau$, for some non-zero probability $p>0$ and some normalized state $\tau \in\Stset_1 (\rA)$.
\end{definition}
In other words, $F_\rho$ is the set of all normalized states that show up in the convex
decompositions of $\rho$.  Clearly, if $\varphi$ is a pure state, then one has $F_\varphi = \{\varphi\}$.
The opposite situation is that of completely mixed states: by definition \ref{def:complemix}, a state $\omega \in\Stset_1 (\rA)$
is completely mixed if every state $\sigma \in\Stset_1 (\rA)$ can stay in its convex
decomposition, that is, if $F_\omega = \Stset_1 (\rA)$.  An equivalent condition for a state to be
completely mixed is the following:
\begin{lemma}\label{lem:completely mixed spanning set}
A state $\omega  \in\Stset_1(\rA)$ is completely mixed if and only if  $\Span(F_\omega) =  \Stset_\Reals (\rA)$. 
\end{lemma}
\Proof The condition is clearly necessary.  It is also sufficient because for a state
$\sigma\in\Stset_1 (\rA)$ the relation $\sigma \in \Span (F_\omega)$ implies $\sigma \in F_\omega$
(see Lemma 16 of Ref.\cite{purification}). \qed

A completely mixed state can never be distinguished from another state with zero error probability: 
\begin{proposition}
  Let $\rho \in \Stset_1(\rA)$ be a completely mixed state and $\sigma \in\Stset_1 (\rA)$ be an
  arbitrary state.  Then, the probability of error in distinguishing $\rho$ from $\sigma$ is
  strictly greater than zero.
  \label{prop: no discrimination from completely mixed state}
\end{proposition}
\Proof By contradiction, suppose that one can distinguish between $\rho$ and $\sigma$ with zero
error probability.  This means that there exists a binary test $\{a_\rho, a_\sigma\}$ such that $
\SC {a_\rho } \sigma = \SC{a_\sigma} \rho = 0 $.  Since $\rho$ is completely mixed there exists a
probability $p>0$ and a state $\tau\in\Stset_1 (\rA)$ such that $\rho = p \sigma + (1-p) \tau$.
Hence, the condition $\SC {a_\sigma} \rho = 0$ implies $\SC {a_\sigma} \sigma= 0$.  Therefore, we
have $\SC {a_\rho} \sigma + \SC{a_\sigma} \sigma =0$.  This is in contradiction with the
normalization of the probabilities in the test $\{a_\rho, a_\sigma\}$, which would require $\SC
{a_\rho} \sigma + \SC {a_\sigma} \sigma = 1$.  \qed

\subsubsection{Perfect distinguishability}

Our second axiom regards the task of state discrimination.    As we saw in proposition \ref{prop: no
  discrimination from completely mixed state}, if a state is completely mixed, then it is impossible
to distinguish it perfectly from any other state.  Axiom \ref{norestr} states the converse:
\begin{axiom}[Perfect distinguishability]
Every state that is not completely mixed can be perfectly distinguished from some other state.   
\label{norestr}
\end{axiom}

Note that the statement of axiom \ref{norestr} holds for quantum and for classical information theory. In quantum theory a
completely mixed state is a density matrix with full rank.  If a density matrix $\rho$ has not full
rank, then it must have a kernel: hence, every density matrix $\sigma$ with support in the kernel of
$\rho$ will be perfectly distinguishable from $\rho$, as stated in Axiom \ref{norestr}. Applying the
same reasoning for density matrices that are diagonal in a given basis, one can easily see that
Axiom \ref{norestr} is satisfied also by classical information theory.

To the best of our knowledge, the perfect distinguishability property is has never been considered in the literature as an axiom, probably because in most works it came for free as a consequence of stronger mathematical assumptions.  For example, one can obtain the perfect distinguishability property from the \emph{no-restriction hypothesis} of Ref. \cite{purification}, stating that for every system $\rA$ any binary probability rule (i.e. any pair of positive functionals $a_0, a_1  \in \Cntset_\Reals(\rA)$ such that $a_0 + a_1 = e_\rA$)  actually describes a measurement allowed by the theory.   This assumption was made e.g. in Ref. \cite{Mas10} in the case of  systems with at most two distinguishable states (see requirement 5 of Ref. \cite{Mas10}).  Note that the difference between the perfect distinguishability Axiom and the no-restriction hypothesis is that the former can be expressed in purely operational terms, whereas the latter requires the notion of ``positive functional" which is not part of the basic operational language.  

\subsubsection{Ideal compression}

The third axiom is about information compression.  An \emph{information source} for system $\rA$ is
a preparation-test $\{\rho_i\}_{i \in \rX}$, where each $\rho_i\in\Stset(\rA)$ is an unnormalized
state and $\sum_{i \in \rX} \SC e {\rho_i} =1$.  A compression scheme is given by an \emph{encoding
  operation} $\tE $ from $\rA$ to a \emph{smaller system} $\rC$, that is, to a system $\rC$ such
that $D_\rC \le D_\rA$.  The compression scheme is \emph{lossless for the source} $\{\rho_i\}_{i \in
  \rX}$ if there exists a \emph{decoding operation} $\tD$ from $\rC$ to $\rA$ such that $\tD \tE
\K{\rho_i} = \K{\rho_i}$ for every value of the index $i \in \rX$.  This means that the decoding
allows one to perfectly retrieve the states $\{\rho_i\}_{i \in\rX}$.  We say that a compression
scheme is \emph{lossless} for the state $\rho$, if it is lossless for every source $\{\rho_i\}_{i
  \in \rX}$ such that $\rho = \sum_{i \in \rX} \rho_i$.  Equivalently, this means that the
restriction of $\tD \tE $ to the face identified by $\rho$ is equal to the identity channel: $\tD\tE
\K \sigma = \sigma$ for every $\sigma \in F_\rho$.

A lossless compression scheme is \emph{maximally efficient} if the encoding system $\rC$ has the
smallest possible size, that is, if the system $\rC$ has no more states than exactly those needed to
compress $\rho$. This happens when every normalized state $\tau \in\Stset_1 (\rC)$ comes from the
encoding of some normalized state $\sigma \in F_\rho$, namely $\K{\tau} = \tE \K{\sigma}$.
 
We say that a compression scheme that is lossless and maximally efficient is \emph{ideal}.  Our
second axiom states that ideal compression is always possible:
\begin{axiom}[Ideal compression]\label{compression}
For every state there exists an ideal compression scheme.  
\end{axiom}
It is easy to see that this statement holds in quantum theory and in classical probability theory.
For example, if  $\rho$ is a density matrix on a $d$-dimensional Hilbert space and
$\mathsf{rank}(\rho) =r$, then the ideal compression is obtained by just encoding $\rho$ in an
$r$-dimensional Hilbert space.  As long as we do not tolerate losses, this is the most efficient one-shot
compression we can devise in quantum theory.  Similar observations hold for classical
information theory.

\subsubsection{Local distinguishability}
The fourth axiom consists in the assumption of local
distinguishability,  here presented in
the formulation of Ref. \cite{purification}.
\begin{axiom}{\bf (Local distinguishability)}
  If two bipartite states are different, then they give different probabilities for at least one
  product experiment. \label{locdisc}
\end{axiom}
In more technical terms: if $\rho, \sigma \in \Stset_1 (\rA \rB)$ are states and $\rho \not =
\sigma$, then there are two effects $a \in \Cntset (\rA)$ and $b \in \Cntset (\rB)$ such that
\begin{equation*}
\begin{aligned} \Qcircuit @C=1em @R=.7em @! R {\multiprepareC{1}{\rho}& \qw \poloFantasmaCn \rA &\measureD a \\
\pureghost\rho & \qw \poloFantasmaCn \rB &\measureD b} \end{aligned}
~\not =~
\begin{aligned}
 \Qcircuit @C=1em @R=.7em @! R {\multiprepareC{1}{\sigma}& \qw \poloFantasmaCn \rA &\measureD a \\
\pureghost\sigma & \qw \poloFantasmaCn \rB &\measureD b} 
\end{aligned}
\end{equation*}

Local distinguishability is equivalent to the fact that two distant parties, holding systems $\rA$
and $\rB$, respectively, can distinguish between the two states $\rho, \sigma \in \Stset_1 (\rA\rB)$
using only local operations and classical communication and achieving an error probability strictly
larger than $p_{ran} = 1/2$, the probability of error in random guess \cite{purification}.  Again,
this statement holds in ordinary quantum theory (on complex Hilbert spaces) and in classical
information theory.

Another equivalent condition to local distinguishability is the  \emph{local tomography axiom}, introduced in Refs.  \cite{maurofirst,barrett}.  The local tomography axiom
state that every bipartite state can be reconstructed from the statistics of local measurements on the component systems.  Technically, local tomography is in turn equivalent to the relation  $D_{\rA\rB}=D_\rA D_\rB$  \cite{Har01} and to the fact that every state $\rho \in\Stset(\rA\rB)$ can be written as
\begin{equation*}
\rho  = \sum_{i=1}^{D_\rA}  \sum_{j=1}^{D_\rB}  \rho_{ij}  ~ \alpha_i \otimes \beta_j,
\end{equation*}
where $\{\alpha_i\}_{i =1}^{D_\rA}$  ($\{\beta_j\}_{j=1}^{D_\rB}$) is a basis for the vector space $\Stset_\Reals (\rA) $  ($\Stset_\Reals (\rB)$). 
The analog condition also holds for effects: every bipartite effect $E \in\Cntset (\rA\rB)$ ben be written as 
\begin{equation*}
E= \sum_{i=1}^{D_\rA}  \sum_{j=1}^{D_\rB}  E_{ij}  ~ a_i \otimes b_j,
\end{equation*}
where $\{a_i\}_{i =1}^{D_\rA}$  ($\{b_j\}_{j=1}^{D_\rB}$) is a basis for the vector space $\Cntset_\Reals (\rA) $  ($\Cntset_\Reals (\rB)$). 
  
An important consequence of local distinguishability, observed in Ref. \cite{purification}, is that a transformation $\tC \in
\Trnset(\rA\rB)$ is completely specified by its action on $\Stset (\rA)$: thanks to local
distinguishability we have the implication
\begin{equation}\label{trasformazioni con local discr}
\tC  \K \rho = \tC' \K{\rho}  \quad \forall  \rho \in\Stset (\rA)  \Longrightarrow  \tC = \tC'.
\end{equation} 
(see Lemma 14 of Ref.\cite{purification} for the proof).  Note that Eq.
(\ref{trasformazioni con local discr}) does not hold for quantum theory on real Hilbert spaces \cite{purification}.

\subsubsection{Pure conditioning}

The fourth axiom states how the outcomes of a measurement on one side of a pure bipartite state can
induce pure states on the other side.  In this case we consider \emph{atomic measurements}, that is,
measurements described by observation-tests $\{a_i\}_{i \in \rX}$ where each effect $a_i$ is atomic. Intuitively, atomic measurement are those with maximum ``resolving power". 

\begin{axiom}[Pure conditioning]
  If a bipartite system is in a pure state, then each outcome of an atomic measurement on one side
  induces a pure state on the other.  \label{purecond}
\end{axiom}

The pure conditioning property holds in quantum theory and in classical information theory as well.
In fact, the statement is trivial in classical information theory,
because the only pure bipartite states are the product of pure states:
no matter which measurement is performed on one side, the remaining
state on the other side will necessarily be pure.

The pure conditioning property, as formulated above, has been recently introduced in Ref. \cite{pcond}. 
A stronger version of axiom 5 is the \emph{atomicity of composition} introduced in Ref.
\cite{maurolast}:
\begin{enumerate} \item[5'] \emph{Atomicity of composition}: the
  sequential composition of two atomic operations is atomic. 
\end{enumerate}
Since pure states and atomic effects are a particular case of atomic transformations, Axiom 5' implies Axiom 5.
In our derivation, however, also the converse implication holds: indeed, thanks to the purification postulate we will be able to show that Axiom 5 implies Axiom 5' (see lemma \ref{lem:atomicity}).

\subsection{The purification postulate}

The last postulate in our list is the purification postulate, which
was introduced and explored in detail in Ref.  \cite{purification}.  While the
previous axioms were also satisfied by classical probability theory,
the purification axiom introduces in our derivation the genuinely
quantum features.  A \emph{purification} of the state $\rho \in
\Stset_1 (\rA)$ is a pure state $\Psi_\rho$ of some composite system
$\rA\rB$, with the property that $\rho$ is the marginal of $\Psi_\rho$, that is,
 \begin{equation*}
\begin{aligned} \Qcircuit @C=1em @R=.7em @! R {\prepareC{\rho}& \qw \poloFantasmaCn \rA &\qw } \end{aligned}
~=~
\begin{aligned}
\Qcircuit @C=1em @R=.7em @! R {\multiprepareC{1}{\Psi_\rho}& \qw \poloFantasmaCn \rA &\qw \\
\pureghost{\Psi_\rho} & \qw \poloFantasmaCn \rB &\measureD {e_\rB}} 
\end{aligned}
\end{equation*}

Here we refer to the system $\rB$ as the \emph{purifying system}.  The
purification axiom states that every state can be obtained as the
marginal of a pure bipartite state in an essentially unique way:
\begin{postulate}[Purification]
  Every state has a purification.  For fixed purifying system, every two purifications of the same
  state are connected by a reversible transformation on the purifying system.
\label{purification}
\end{postulate}
Informally speaking, our postulate states that the ignorance about a
part is always compatible with a maximal knowledge of the whole.  The
existence of pure bipartite states with mixed marginal was already
recognized by Schr\"odinger as the characteristic trait of quantum
theory \cite{Schr35}.  Here, however, we also emphasize the importance
of the uniqueness of purification up to reversible transformations:
this property sets up a relation between pure states and reversible
transformations that generates most of the structure of quantum
theory.  As shown in Ref. \cite{purification}, an impressive number of
quantum features are actually direct consequences of purification. In
particular, purification implies the possibility of simulating any
irreversible process through a reversible interaction of the system
with an environment that is finally discarded.

\section{First consequences of the principles}\label{sec:consequences}

\subsection{Results about ideal compression}  
Let $\rho \in\Stset_1 (\rA)$ be a state and let $\tE \in\Trnset(\rA,\rC)$ (resp.
$\tD\in\Trnset(\rC,\rA)$) be its encoding (resp. decoding) in the ideal compression scheme of Axiom
\ref{compression}.

Essentially, the encoding operation $\tE\in\Trnset (\rA,\rC)$ identifies the face $F_\rho$ with the
state space $\Stset_1 (\rC)$.  In the following we provide a list of elementary lemmas showing that all
statements about $F_\rho$ can be translated into statements about $\Stset_1(\rC)$ and vice-versa.


\begin{lemma}
  The composition of decoding and encoding is the identity on $\rC$, namely $ \tE  \tD  =   \tI_\rC$.
  \label{lem:EDeqI}
\end{lemma}

\Proof Since the compression is maximally efficient, for every state $\tau\in\Stset_1(\rC)$ there is
a state $\sigma\in F_\rho$ such that $\tE \sigma=\tau$. Using the fact that $\tD\tE \sigma = \sigma
$ (the compression is lossless) we then obtain $\tE\tD\tau=\tE\tD\tE\sigma=\tE\sigma=\tau$. By local
distinguishability [see Eq. (\ref{trasformazioni con local discr})], this implies $\tE\tD =
\tI_\rC$. \qed

\begin{lemma}
  The image of $\Stset_1(\rC)$ under the decoding operation $\tD $ is $F_\rho$.
\label{lem:rangedec}
\end{lemma}

\Proof Since the compression is maximally efficient, for all
$\tau\in\Stset_1(\rC)$ there exists $\sigma\in F_\rho$ such that
$\tau=\tE\sigma$. Then, $\tD\tau=\tD\tE\sigma=\sigma$. This implies
that $\tD (\Stset_1(\rC)) \subseteq F_\rho$. On the other hand, since
the compression is lossless, for every state $\sigma \in F_\rho$ one
has $\tD \tE \sigma = \sigma $. This implies the inclusion $F_\rho
\subseteq \tD (\Stset_1 (\rC))$.  \qed

\begin{lemma}
  If the state $\varphi \in F_\rho$ is pure, then the state $\tE
  \varphi \in\Stset_1 (\rC)$ is pure.  If the state $\psi\in\Stset_1
  (\rC)$ is pure, then the state $\tD \psi \in F_\rho$ is pure.
  \label{lem:pureenc}
\end{lemma}

\Proof Suppose that $\varphi \in F_\rho$ is pure and that $\tE
\varphi$ can be written as $\tE \varphi = p \sigma + (1-p) \tau$ for
some $p>0$ and some $\sigma, \tau \in\Stset_1 (\rC)$.  Applying $\tD$
on both sides we obtain $\varphi = p \tD \sigma + (1-p) \tD \tau$.
Since $\varphi$ is pure we must have $\tD \sigma = \tD \tau =
\varphi$.  Now, applying $\tE$ on all terms of the equality and using
lemma \ref{lem:EDeqI} we obtain $\sigma = \tau = \tE \varphi $. This
proves that $\tE\varphi$ is pure.  Conversely, suppose that
$\psi\in\Stset_1(\rC)$ is pure and $\tD \psi = p\sigma + (1-p) \tau$
for some $p>0$ and some $\sigma, \tau\in \Stset_1 (\rA)$. Since $\tD
\psi$ is in the face $F_\rho$ (lemma \ref{lem:rangedec}), also
$\sigma$ and $\tau$ are in the same face.  Applying $\tE$ on both
sides of the equality $\tD \psi = p\sigma + (1-p) \tau$ and using
lemma \ref{lem:EDeqI} we obtain $\psi = \tE \tD \psi = p \tE\sigma
+(1-p) \tE \tau$. Since $\psi$ is pure we must have $\tE \sigma = \tE
\tau = \psi$. Applying $\tD$ on all terms of the equality we then have
$\sigma = \tau = \tD \psi$, thus proving that $\tD \psi$ is pure.
\qed

We say that a state $\sigma \in F_\rho$ is \emph{completely mixed relative to the face $F_\rho$} if
every state $\tau \in F_\rho$ can stay in the convex decomposition of $\sigma$.  In other words,
$\sigma$ is completely mixed relative to $F_\rho$ if one has $F_\sigma = F_\rho$.  Note that in
general $\sigma \in F_\rho$ implies $F_\sigma \subseteq F_\rho$.

We then have the following:
\begin{lemma}\label{lem:encrhointernal}
If the state $\omega \in F_\rho$ is completely mixed relative to $F_\rho$, then the state $\tE \omega \in \Stset_1(\rC)$ is completely mixed.  
If the state $\upsilon \in \Stset_1(\rC)$ is completely mixed, then the state $\tD\upsilon \in F_\rho$ is completely mixed relative to $F_\rho$. 
\end{lemma}
\Proof Suppose that $\omega$ is completely mixed relative to $F_\rho$.
Then every state $\sigma \in F_\rho$ can stay in its convex
decomposition, say $\omega =p \sigma + (1-p) \sigma'$ with $p>0$ and
$\sigma'\in F_\rho$.  Applying $\tE$ we have 
\begin{equation}\label{eq:pedantissimo} 
\tE \omega = p \tE
\sigma + (1-p) \tE \sigma'.
\end{equation} 
Since the compression is maximally efficient, for every state $\tau\in\Stset_1(\rC)$ there exists a state
$\sigma\in F_\rho$ such that $\tau=\tE\sigma$. Choosing the suitable $\sigma \in F_\rho$ and substituting $\tau$ to $\tE \sigma$ in   Eq. (\ref{eq:pedantissimo})  we then obtain that for every state $\tau \in \Stset_1 (\rC)$  there exists probability $p>0$ and a state $\sigma' \in F_\rho$ such that  
\begin{equation*} \tE \omega = p \tau + (1-p)
\tE \sigma' \,  .
\end{equation*}   This implies that
$\tE \omega$ is completely mixed. Suppose now that $\upsilon\in
\Stset_1(\rC)$ is completely mixed. Then every state $\tau \in\Stset_1
(\rC)$ can stay in its convex decomposition, say $\upsilon = p \tau +
(1-p) \tau'$.  with $p>0$ and $\tau'\in\Stset_1 (\rC)$. Applying $\tD$
on both sides we have 
\begin{equation}\label{eq:pedantissimo2}
\tD \upsilon = p \tD \tau + (1-p)\tD \tau'.
\end{equation}
Now, using lemma \ref{lem:rangedec} we have that every state $\sigma
\in F_\rho$ can be written as $\sigma =\tD \tau$ for some $\tau \in
\Stset_1 (\rC)$.  Choosing the  suitable $\tau \in \Stset_1 (\rC) $ and substituting $\sigma$ to $\tD \tau$ in Eq. (\ref{eq:pedantissimo2}) we then obtain that for evert state $\sigma\in F_\rho$ there exists a probability $p>0$ and a state $\tau' \in\Stset_1 (\rC)$ such that $\tD \upsilon = p \sigma+ (1-p)\tD \tau'$.     Therefore, $\tD \upsilon$ is completely mixed
relative to $F_\rho$.  \qed

We now show that the system $\rC$ used for ideal compression of the state $\rho$ is unique up to operational equivalence:

\begin{lemma}
  If two systems $\rC$ and $\rC'$ allow for
  ideal compression of a state $\rho\in\Stset_1(\rA)$, then $\rC$ and $\rC'$ are
  operationally equivalent.
  \label{lem:equiva}
\end{lemma}

\Proof Let $\tE, \tD$ and $\tE', \tD'$ denote the encoding/decoding schemes for systems $\rC$ and
$\rC'$, respectively.  Define the transformations $\tU := \tE'\tD \in\Trnset (\rC, \rC')$ and $\tV =
\tE \tD' \in\Trnset (\rC',\rC)$.  It is easy to see that $\tU$ is reversible and $\tU^{-1} = \tV$.
Indeed, since the restriction of $\tD'\tE'$ and $\tD\tE$ to the face $F_\rho$ is the identity, using
Lemma \ref{lem:rangedec} one has $\tD'\tE'\tD=\tD$ and similarly $\tD\tE\tD'=\tD'$. Hence, we have
$\tU\tV = \tE' \tD \tE\tD' =\tE'\tD' =\tI_{\rC'}$ and $\tV\tU=\tE\tD' \tE'\tD = \tE\tD =
\tI_{\rC}$.\qed

It is useful to introduce the notion of \emph{equality upon input of $\rho$}. We say that two
transformations $\tA , \tA' \in \Trnset (\rA,\rB)$ are \emph{equal upon input of} $\rho \in\Stset
(\rA)$ if their restrictions to the face identified by $\rho$ are equal, that is, if $\tA \sigma =
\tA' \sigma$ for every $\sigma\in F_\rho$.  If $\tA$ and $\tA'$ are equal upon input of $\rho$ we
write $\tA =_\rho \tA'$.

Using the notion of equality upon input of $\rho$ we can rephrase the fact that the compression is
lossless for $\rho$ as $\tD\tE =_\rho \tI_\rA$.  Similarly, we can state the following:
\begin{lemma}\label{lem:Edetuponinput}
The encoding $\tE$ is deterministic upon input of $\rho$, that is $\B {e_\rC}  \tE  =_\rho \B {e_\rA} $. 
\end{lemma}
\Proof For every $\sigma \in F_\rho$ we have $\B {e_\rC} \tE \K\sigma \ge\B {e_\rA} \tD\tE \K \sigma
= \SC {e_\rA} \sigma = 1$, having used Eq. (\ref{channel normalization}) and the fact that the
compression is lossless.  Since probabilities are bounded by 1, this implies $\B {e_\rC} \tE \K
\sigma = \SC {e_\rA} \sigma$ for every $\sigma\in F_\rho$, that is, $\B {e_\rC} \tE =_\rho \B{ e_\rA}$. \qed 
A similar result holds for the decoding:
\begin{lemma}
The decoding $\tD$ is deterministic, that is  $\B {e_\rA}  \tD  = \B {e_\rC} $. 
\end{lemma}

\Proof For every $\tau\in\Stset_1(\rA)$ we have $\B {e_\rA} \tD \K \tau \ge \B {e_\rC} \tE \tD \K
\tau = \SC{e_\rC} \tau$, having used Eq. (\ref{channel normalization}) and lemma \ref{lem:EDeqI}.
Hence, $\B {e_\rA} \tD = \B {e_\rC} $. \qed


\subsection{Results about purification}

The purification postulate \ref{purification} implies a large number of quantum features, as it was
shown in Ref. \cite{purification}.  Here we review only the facts that are useful for our
derivation, referring to Ref. \cite{purification} for the proofs.

An elementary consequence of the uniqueness of purification is that the group $\grp G_\rA$ of
reversible transformations on $\rA$ acts transitively on the set of pure states:
\begin{lemma}[Transitivity on pure states]\label{lem:trans}
  For every couple of pure states $\varphi, \varphi' \in\Stset_1 (\rA)$ there is a reversible
  transformation $\tU \in\grp G_\rA$ such that $\varphi' = \tU \varphi$.
\end{lemma} 
\Proof See Lemma 20 of Ref. \cite{purification}.  \qed 

Transitivity implies that for every system $\rA$ there is a unique state $\chi_\rA \in\Stset_1(\rA)$
that is invariant under reversible transformations, that is, a unique state such that $\tU \chi_\rA
= \chi_\rA$ for every $\tU \in \grp G_\rA$:
\begin{lemma}[Uniqueness of the  invariant state]\label{lem:uniquenessinvariant} 
  For every system $\rA$, there is a unique state $\chi_\rA$ invariant
  under all reversible transformations in $\grp G_\rA$. The invariant state has the following properties: 
  \begin{enumerate}
   \item $\chi_\rA$ is completely mixed
   \item $\chi_{\rA\rB}  = \chi_\rA \otimes \chi_\rB$.
   \end{enumerate}
\end{lemma}
\Proof See Corollary 34 and Theorem 4 of Ref. \cite{purification}.  The proof of item 2 uses  the local distinguishability axiom.   \qed
When there is no ambiguity we will drop the subindex $\rA$ and simply write $\chi$.  

The uniqueness of purification in postulate \ref{purification} requires that if $\Psi_\rho ,
\Psi_\rho' \in\Stset_1(\rA\rB)$ are two purifications of $\rho\in \Stset_1 (\rA)$, then there exists
a reversible transformation $\tU \in \grp G_\rB$ such that $\Psi_\rho' = (\tI_\rA \otimes \tU)
\Psi_\rho$.  The following lemma extends the uniqueness property to purifications with different
purifying systems:
\begin{lemma}\label{lem:purichan}{\bf (Uniqueness of the purification up to channels on the purifying systems)}
  Let $\Psi\in\Stset_1(\rA\rB)$ and $\Psi'\in\Stset_1(\rA\rC)$ be two purifications of $\rho \in
  \Stset_1 (\rA)$. Then there exists a channel $\tC \in \Trnset (\rB , \rC)$ such
  that\begin{equation*}
\begin{aligned}\Qcircuit @C=1em @R=.7em @! R {
\multiprepareC{1}{\Psi'}&\qw\poloFantasmaCn{\rA}&\qw\\
\pureghost{\Psi'}&\qw\poloFantasmaCn{\rC}&\qw} \end{aligned}
 ~=~
\begin{aligned}
\Qcircuit @C=1em @R=.7em @! R {
\multiprepareC{1}{\Psi}&\qw&\qw&\qw\poloFantasmaCn{\rA}&\qw\\
\pureghost{\Psi}&\qw\poloFantasmaCn{\rB}&\gate{\tC} &\qw \poloFantasmaCn \rC &\qw}
\end{aligned}
\end{equation*}
\end{lemma}
\Proof See Lemma 21  of Ref. \cite{purification}. \qed 

Another consequence of the uniqueness of purification is the fact that any ensemble decomposition of
a given mixed state can be obtained by performing a measurement on the purifying system:
\begin{lemma}{\bf (Purification of preparation-tests)} \label{lem:purificami-ens}  
Let $\rho\in\Stset_1 (\rA)$ be a state and $\Psi_\rho \in\Stset_1 (\rA\rB)$ be a purification of $\rho$. If $\{\rho_i\}_{i \in \rX}$ be a preparation-test such that $\sum_{i\in\rX}  \rho_i = \rho$, then there exists an observation-test $\{a_i\}_{i\in\rX}$ on the purifying system such that   \begin{align*}\label{purificami-ens}
 \begin{aligned} \Qcircuit @C=1em @R=.7em @! R {
    \prepareC{\rho_i}&\qw\poloFantasmaCn{\rA}&\qw}& \end{aligned}~=~ 
\begin{aligned}\Qcircuit @C=1em @R=.7em @! R {
    \multiprepareC{1}{\Psi_{\rho}}&\qw\poloFantasmaCn{\rA}&\qw&\\
    \pureghost{\Psi_{\rho}}&\qw\poloFantasmaCn{\rB}&\measureD{b_i}} 
\end{aligned}
\end{align*}
\end{lemma} 
\Proof See  lemma 8 of Ref. \cite{purification}. \qed

An easy consequence is the following: 
\begin{corollary}\label{cor:steerface}
  If $\Psi_\rho \in\Stset_1 (\rA\rB)$ is a purification of $\rho\in\Stset_1 (\rA)$ and $\sigma$
  belongs to the face $F_\rho$, then there exists an effect $b$ and a non-zero probability $p>0$
  such that
 \begin{align*}
 p \begin{aligned} \Qcircuit @C=1em @R=.7em @! R {
    \prepareC{\sigma}&\qw\poloFantasmaCn{\rA}&\qw}& \end{aligned}~=~ 
\begin{aligned}\Qcircuit @C=1em @R=.7em @! R {
    \multiprepareC{1}{\Psi_{\rho}}&\qw\poloFantasmaCn{\rA}&\qw&\\
    \pureghost{\Psi_{\rho}}&\qw\poloFantasmaCn{\rB}&\measureD{b}} 
\end{aligned}
\end{align*}
\end{corollary}
An important consequence of purification and local distinguishability is the relation between
equality upon input of $\rho $ and equality on the purifications of $\rho$:
\begin{theorem}\label{theo:uponinput}{\bf (Equality upon input of $\rho$ vs equality on purifications of $\rho$)}
  Let $\Psi \in \Stset_1(\rA \rC)$ be a purification of $\rho \in \Stset_1
  (\rA)$, and let $\tA, \tA' \in \Trnset (\rA,\rB)$ be two
  transformations. Then one has
\begin{equation*} 
  (\tA  \otimes \tI_\rC ) {\Psi_\rho} = (\tA'\otimes \tI_\rC ) \Psi_\rho   \qquad \Longleftrightarrow \qquad \tA =_{\rho}  \tA'~.
\end{equation*} 
\end{theorem}
\Proof See theorem 1 of Ref. \cite{purification}.  The proof of the direction $\Longleftarrow$  uses the local distinguishability axiom. \qed

As a consequence, the purification of a completely mixed state allows for the tomography of transformations:   
\begin{corollary}\label{cor:faithful1}  
  Let $\omega \in\Stset_1 (\rA)$ be completely mixed and $\Psi_\omega \in\Stset_1 (\rA\rC)$ is a
  purification of $\omega$. Then, for all transformations $\tA, \tA' \in\Trnset (\rA,\rB)$ one has
\begin{equation*}
(\tA \otimes \tI_\rC)  \Psi_\omega  = (\tA' \otimes \tI_\rC)  \Psi_\omega   \qquad \Longleftrightarrow \qquad  \tA  = \tA'. 
\end{equation*}
\end{corollary}
\Proof By theorem \ref{theo:uponinput} the first condition is equivalent to $\tA =_\omega \tA'$.
Since $\omega$ is completely mixed, this means $\tA \sigma = \tA' \sigma$ for every
$\sigma\in\Stset_1(\rA)$.  By local distinguishability [see Eq.  (\ref{trasformazioni con local
  discr})] this implies $\tA = \tA'$.  \qed

Corollary \ref{cor:faithful1} shows that the state $(\tA \otimes \tI_\rC) \Psi_\omega$ characterizes
the transformation $\tA$ completely.  We will express this fact by saying that the state
$\Psi_\omega$ is \emph{dynamically faithful} \cite{maurolast}, or just \emph{faithful}, for short.
Using this notion we can rephrase corollary \ref{cor:faithful1} as:
\begin{corollary}\label{cor:faithful}  
  If $\Psi \in\Stset_1 (\rA\rC)$ is pure and its marginal on system $\rA$ is completely mixed, then
  $\Psi$ is dynamically faithful for system $\rA$.
\end{corollary}

Let us choose a fixed faithful state for system $\rA$, say $\Psi\in\Stset_1(\rA\rC)$.  Then for
every transformation $\tC\in\Trnset (\rA,\rB)$ we can define the \emph{Choi state}
$R_\tC\in\Stset(\rB\rC)$ as
\begin{equation*}
\begin{aligned}  
\Qcircuit @C=1em @R=.7em @! R {
\multiprepareC{1}{R_{\tC}} &\qw \poloFantasmaCn{\rB} &\qw  \\
\pureghost{R_{\tC}} &\qw \poloFantasmaCn{\rC} &\qw}
\end{aligned} := 
\begin{aligned}  
\Qcircuit @C=1em @R=.7em @! R {
\multiprepareC{1}{\Psi} &\qw \poloFantasmaCn{\rA} &\gate \tC  & \qw \poloFantasmaCn{\rB}&\qw \\
\pureghost{\Psi} &\qw \poloFantasmaCn{\rC} &  \qw&\qw&\qw}
\end{aligned}
\end{equation*}

We then have the following: 
\begin{theorem}\label{theo:iso} {\bf (Choi isomorphism)}
  For a given faithful state $\Psi\in\Stset_1 (\rA\rC)$ the map $\tC \mapsto R_{\tC} := (\tC\otimes
  \tI_\rC) \Psi$ has the following properties:
\begin{enumerate}
\item it defines a bijective correspondence between tests $\{\tC_i \}_{i\in \rX}$ from $\rA$ to
  $\rB$ and collections of states $\{R_i\}_{i\in \rX}$ for $\rB \rC$ satisfying
 \begin{equation*} 
 \sum_{i \in \rX} \B{e}_{\rB} \K{R_{i}}_{\rB
      \rC} =  \B{e}_{\rA} \K{\Psi}_{\rA\rC}.
 \end{equation*}
\item the transformation $\tC$ is atomic if and only if the
 corresponding state $R_\tC$ is pure.
\end{enumerate}
\end{theorem}  
\Proof See Theorem 17 of Ref. \cite{purification}.  

A simple consequence of the Choi isomorphism is the following:
\begin{corollary}\label{cor:EVM=test}
  Let $\{\tC_i\}_{i\in \rX} \subset \Trnset(\rA,\rB)$ be a collection
  of transformations.  Then, $\{\tC_i\}_{i\in \rX}$ is a test if and
  only if
\begin{equation*}
\sum_{i\in \rX}  \B{e}_\rB  \tC_i =\B{ e}_\rA.
\end{equation*}
In particular, let $\{a_i\}_{i\in \rX} \subset \Cntset(\rA)$ be a collection of effects.  Then,
$\{a_i\}_{i\in \rX}$ is an observation-test if and only if
\begin{equation}\label{eq:normalizzazionepovm}
\sum_{i\in \rX}  \B{a_i}  =\B{ e}.
\end{equation}
\end{corollary}

\Proof Apply item 1 of theorem \ref{theo:iso} to the collection of states $\{R_i\}_{i \in \rX}$
defined by $R_i := (\tC_i \otimes \tI_\rC) {\Psi}$. \qed

A much deeper consequence of the Choi isomorphism is the following theorem:  
\begin{theorem}{\bf (States specify the theory)}\label{theo:statesspecify}
  Let $\Theta, \Theta'$ be two theories satisfying the purification postulate. If $\Theta$ and
  $\Theta'$ have the same sets of normalized states, then $\Theta ' = \Theta$.
\end{theorem} 
\Proof See Theorem 19 of Ref.\cite{purification}.\qed
Thanks to theorem \ref{theo:statesspecify} to derive quantum theory we will only need to prove that
our principles imply that for every system $\rA$ the normalized states $\Stset_1 (\rA)$ can be
described as positive Hermitian matrices with unit trace. Once this is proved, theorem
\ref{theo:statesspecify} automatically ensures that all the dynamics and all the measurements allowed
by the theory are exactly the dynamics and the measurements allowed in quantum theory.  

Note that in the definition of the Choi state we left the freedom to choose the
faithful state $\Psi \in\Stset_1 (\rA\rC)$.  Among many possibilities, one convenient choice is to
take a faithful state $\Phi \in\Stset_1 (\rA\rC)$ obtained as a purification of the invariant state
$\chi \in\Stset_1 (\rA)$. Moreover, as we will see in the next paragraph, we can always choose the
purifying system $\rC$ in such a way that the marginal on $\rC$ is completely
mixed.

\subsection{Results about the combination of compression and purification}

An important consequence of the combination of the purification postulate with the compression axiom
is the fact that one can always choose a purification of $\rho$ such that the marginal state on the
purifying system is completely mixed.  To prove this result we need the following lemma:

\begin{lemma}
  Let $\rho \in\Stset_1 (\rA)$ be a state and let $\Psi_\rho \in\Stset_1(\rA\rB)$ be a purification of
  $\rho$.  If $\tE\in\Trnset(\rA,\rC)$ is the encoding operation in the compression scheme of axiom
  \ref{compression}, then the state $\Psi_\rho' := (\tE\otimes\tI_\rB) \Psi_\rho$ is pure.
  \label{lem:pure}
\end{lemma}

\Proof Let $\tD \in \Trnset (\rC, \rA)$ be the decoding operation.  Since the compression is
lossless for $\rho$ we know that $\tD\tE =_\rho \tI_\rA$.  By theorem \ref{theo:uponinput} this is
equivalent to the condition $(\tD\tE\otimes\tI_\rB)\Psi_\rho=\Psi_\rho$.  Now, suppose that
$(\tE\otimes\tI_\rB )\Psi_\rho=\sum_{i\in\rX} \Gamma_i$. Applying $\tD$ on both sides we then obtain
$ \Psi_\rho = \sum_{i \in \rX} (\tD \otimes \tI_\rB) \Gamma_i $, and, since $\Psi_\rho$ is pure, for
every $i \in \rX$ we must have $(\tD\otimes\tI_\rB)\Gamma_i=p_i \Psi_\rho$, where $p_i\ge0$ is some
probability.  Finally, since $\tE\tD=\tI_{\rC}$ (lemma \ref{lem:EDeqI}), one has $\Gamma_i=p_i(
\tE\otimes\tI_\rB)\Psi_\rho$. Hence, $(\tE\otimes\tI_\rB)\Psi_\rho$ admits only decompositions with
$\Gamma_i=p_i (\tE\otimes\tI_\rB) \Psi_\rho$, that is, $(\tE\otimes\tI_\rB) \Psi_\rho$ is pure.\qed

 We are now in position to prove the desired result: 
\begin{theorem}
  For every state $\rho\in\Stset_1(\rA)$ there exists a system $\rC$ and a purification
  $\Psi_\rho\in\Stset_1(\rA \rC)$ of $\rho$ such that the marginal state on system $\rC$ is
  completely mixed. Moreover, the system $\rC$ is unique up to operational equivalence.
  \label{theo:conjusys}
\end{theorem}

\Proof Take an arbitrary purification of $\rho$, say $\Phi_\rho\in\Stset_1(\rA\rB)$ for some
purifying system $\rB$.  Define the marginal state on system $\rB$ as $|\theta)_\rB
:=(e|_{\rA}|\Phi_\rho)_{\rA\rB}$ and define the state $\Psi_\rho:= (\tI_\rA\otimes\tE) \Phi_\rho$,
where $\tE\in\Trnset (\rB, \rC)$ the encoding operation for state $\theta$.  By Lemma \ref{lem:pure}
we know that $\Psi_\rho\in\Stset(\rA\rC)$ is pure.  Using lemma \ref{lem:Edetuponinput} and theorem
\ref{theo:uponinput} we obtain $\B{e_\rC} \K{\Psi_\rho} = [ \B{ e_\rC} \tE ] \K{ \Phi_{\rho}} = \B
{e_\rB} \K{ \Phi_\rho} = \K \rho$, that is, $\Psi_\rho$ is a purification of $\rho$.  Finally, the
marginal on system $\rC$ is given by $\tilde \rho = \tE \theta$, which by Lemma
\ref{lem:encrhointernal} is completely mixed.  This proves the first part of the thesis. It remains
to show that the system $\rC$ is uniquely defined up to operational equivalence.  Suppose that
$\Psi_\rho' \in\Stset(\rA \rC')$ is another purification of $\rho$ with the property that the
marginal on system $\rC'$ is completely mixed.  Since $\Psi_\rho$ and $\Psi_\rho'$ are two
purifications of the same state, there must be two channels $\tC\in \Trnset(\rC, \rC')$ and $\tR
\in\Trnset( \rC', \rC)$ such that $\Psi'_{\rho} = (\tI_\rA \otimes \tC) \Psi_\rho$ and $\Psi_\rho =
(\tI_\rA \otimes \tR) \Psi_\rho'$ (lemma \ref{lem:purichan}).  Combining the two equalities one
obtains $\Psi_{\rho} = (\tI_\rA \otimes \tR \tC) \Psi_\rho$. Now, the marginal of $\Psi_\rho$ on
system $\rC$ is completely mixed, and this implies that $\Psi_\rho$ is faithful for system $\rC$
(corollary \ref{cor:faithful}).  Hence, we have $\tR \tC = \tI_{ \rC}$.  Repeating the same argument
for $\Psi_\rho'$ we obtain $\tC \tR = \tI_{\rC'}$.  Therefore, $\tC$ is reversible and $\tR =
\tC^{-1}$.  This proves that $\rC$ and $\rC'$ are operationally equivalent.  \qed


The following facts will also be useful

\begin{corollary}
  Let $\Psi_\rho\in\Stset_1(\rA\rB)$ be a purification of $\rho \in\Stset_1 (\rA)$ and let $\tE\in
  \Trnset(\rA, \rC)$ be the encoding for $\rho$.  Then, the state $(\tE\otimes\tI_\rB) \Psi_\rho
  \in\Stset_1(\rC\rB)$ is dynamically faithful for $\rC$.
 \label{cor:faithcomp}
\end{corollary}
\Proof The marginal of $(\tE \otimes \tI_\rB) \Psi_\rho$ on system $\rC$ is $\tE \rho$, which is completely mixed by
lemma \ref{lem:encrhointernal}. Hence, $(\tE \otimes \tI_\rB) \Psi_\rho$ is dynamically faithful by
corollary \ref{cor:faithful}. \qed

\begin{lemma}
  The decoding transformation $\tD \in \Trnset(\rC, \rA)$ in the ideal compression for $\rho\in\Stset_1(\rA)$ is atomic.
  \label{lem:atomicdec}
\end{lemma}

\Proof Let $\Psi_\rho  \in\Stset_1 (\rA\rB)$ be a purification of $\rho$, for some purifying system $\rB$.  Since $\tD\tE =_\rho \tI_\rA$ (the compression is lossless), we have
$(\tD\tE\otimes\tI_{\rB})|\Psi_\rho)=|\Psi_\rho)$ (theorem \ref{theo:uponinput}).  Now, by corollary
\ref{cor:faithcomp} $(\tE\otimes\tI_\rB)|\Psi_\rho)$ is faithful for $\rC$ and by lemma \ref{lem:pure} $(\tE\otimes\tI_\rB)|\Psi_\rho)$ is pure. Using the Choi  isomorphism with the faithful state $\Psi: = (\tE\otimes\tI_\rB)\Psi_\rho $ we then obtain
that $\tD$ is atomic. \qed


\subsection{Teleportation and the link product}

For every system $\rA$ one can choose a completely mixed state $\omega_{\rA}$ and a purification
$\Psi^{(\rA)} \in\Stset(\rA\tilde\rA)$ such that the marginal on system $\tilde \rA$ is completely
mixed (cf. theorem \ref{theo:conjusys}).
Any such purification allows for a probabilistic teleportation
scheme:
\begin{lemma}[Probabilistic teleportation]\label{lem:probtele} There exists an atomic effect
  $E^{(\rA)} \in \Cntset(\tilde
\rA \rA)$ and a non-zero probability $p_\rA$ such that
\begin{equation*}
\begin{aligned}  
\Qcircuit @C=1em @R=.7em @! R {
\multiprepareC{1}{\Psi^{(\rA)}} &\qw \poloFantasmaCn{\rA} &\qw \\
\pureghost{\Psi^{(\rA)}} &\qw \poloFantasmaCn{\tilde \rA} & \multimeasureD{1}{E^{(\rA)}} \\
   &  \qw  \poloFantasmaCn{\rA}  & \ghost{E^{(\rA)}}  }
\end{aligned}
 ~= ~ p_{\rA} ~ 
\begin{aligned}
\Qcircuit @C=1em @R=.7em @! R { &\qw  \poloFantasmaCn{\rA}  &  \gate{ \tI} &\qw
  \poloFantasmaCn{\rA}&\qw } 
\end{aligned}
\end{equation*}
and
\begin{equation*} 
 \begin{aligned} \Qcircuit @C=1em @R=.7em @! R {
     &\qw \poloFantasmaCn{\tilde \rA} & \multimeasureD{1}{E^{(\rA)}} \\
    \multiprepareC{1} {\Psi^{(\rA)}} &\qw  \poloFantasmaCn{\rA} &\ghost{E^{(\rA)}} \\
 \pureghost{\Psi^{(\rA)}} &\qw \poloFantasmaCn{\tilde \rA} &\qw }
\end{aligned}
~=  ~ p_{\rA} ~
\begin{aligned} \Qcircuit @C=1em @R=.7em @! R { & \poloFantasmaCn{\tilde \rA} \qw & \gate{ \tI} &\qw \poloFantasmaCn{\tilde \rA}&\qw }
\end{aligned}
\end{equation*}
   \end{lemma}
\Proof See Corollary 19 of Ref. \cite{purification}. \qed

Let us choose $\Psi^{(\rA)}$ to be the faithful state in the definition of the Choi
isomorphism. Then the sequential composition of transformation induces a composition of Choi  states in following way:

\begin{corollary}[Link product]
  For two transformations $\tC \in \Trnset (\rA , \rB)$ and $\tD\in \Trnset (\rB, \rC)$ the
  Choi state of $\tD\tC\in\Trnset(\rA,\rC)$ is given by the \emph{link product}
\begin{equation}\label{linkprod}
\begin{aligned}
  \Qcircuit @C=1em @R=.7em @! R { \multiprepareC{1} {R_{\tD \tC}} &\qw \poloFantasmaCn{\rC} &\qw\\
    \pureghost{R_{\tD\tC}} &\qw \poloFantasmaCn {\tilde \rA}&\qw}
\end{aligned}  = ~\frac 1 {p_{\rB}}~
\begin{aligned} 
\Qcircuit @C=1em @R=.7em @! R { \multiprepareC{1}{R_\tD} &\qw \poloFantasmaCn{\rC} &\qw \\
    \pureghost{R_\tD} &\qw \poloFantasmaCn{\tilde \rB} & \multimeasureD{1}{E^{(\rB)}} \\
    \multiprepareC{1}{R_\tC} &  \qw  \poloFantasmaCn{\rB}  &  \ghost{E^{(\rB)}}\\
    \pureghost{R_\tC} &\qw \poloFantasmaCn{\tilde \rA} & \qw }
\end{aligned}
 \end{equation}
\end{corollary}
\Proof See Corollary 22 of Ref. \cite{purification}. \qed

We conclude this paragraph with an important result that follows from the combination of the link
product structure with the pure conditioning axiom:

\begin{lemma}[Atomicity of composition]\label{lem:atomicity}
The composition of two atomic transformations is atomic.
\end{lemma}
\Proof Let $\tC\in\Trnset(\rA , \rB)$ and $\tD \in\Trnset(\rB,\rC)$ be two atomic transformations.
By the Choi isomorphism, the (unnormalized) states $R_{\tC}$ and $R_{\tD}$ are pure.
Since the teleportation effect $E^{(\rB)}$ in Eq. \eqref{linkprod} is atomic (lemma \ref{lem:probtele}), the pure conditioning
axiom \ref{purecond} implies the state $R_{\tD\tC}$ is pure.  By the Choi isomorphism
this means that $\tD\tC$ is atomic. \qed

\subsection{No information without disturbance}

We say that a test $\{\tC_i\}_{i \in \rX} \subset \Trnset(\rA)$ is \emph{non-disturbing upon input
  of $\rho$} if $\sum_{i \in \rX} \tC_i =_\rho \tI_\rA$.  If $\rho$ is completely mixed, we simply
say that the test is \emph{non-disturbing}.

A consequence of the purification postulate is the following ``no-information without disturbance"
result:
\begin{lemma}[No information without disturbance]\label{lem:nodisturbance}
A test $\{\tC_i\}_{i \in \rX}\subset \Trnset(\rA)$ is non-disturbing upon input of $\rho$ if and only if there is a set of probabilities $\{p_i\}_{i \in \rX}$ such that  $\tC_i =_\rho p_i \tI_\rA$ for every $i \in \rX$.
\end{lemma}
\Proof See Theorem 10 of Ref. \cite{purification}. \qed 

The no-information without disturbance result implies the following geometrical limitation  
\begin{corollary}\label{cor:nosegment}
For every system $\rA$ the convex set of states $\Stset_1 (\rA)$ is not a segment.
\end{corollary}
\Proof The proof is by contradiction. Suppose that for some system $\rA$ the set $\Stset_1 (\rA)$ is
a segment. The segment has only two pure states, say $\varphi_1$ and $\varphi_2$, and every other
state $\rho\in\Stset_1 (\rA)$ is completely mixed.  Then the distinguishability axiom \ref{norestr}
imposes that $\varphi_1$ and $\varphi_2$ are perfectly distinguishable.  Take the binary test
$\{a_1, a_2\}$ such that $\SC {a_i} {\varphi_j} = \delta_{ij}$ and define the ``measure-and-prepare"
test $\{\tC_1,\tC_2\}$ as $\tC_i = \K {\varphi_i} \B {a_i}$, $i=1,2$ (the possibility of preparing a state depending on the outcome of a previous measurement is guaranteed by causality \cite{purification}).  Since every state $\rho$ in
the segment can be written as convex combination of the two extreme points, we have that the test
$\{\tC_1, \tC_2\}$ is non-disturbing: $(\tC_1 + \tC_2) \rho = \rho$ for every $\rho$.  This is in
contradiction with lemma \ref{lem:nodisturbance} because $\tC_1$ and $\tC_2$ are not proportional to the identity. \qed

We know that no information can be extracted without disturbance. In the following we will prove a
result in the converse direction: if a measurement extracts no information, than it can be realized
in a non-disturbing fashion.  To show this result we first need the following
\begin{lemma}
 For every observation test $\{a_{i}\}_{i\in \rX}  \subset \Cntset(\rA)$ with finite outcome set $\rX$
 there is a system $\rC$ and a test $\{\tA_{i}\}_{i\in \rX}   \subset \Trnset(\rA,\rC)$ consisting of atomic transformations such that $\B
 {a_{i}} = \B {e_\rC}\tA_i $.
  \label{lem:atomrealiz}
\end{lemma}

\Proof Let $\K\Psi_{\rA\rB}$ be a pure faithful state for system $\rA$ and let $ \K {R_i}_\rB = \B {a_{i}}_\rA
\K\Psi_{\rA\rB}$ the Choi state of $a_{i}$.  Take a purification of $R_{i}$, say
$\K{\Psi_i}_{\rB\rC}$ for some purifying system $\rC$ \cite{singlepurifying}.  Then, by the
Choi isomorphism there is a test $\{\tA_i\}_{i\in \rX}$, with input $\rA$ and output $\rC$,   such that
\begin{equation*}
  \begin{aligned}
    \Qcircuit @C=1em @R=.7em @! R
    {& \multiprepareC{1}{\Psi_i}   & \qw\poloFantasmaCn{\rC}&\qw \\
    & \pureghost{\Psi_i}   & \qw \poloFantasmaCn{\rB} & \qw }
  \end{aligned}=
  \begin{aligned}
    \Qcircuit @C=1em @R=.7em @! R {&\multiprepareC{1}{\Psi}&\qw\poloFantasmaCn{\rA}&\gate {\tA_{i}} &  \qw \poloFantasmaCn{\rC} &\qw \\
      &\pureghost{\Psi}&\qw\poloFantasmaCn{\rB}& \qw &\qw&\qw}
  \end{aligned}
\end{equation*}
[see item 1 of theorem \ref{theo:iso}]
Moreover,   each transformation $\tA_{i}: \rA \to \rC$ is atomic (item 2 of theorem \ref{theo:iso}).    
Applying the deterministic effect $\B {e_\rC}$ on both sides we then obtain $\K{R_i}_\rB  =   \B {e_\rC} \K{\Psi_i}_{\rC\rA} = \B {e_\rC} \tA_i  \K{\Psi}_{\rA\rB}$.  
By definition of $R_{i}$, this implies $\B {a_{i}}_{\rA} \K{\Psi}_{\rA\rB}  = \B {e_\rC} \tA_{i}  \K{\Psi}_{\rA\rB} $, and, since $\Psi$ is dynamically faithful, $\B {a_{i}}_{\rA} = \B {e_\rC} \tA_i$.\qed

\begin{theorem}
  Let $\rho\in\Stset_1(\rA)$ be a state,  $a\in\Cntset(\rA)$ be an effect, and $\tA \in \Trnset(\rA,\rB)$ be an atomic transformation such that $\B a _\rA=  \B e_\rB  \tA$.  If $ \B a=_\rho p \B e$ for some $p\ge 0$, then there exists a channel
  $\tC \in\Trnset(\rB, \rA)$ such that $\tC \tA =_\rho p \tI_\rA$.
  \label{theo:utile ma senza nome}
\end{theorem}

\Proof Consider a purification of $\rho$, say $\Psi_\rho\in\Stset(\rA\rC)$, and define the state
$\Sigma\in\Stset_1 (\rB\rC)$ by $|\Sigma):=\frac1p (\tA\otimes\tI_\rC)|\Psi_\rho)$. By the atomicity of composition  \ref{lem:atomicity} the state $\Sigma$ is pure.  Moreover,  we have 
\begin{align*} \B {e_\rB} \K
\Sigma_{\rB\rC} &= \frac 1p (a|_\rA |\Psi_\rho)_{\rA\rB}\\
&=(e_\rA|  |\Psi_\rho)_{\rA\rC} , 
\end{align*}
having used theorem \ref{theo:uponinput} in the last equality. This implies
that $\Psi_\rho$ and $\Sigma$ are different purifications of the same mixed state on system $\rC$.
Then, by lemma \ref{lem:purichan} there exists a channel $\tC\in\Trnset(\rB,\rA)$ such that
$\K{\Psi_\rho} = (\tC\otimes\tI_\rC)|\Sigma)=\frac 1 p(\tC\tA\otimes\tI_\rC) |\Psi_\rho)$. By
theorem \ref{theo:uponinput}, the last equality implies $\tC \tA =_\rho p \tI_\rA$.  \qed

We now make a simple observation that combined with  Theorem \ref{theo:utile ma senza nome} will lead to  some interesting consequences: 
\begin{lemma}
  If $(a|\rho)=\|a\|$, then $a=_\rho
  \|a\|e$. Similarly, if $(a|\rho)=0$, then $a=_\rho0$.
  \label{lem:detuponrho}
\end{lemma}

\Proof By definition, $\sigma \in F_\rho$ iff there exists $p>0$ and $\tau \in \Stset_1(\rA)$ such
that $\rho=p\sigma+(1-p)\tau$. If $\SC a \rho = \| a\|$, then we have $\| a \| = p \SC a{\sigma} +
(1-p) \SC a {\tau} $. Since $\SC a {\sigma}$ and $\SC a {\tau}$ cannot be larger than $\| a\|$, the
only way to have the equality is to have $\SC a {\sigma} = \SC a{\tau} = \| a\|$.  By definition,
this amounts to say $a =_\rho \| a \| e$.  Similarly, if $\SC a \rho = 0$, one has $0 = p \SC
a{\sigma} + (1-p) \SC a{\tau}$, which is satisfied only if $\SC a {\sigma} = \SC{a}{\tau} =0$, that
is, if $a=_\rho 0$.  \qed

As  consequence, we have the following:

\begin{corollary}
  Let $\rho\in\Stset_1(\rA)$ be a state, $a\in\Cntset(\rA)$ be an effect, and  $\tA\in\Trnset(\rA,\rB)$ be an atomic transformation such that  $\B a_{\rA}=(e|_\rB\tA$. 
  If $(a|\rho)=1$, then $\tA$ is correctable upon input of $\rho$, that is, there exists a correction operation $\tC \in\Trnset(\rB, \rA)$ such that $\tC \tA =_\rho \tI_\rA$.
  \label{cor:revuponrho}
\end{corollary}

\Proof  If $\SC a \rho = 1$, then clearly $\| a\| = 1$.  Lemma \ref{lem:detuponrho} then implies $\B a =_\rho \B e$.  Applying theorem \ref{theo:utile ma senza nome} we finally obtain the thesis. \qed

\begin{corollary}
  Let $\rho\in\Stset_1(\rA)$ be a state, $a\in\Cntset(\rA)$ be an effect such that   $(a|\rho)=1$. Then there exists a transformation $\tC\in\Trnset(\rA)$ such that $\B a = \B e \tC$ and $\tC =_\rho  \tI$.
  \label{cor:se fa uno non disturba}
\end{corollary}

\Proof  Straightforward consequence of lemma \ref{lem:atomrealiz} and of corollary \ref{cor:revuponrho}. \qed

Finally, we say that an observation-test $\{a_i\}_{i \in \rX}$ is \emph{non-informative upon input
  of $\rho$} if we have $\B {a_i }=_\rho p_i ~\B e$ for every $i \in\rX$.  This means that the test
$\{a_i\}_{i \in \rX}$ is unable to distinguish the states in the face $F_\rho$.  As a consequence of
theorem \ref{theo:utile ma senza nome} we have the following ``no disturbance without information"
result:
\begin{corollary}[No disturbance without information]
  If the test $\{a_i\}_{i\in \rX}$ is non-informative upon input of
  $\rho$ then there is a test $\{ \tD_i\}_{i \in \rX}\subset
  \Trnset(\rA)$ that is non-disturbing upon input of $\rho$ and
  satisfies $\B e \tD_i = \B{a_i}$ for every $i \in \rX$.
\end{corollary}
\Proof By lemma \ref{lem:atomrealiz} there exists a test $\{\tA_i\} \subset \Trnset(\rA,\rB)$ such
that each transformation $\tA_i$ is atomic and $\B e \tA_i = \B{a_i}$.  By theorem \ref{theo:utile
  ma senza nome}, for each $\tA_i$ there is a correction channel $\tC_i$ such that $\tC_i \tA_i
=_\rho p_i \tI_\rA$.  Defining $\tD_i := \tC_i \tA_i$ we then obtain the thesis. \qed

\section{Perfectly distinguishable states}\label{sec:distinguishable}
In this section we prove some basic facts about perfectly distinguishable states.  Let us start from the definition:
\begin{definition}[Perfectly distinguishable states]
  The normalized states $\{\rho_i\}_{i=1}^N \subseteq \Stset_1 (\rA)$  are  {\em
    perfectly distinguishable} if there exists an observation-test
  $\{a_i\}_{i=1}^N$ such that
  $(a_j|\rho_i)=\delta_{ij}$. The observation-test
  $\{a_i\}_{i=1}^N$ is called {\em perfectly distinguishing}.
\end{definition}

From the distinguishability axiom \ref{norestr} it is clear that every nontrivial system has at
least two perfectly distinguishable states:
\begin{lemma}\label{lem:atleasttwo}   
For every nontrivial system $\rA$ there are at least two perfectly distinguishable states.
\end{lemma}
\Proof Let $\varphi$ be a pure state of $\rA$.  Obviously, $\varphi$ is not completely mixed (unless the system $\rA$ has only one state, that is, unless $\rA$ is trivial).  Hence, by axiom \ref{norestr} there exists at least a state $\sigma$ that is perfectly distinguishable from $\varphi$.\qed

An equivalent condition for perfect distinguishability is the following:

\begin{lemma}\label{lem:simpler}
  The states $\{\rho_i\}_{i=1}^N \subset \Stset_1 (\rA)$ are perfectly distinguishable if and only
  if there exists an observation-test $\{a_i\}_{i=1}^N$ such that $(a_i|\rho_i)=1$ for every $i$.
\end{lemma}

\Proof The condition $\SC {a_i} {\rho_i}  =1, \quad \forall i =1, \dots, N$  is clearly necessary. On the other hand, the condition $\SC {a_i} {\rho_i}  =1, \quad \forall i =1, \dots, N$ implies 
\begin{equation*}
  (a_i|\rho_i)= 1=\sum_{j=1}^N(a_j|\rho_i)=(a_i|\rho_i)+\sum_{i\neq j}(a_j|\rho_i).
\end{equation*}
Since all probabilities are non-negative, we must have $(a_j|\rho_i)=0$ for $i\neq j$, and therefore, $(a_j|\rho_i)=\delta_{ij}$. \qed

A very general fact about state discrimination is expressed by the following:
\begin{lemma}
 If $\rho$ is perfectly distinguishable from $\sigma$ and $\rho' $ (resp. $\sigma'$) belongs to the
 face identified by $\rho$ (resp.  $\sigma$), then $\rho'$ is perfectly distinguishable from
 $\sigma'$.
  \label{lem:inclusiondiscrimination}
\end{lemma}

\Proof Let $\{a, e- a\}$ be the binary observation-test that distinguishes perfectly between $\rho$ and $\sigma$.  By definition, $a\in\Cntset(\rA)$ is such that $(a|\rho)=1$ and $(a|\sigma)=0$.
Now, by lemma \ref{lem:detuponrho}, $(a|\rho')=1$ and $(a|\sigma')=0$ for all $\rho'\in F_\rho$ and
$\sigma'\in F_\sigma$.\qed


Thanks to purification and to the local distinguishability axiom \ref{locdisc}, we are also in position to show a much stronger result: 

\begin{lemma}\label{lem:refinement}
  Let $\{\rho_i\}_{i=1}^N \subset F_\rho$ and
  $\{\rho_j\}_{j=N+1}^{N+M} \subset F_\sigma$ be two sets of
  perfectly distinguishable states. 
   If $\rho$ is perfectly distinguishable from $\sigma$, then the states $\{\rho_i\}_{i=1}^{N+M}$ are perfectly
  distinguishable.
\end{lemma}

\Proof Let $\{a,e_\rA-a\}$ be the observation-test such that $\SC {a} \rho =1$ and $(a|\sigma)=0$.
Now, by corollary \ref{cor:se fa uno non disturba} there is a transformation $\tC \in\Trnset(\rA)$ such that $\B {e_\rA}  \tC = \B a$ and $\tC =_\rho \tI_\rA$. Similarly, there exists a
transformation $\tC' \in\Trnset(\rA)$  such that $\B {e_\rA} \tC'=\B {e_\rA}-\B a$ and $\tC' =_\sigma \tI_\rA$.
We can then define the following
observation-test
\begin{equation*}
  \B {c_i}  =   \left\{  
    \begin{array}{ll}
      \B {a_i}  \tC  & i \le N   \\  
      \B{b_i}  \tC'  \qquad &  N+1\le i \le N+M 
    \end{array}
  \right. 
\end{equation*}      
where $\{a_i\}_{i=1}^N$ (resp. $\{b_j\}_{j=N+1}^{N+M}$) is the observation-test that perfectly
distinguishes among the states $\{\rho_i\}_{i=1}^N$ (resp. $\{\rho_j\}_{j=N+1}^{N+M}$). By corollary
\ref{cor:EVM=test} [see in particular Eq. (\ref{eq:normalizzazionepovm})], $\{c_i\}_{i=1}^{N+M}$ is indeed an observation-test: each $c_i$ is an effect and
one has the normalization
\begin{equation*}
  \begin{split}
    \sum_{i=1}^{N+M}  \B{c_i}   &=   \sum_{i=1}^N  \B{a_i}  \tC   + \sum_{i=N+1}^{N+M}\B{b_i} \tC' \\
    &  =  \B {e_\rA}   \tC  +  \B {e_\rA}\tC' \\
    &= \B {a}+ \B {e_\rA}-\B a = \B {e_\rA}.
  \end{split}
\end{equation*}
Moreover, since $\tC=_\rho\tI_\rA$ and $\tC'=_\sigma\tI_\rA$, one has $\SC {c_i} {\rho_i}=1$ for every $i = 1, \dots, M+N$.
By lemma \ref{lem:simpler}, this implies that the states $\{\rho_i\}_{i=1}^{N+M}$ are perfectly
distinguishable. \qed

\begin{definition}
  A set of perfectly distinguishable states $\{\rho_i\}_{i=1}^N$ is \emph{maximal} if there is no
  state $\rho_{{N+1}}\in \Stset_1 (\rA)$ such that the states $\{\rho_i\}_{i=1}^{N+1}$ are perfectly
  distinguishable
\end{definition}

\begin{theorem}
  A set of perfectly distinguishable states $\{\rho_i\}_{i=1}^N$ is maximal if and only if the state
  $\omega =\sum_{{i=1}}^{N} \rho_i/N$ is completely mixed.
  \label{theo:interndiscr}
\end{theorem}
\Proof We first prove that if $\omega$ is completely mixed, then the set
$\{\rho_{i}\}_{i=1}^{N}$ must be maximal. Indeed, if there existed a
state $\rho_{N+1}$ such that $\{\rho_{i}\}_{i=1}^{N+1}$ are perfectly
distinguishable, then clearly $\rho_{N+1}$ would be distinguishable
from $\omega$. This is absurd because by proposition \ref{prop: no
  discrimination from completely mixed state} no state can be
perfectly distinguished from a completely mixed state.  Conversely, if
$\{\rho_i\}_{i=1}^N$ is maximal, then $\omega$ is completely mixed. If
it were not, by the distinguishability axiom \ref{norestr}, $\omega$
would be perfectly distinguishable from some state $\rho_{N+1}$.  By
lemma \ref{lem:refinement}, this would imply that the states
$\{\rho_i\}_{i=1}^{N+1}$ are perfectly distinguishable, in
contradiction with the hypothesis that the set $\{\rho_i\}_{i=1}^N$ is
maximal.  \qed

\begin{lemma}\label{lem:extensiontomaximal}
  Every set of perfectly distinguishable pure states can be extended to a maximal set of perfectly
  distinguishable pure states.
\end{lemma}

\Proof Let $\{\varphi_i\}_{i=1}^N$ be a non-maximal set of perfectly
distinguishable pure states.  By definition, there exists a state
$\sigma$ such that $\{\varphi_i\}_{i=1}^N\cup\{\sigma\}$ is perfectly
distinguishable.  Let $\varphi_{N+1}$ be a pure state in $F_\sigma$. By
Lemma \ref{lem:detuponrho}, the states $\{\varphi_i\}_{i=1}^{N+1}$
will be perfectly distinguishable. Since the dimension of
$\Stset_\Reals (\rA)$ is finite and distinguishable states are
linearly independent, iterating this procedure one finally obtains a
maximal set of pure states in a finite number of steps.  \qed

\begin{corollary}\label{cor:purebelongs}
  Any pure state belongs to a maximal set of perfectly distinguishable
  pure states.
\end{corollary}

We conclude this section with a few elementary facts about how the ideal compression of axiom
\ref{compression} preserves the distinguishability properties. In the following we will choose a
state $\rho\in\Stset_1 (\rA)$ and $\tE \in\Trnset(\rA,\rC)$ (resp. $\tD \in\Trnset(\rC,\rA)$) will
be the encoding (resp. decoding) in the ideal compression scheme for $\rho$.

\begin{lemma}
  If the states $\{\rho_i\}_{i=1}^k\subset F_\rho$ are perfectly distinguishable, then the states
  $\{\tE\rho_i\}_{i=1}^k\subset\Stset_1(\rC)$ are perfectly distinguishable.  Conversely, if the
  states $\{\sigma_i\}_{i=1}^k\subset\Stset_1(\rC)$ are perfectly distinguishable, then the
  states $\{\tD\sigma_i\}_{i=1}^k\subset F_\rho$ are perfectly distinguishable.
  \label{lem:distienc}
\end{lemma}

\Proof Let $\{a_i\}_{i=1}^k$ be the observation-test such that $(a_i|\rho_i)=1$ for every $i=1,
\dots, k$.  Since the compression is lossless, we have $\tD\tE\K{\rho_i}=\K{\rho_i}$ and
$(a_i|\tD\tE|\rho_i)=1$.  Now, consider the test $\{ c_i\}_{i=1}^k$ defined by $\B {c_i} = \B{a_i}
\tD$.  Clearly we have $\B {c_i} \tE \K{\rho_i} = 1$ for every $i=1, \dots, k$.  By lemma
\ref{lem:simpler} this means that the states $\{\tE\rho_i\}_{i=1}^k$ are perfectly distinguishable.
Similarly, let $\{b_i\}_{i=1}^k$ the observation-test that distinguishes the set $\{\sigma_i\}_{i=1}^k$. Since
$\tE\tD=\tI_\rC$ (lemma \ref{lem:EDeqI}), we can conclude by the same argument that the states
$\{\tD \sigma_i\}_{i=1}^k$ are perfectly distinguishable. \qed

We say that a set of perfectly distinguishable states $\{\rho_i\}_{i=1}^k \subset F_\rho$ is
\emph{maximal in the face $F_\rho$} if there is no state $\rho_{k+1}\in F_\rho$ such that the states
$\{\rho_i\}_{i=1}^{k+1}$ are perfectly distinguishable.  We then have the following:
  \begin{corollary}
    If $\{\rho_i\}_{i=1}^k \subset F_\rho$ is a maximal set of perfectly distinguishable states in
    the face $F_\rho$, then $\{\tE\rho_i\}_{i=1}^k\in\Stset_1 (\rC)$ is a maximal set of perfectly
    distinguishable states. Conversely, if $\{\sigma_i\}_{i=1}^k \Stset_1 (\rC)$ is a maximal set of
    perfectly distinguishable states, then $\{\tD\sigma_i\}_{i=1}^k$ is a maximal set of perfectly
    distinguishable states in the face $F_\rho$.
  \label{cor:maxienc}
\end{corollary}
\Proof Distinguishability of the states $\{\tE \rho_i\}_{i=1}^k$ and $\{\tD\sigma_i\}_{i=1}^k$ is
proved by lemma \ref{lem:distienc}.  Let us now prove maximality.  By contradiction, suppose that the set $\{\rho_i\}_{i=1}^k$ is maximal in the face $F_\rho$ while the set $\{\sigma_i\}_{i=1}^k$, $\sigma_i: =\tE \rho_i$ is not maximal.  This means that there exists a  state $\sigma_{k+1} \in\Stset_1 (\rC)$ such that the states $\{\sigma_i\}_{i=1}^{k+1}$ are perfectly distinguishable.   By lemma \ref{lem:distienc} the states  $\{\tD  \sigma_i\}_{i=1}^{k+1}$ are perfectly distinguishable. Since $\tD \tE  \rho_i  =  \rho_i$ for every $i=1, \dots, k$, this means that the states $\{\rho_i\}_{i=1}^k \cup  \{ \tD \sigma_{k+1} \}$ are perfectly distinguishable, in contradiction with the fact that $\{\rho_i\}_{i=1}^k$ is maximal.  This proves that the set $\{\tE \rho_i\}_{i =1}^k$  must be maximal.   Conversely, if the set $\{\sigma_i\}  \subset \Stset_1 (\rC)$ is maximal, using the same argument we can prove that the set $\{\tD  \sigma_i\}_{i =1}^k$ must be maximal in $ F_\rho$.      \qed

\section{Duality between pure states and atomic effects}\label{sec:duality}

We now show the existence of a one-to-one correspondence between states and effects of any system
$\rA$ in the theory.  Let us start from a simple observation:

\begin{lemma}\label{lem:atomiccontainspure}
  If $a$ is atomic and $\SC a \rho =\|a\|$ for $\rho\in\Stset_1(\rA)$,
  then $\rho$ must be pure.
\end{lemma}

\Proof By lemma \ref{lem:detuponrho}, the condition $\SC a \rho
=\|a\|$ implies $a =_\rho  \|  a\|   ~e$. By theorem \ref{theo:uponinput}, the condition $a =_\rho  \|  a\|   ~e$ implies
\begin{equation*}
  \begin{aligned}
    \Qcircuit @C=1em @R=.7em @! R
    {& \multiprepareC{1}{\Psi_\rho}   & \qw\poloFantasmaCn{\rA}& \measureD a\\
      & \pureghost{\Psi_\rho} & \qw \poloFantasmaCn{\rB} & \qw }
  \end{aligned} =\|a\|
  \begin{aligned}
    \Qcircuit @C=1em @R=.7em @! R {&\multiprepareC{1}{\Psi_\rho}&\qw\poloFantasmaCn{\rA}&\measureD e \\
      &\pureghost{\Psi_\rho}&\qw\poloFantasmaCn{\rB}& \qw }
  \end{aligned}
\end{equation*}
where $\Psi_\rho\in\Stset_1 (\rA\rB)$ is any purification of $\rho$.  Since $a$ is atomic, the pure
conditioning axiom \ref{purecond} implies that the marginal state $\K{\tilde \rho}_\rB = \B e_\rA
\K{\Psi_\rho}_{\rA\rB}$ is pure.   Since the marginal of $\Psi_\rho$ on system $\rB$ is pure,  $\Psi_\rho$ must be factorized, i.e.  $\Psi_\rho = \rho
\otimes \tilde \rho$  (see lemma 19 of Ref. \ref{purification}).  Hence, $\rho$ must be pure, otherwise we would have a non-trivial convex decomposition of the pure state $\Psi_\rho$.  \qed
 
 We are now in position to show that every atomic effect is associated to a unique pure state.
\begin{theorem}\label{theo:dualeeffetto}
  For every atomic effect $a\in \Cntset(\rA)$, there exists a unique pure state $\varphi \in
  \Stset_1(\rA)$ such that $\SC a {\varphi} =\|a\|$.
\end{theorem}

\Proof Let $\rho$ be a state such that $\SC a \rho = \|a\|$. By lemma \ref{lem:atomiccontainspure},
$\rho$ must be pure. Moreover, this pure state must be unique: suppose that $\varphi$ and $\varphi'$
are pure states such that $\SC a {\varphi} = \SC a{\varphi'}=\|a\|$.  Then for $\omega= 1/2 (\varphi
+ \varphi')$ one has $\SC a \omega =\|a\|$. Since $\omega$ must be pure, one has $\varphi =
\varphi'$. \qed
 
We now show the converse result: for every pure state $\varphi\in\Stset_1(\rA)$ there exists a unique
atomic effect $a$ such that $\SC a \varphi =1$.  Let us start from the existence:
\begin{lemma}\label{lem:atomicobstest}
  Let $\{\varphi_i\}_{i=1}^N \subset \Stset_1 (\rA)$ be a maximal set of perfectly distinguishable pure states and let
  $\{a_i\}_{i=1}^N$ be the observation-test such that $\SC {a_i} {\varphi_j} = \delta_{ij}$.  Then
  each effect $a_i$ is atomic with $\|a_i\|=1$.
\end{lemma}
\Proof It is obvious that $\|a_i\|=1$, because of the condition $\SC {a_i} {\varphi_i} =1$.  It
remains to prove atomicity. Consider the state $\omega =\sum_{i=1}^N \varphi_i/N$, which is
completely mixed by theorem \ref{theo:interndiscr}. Let $\Psi_\omega \in \Stset_1 (\rA\rB)$ be a
purification of $\omega$, chosen in such a way that the marginal on system $\rB$ is completely mixed
(theorem \ref{theo:conjusys}). As a consequence of purification (lemma \ref{lem:purificami-ens}), there exists an observation-test
$\{b_i\}_{i=1}^N$ on system $\rB$ such that $\B {b_i}_\rB \K {\Psi_\omega}_{\rA\rB} = 1/N \K
{\varphi_i}_\rA$.  Since $\Psi_\omega$ is dynamically faithful on system $\rB$, each effect $b_i$
must be atomic.  Now, define the normalized states $\{\rho_i\}_{i=1}^N \subset \Stset_1 (\rB)$ and the probabilities
$\{p_i\}_{i=1}^N$ by
 \begin{equation} 
  \begin{aligned}
   \Qcircuit @C=1em @R=.7em @! R
    {& \multiprepareC{1}{\Psi_\omega}   & \qw\poloFantasmaCn{\rA}& \measureD{a_i}\\
    & \pureghost{\Psi_\omega}   & \qw \poloFantasmaCn{\rB} & \qw}
  \end{aligned}=
  \begin{aligned}
    \Qcircuit @C=1em @R=.7em @! R { &p_i ~&\prepareC {\rho_i} &\qw \poloFantasmaCn{\rB}&\qw  } \end{aligned}
 \end{equation}
 Applying the deterministic effect $e_\rB$ on both sides one has $p_i = \SC {a_i} \omega = 1/N$.  On
 the other hand, applying the effect $b_j$ one has instead $1/N \SC {b_j} {\rho_i}_\rB = 1/N
 \SC{a_i} {\varphi_j} = \delta_{ij}/N$.  This implies $\SC {b_i} {\rho_i} =1$ for every $i$.  Since
 $b_i$ is atomic, lemma \ref{lem:atomiccontainspure} forces each $\rho_i$ to be pure.  Finally, each
 $a_i$ must be atomic since its Choi state $p_i \K{\rho_i}_\rB = \B {a_i}_\rA
 \K{\Psi_{\omega}}_{\rA\rB}$ is pure (theorem \ref{theo:iso}). \qed
 
 As a consequence, we can prove the following existence result:
 \begin{lemma}\label{lem:thereexistsaneffect}
   For every pure state $\varphi\in\Stset_1 (\rA)$ there exists an atomic effect such that $\SC a
   \varphi =1$.
 \end{lemma}
 \Proof By corollary \ref{cor:purebelongs}, every pure state belongs to a maximal set of perfectly
 distinguishable pure states $\{\varphi_i\}_{i=1}^N$, say $\varphi = \varphi_1$.  The thesis then
 follows from lemma \ref{lem:atomicobstest}. \qed
 
 We now prove that the atomic effect $a$ such that $(a|\varphi)=1$ is unique. For this purpose we
 need two auxiliary lemmas:
 \begin{lemma}\label{lem:samepmax}
   Let $\varphi\in\Stset_1 (\rA)$ be an arbitrary pure state and let $p_{\varphi}$ be the
   probability defined by
\begin{equation}
p_{\varphi} = \max\left\{ p  :    \exists \sigma , \chi = p \varphi + (1-p) \sigma\right\}
\end{equation}
where $\chi$ is the invariant state of system $\rA$. Then the value of the probability $p_{\varphi}$ is independent of $\varphi$. 
 \end{lemma}
  
 \Proof Since for every couple of pure states $\varphi$ and $\psi$ one has $\psi = \tU \varphi$ for
 some reversible channel $\tU $ (lemma \ref{lem:trans}), and since $\chi$ is invariant, one has $\chi = p \varphi + (1-p)
 \sigma$ if and only if $\chi = p \psi + (1-p) \tU \sigma$.  The maximum probabilities for $\varphi$
 and $\psi$ are then equal.\qed Since $p_\varphi = p_\psi$ for every couple of pure states, from now
 on we will write $p_{\max}$ in place of $p_\varphi$.
 
 \begin{lemma}\label{lem:prepver} Let $\varphi\in\Stset_1 (\rA)$ be a pure state and
   $a\in\Cntset(\rA)$ be an atomic effect such that $\SC a \varphi  =1$.   
   Let $\K\Phi_{\rA\rB}$ be a purification of the invariant state $\K \chi_\rA$, chosen in such a
   way that the marginal on system $\rB$ is completely mixed, and let $b$ be the unique atomic effect on
   $\rB$ such that
 \begin{equation} \label{eqprovata}
  \begin{aligned}
   \Qcircuit @C=1em @R=.7em @! R
    {& \multiprepareC{1}{\Phi}   & \qw\poloFantasmaCn{\rA}& \qw \\
    & \pureghost{\Phi}   & \qw \poloFantasmaCn{\rB} & \measureD b}
  \end{aligned}=p_{\max} 
  \begin{aligned}
    \Qcircuit @C=1em @R=.7em @! R {\prepareC \varphi &\qw \poloFantasmaCn{\rA}&\qw  } 
  \end{aligned}
\end{equation}
[note that $b$ exists by lemma \ref{lem:purificami-ens}is uniquely defined by Eq. (\ref{eqprovata}) because $\Phi$ is faithful for system $\rB$].
Then one has
\begin{equation}\label{eqdaprovare}
  \begin{aligned}
    \Qcircuit @C=1em @R=.7em @! R
    {& \multiprepareC{1}{\Phi}   & \qw\poloFantasmaCn{\rA}& \measureD a \\
      & \pureghost{\Phi} & \qw \poloFantasmaCn{\rB} & \qw }
  \end{aligned}= p_{\max}
  \begin{aligned}
    \Qcircuit @C=1em @R=.7em @! R {\prepareC {\psi} &\qw \poloFantasmaCn{\rB}&\qw  } 
  \end{aligned}
\end{equation}
where $\psi$ is the unique pure state such that $\SC b \psi =1$.
\end{lemma}

\Proof Define the normalized pure state $\psi$ and the probability $q$
by
\begin{equation}
  q\ 
  \begin{aligned}
    \Qcircuit @C=1em @R=.7em @! R {\prepareC {\psi} &\qw \poloFantasmaCn{\rB}&\qw  } 
  \end{aligned} =
  \begin{aligned}
    \Qcircuit @C=1em @R=.7em @! R
    {& \multiprepareC{1}{\Phi}   & \qw\poloFantasmaCn{\rA}& \measureD a \\
      & \pureghost{\Phi} & \qw \poloFantasmaCn{\rB} & \qw }
    \label{steerata}
  \end{aligned}
\end{equation}
In order to prove the thesis we have to show that $q = p_{\max}$ and $\SC b \psi =1$.  Applying $b$
on both sides of Eq.~\eqref{steerata} and using Eq.~\eqref{eqprovata} we obtain $ q \SC b \psi  =
p_{\max} \SC a \varphi  = p_{max}$.  This implies
\begin{equation}\label{diventauguale}
  q\ge p_{\max},
\end{equation} 
with the equality if and only if $\SC b \psi =1$.  Let $b'$ be an atomic effect such that $\SC
{b'} {\psi}  =1$ (such an effect exists because of lemma \ref{lem:thereexistsaneffect} ). Define
the normalized pure state $\varphi'$ and the probability $p'$ by
\begin{equation*}
  \begin{aligned}
    \Qcircuit @C=1em @R=.7em @! R { &p' ~&\prepareC {\varphi'} &\qw
      \poloFantasmaCn{\rA}&\qw } \end{aligned}=\begin{aligned}
    \Qcircuit @C=1em @R=.7em @! R
    {& \multiprepareC{1}{\Phi}   & \qw\poloFantasmaCn{\rA}& \qw \\
      & \pureghost{\Phi} & \qw \poloFantasmaCn{\rB} & \measureD {b'}}
  \end{aligned}
\end{equation*}
Applying $a$ on both sides and using Eq. (\ref{steerata}) we obtain $p' \SC {a} {\varphi'} = q
\SC {b'} \psi  = q$, which implies $p'\ge q$, with the equality if and only if $\SC a {\varphi'}
= 1$. Combining this with the inequality (\ref{diventauguale}) we have $p' \ge q\ge p_{\max}$.  On
the other hand, by Lemma \ref{lem:samepmax} one has $p'\leq p_{\max}$, and consequently $p' = q=
p_{\max}$. This also implies that $\SC{b} \psi =1$ and $\SC a {\varphi'}=1$.\qed

\begin{theorem}\label{theo:dualestato}
  For every pure state $\varphi\in\Stset_1(\rA)$ there is a unique atomic effect $a\in\Cntset(\rA)$
  such that $\SC a \varphi =1$.
\end{theorem}
 
\Proof Existence has been already proved in lemma \ref{lem:thereexistsaneffect}.  Let us prove
uniqueness: suppose that $a$ and $a'$ are two atomic effects such that $\SC a \varphi = \SC {a'} \varphi =1$.  Then,
applying lemma \ref{lem:prepver} to $a$ and $a'$ we obtain
\begin{equation*}
  \begin{aligned}
    \Qcircuit @C=1em @R=.7em @! R
    {& \multiprepareC{1}{\Phi}   & \qw\poloFantasmaCn{\rA}& \measureD a \\
      & \pureghost{\Phi} & \qw \poloFantasmaCn{\rB} & \qw }
  \end{aligned}=
  \begin{aligned}
    \Qcircuit @C=1em @R=.7em @! R
    {& \multiprepareC{1}{\Phi}   & \qw\poloFantasmaCn{\rA}& \measureD {a'} \\
      & \pureghost{\Phi} & \qw \poloFantasmaCn{\rB} & \qw }
  \end{aligned}
\end{equation*}
Since $\Phi$ is dynamically faithful, this implies $a=a'$.\qed 

Finally, an important consequence of theorem \ref{theo:dualestato} is
\begin{corollary}\label{cor:orbiatom}
  If $a, a' \in\Cntset(\rA)$ are two atomic effects with $\|a \| =
  \|a'\|=1$, then there is a reversible channel $\tU\in\grp G_\rA$ such
  that $ \B {a'}_\rA = \B a_\rA \tU$.
\end{corollary} 
\Proof Let $\varphi$ and $\varphi'$ be the (unique) normalized states
such that $\SC a \varphi =1$ and $\SC {a'}{\varphi'} = 1$,
respectively.  Now, there is a reversible channel $\tU \in\grp G_\rA$ such that $\K
{\varphi}_\rA = \tU \K{\varphi'}_\rA$. Hence, $ \SC{a'} {\varphi'} =\SC {a} {\varphi}_\rA =
\B {a} \tU \K{\varphi'}$. By theorem
\ref{theo:dualestato}, one has $\B{a'}_\rA = \B a_\rA \tU$.\qed

We conclude this section with an elementary result that will be used later in the paper:
\begin{lemma}
  Let $\tE \in\Trnset (\rA, \rC)$ and $\tD \in\Trnset (\rC, \rA)$ be the encoding and the decoding
  in the ideal compression scheme for $\rho \in\Stset_1 (\rA)$.  If $\K\varphi \in F_\rho$ is a pure state
  and $\B a \in\Cntset (\rA)$ is the atomic effect such that $\SC a \varphi =1$, then $\K {\gamma}
  := \tE \K \varphi \in \Stset_1 (\rC)$ is a pure state and $\B c :=\B a \tD \in\Cntset(\rC)$ is the
  atomic effect such that $\SC c \gamma =1$.
  \label{lem:dualdec}
\end{lemma}

\Proof The state $\K \gamma:= \tE\K \varphi$ is pure by lemma \ref{lem:pureenc}.  The effect $\B c :
= \B a \tD$ is atomic by lemmas \ref{lem:atomicdec} and \ref{lem:atomicity}. Since
$\tD\tE=_\rho\tI_\rA$, one has $ \SC c \gamma = \B a \tD\tE|\varphi)=(a|\varphi)=1$.  \qed

\section{Dimension}\label{sec:dimension}
In this section we show that each system in our theory has given \emph{informational dimension}, defined as the maximum number of perfectly distinguishable pure states available in the system.  In the Hilbert space framework, the informational dimension will be the dimension of the Hilbert space.

\begin{lemma}\label{lem:samecard}
All maximal sets of perfectly distinguishable pure states have the same number of elements.
\end{lemma}
 
\Proof Let $\{\varphi_i\}_{i=1}^N$ be a maximal set of perfectly distinguishable pure states for
system $\rA$, and let $\{a_i\}_{i=1}^N$ the observation-test such that $\SC {a_i}{\varphi_j}=
\delta_{ij}$.  By lemma \ref{lem:atomicobstest}, each $a_i$ is atomic and $\|a_i\|=1$.  Then, by
corollary \ref{cor:orbiatom}, one has $\B {a_i}_\rA = \B{a_0} \tU_i$, where each $\tU_i$ is a
reversible channel and $a_0$ is a fixed atomic effect with $\|a_0\|=1$.  By the invariance of $\chi$
we then obtain $(a_i|\chi_\rA)=(a_0|\tU_i|\chi_\rA)=(a_0|\chi_\rA)$. On the other hand, one has
$\sum_{i=1}^N \SC {a_i} {\chi_\rA} =1$, which implies $N = 1/\SC {a_0} {\chi_\rA}$.  Since $a_0$ is
arbitrary, $N$ is independent of the choice of the set $\{\varphi_i\}_{i=1}^N$. \qed

As a consequence, the number of perfectly distinguishable pure states in a maximal set is a property
of the system $\rA$. We will call this number the \emph{informational dimension} (or simply the
dimension) of system $\rA$, and denote it with $d_\rA$. The informational dimension $d_\rA$ has not to be confused with the size $D_\rA$ of the state space $\Stset(\rA)$: recall that $D_\rA$ was defined as the dimension of the real vector space $\Stset_{\Reals} (\rA)$.   In quantum theory one has $D_\rA  = d_\rA^2$.

An immediate consequence of the proof of lemma \ref{lem:samecard} is
 \begin{corollary}\label{cor:uno degli ultimi corollari}
   For every atomic effect $a$ with $\|a\|=1$ one has $\SC a {\chi_\rA}
   = 1/d_\rA$.
 \end{corollary}

 This simple fact has two very important consequences. The first is that the dimension of a
 composite system is the product of the dimensions of the components:

\begin{corollary}\label{cor:dimprod}  
The dimension of the composite system $\rA\rB$ is the product of the dimensions of $\rA$ and $\rB$, namely $d_{\rA\rB} = d_\rA d_\rB$.
\end{corollary} 

\Proof 
From lemma \ref{lem:uniquenessinvariant} we know that $\chi_\rA \otimes \chi_\rB$ is the unique invariant state of system $\rA\rB$.  Now, if $a \in\Cntset(\rA)$ and $b \in \Cntset (\rB)$ are such that $\| a\| = \| b\| =1$, then $a\otimes b$ is such that $\| a \otimes b \| = 1$.  Hence we have  $1/d_{\rA\rB}  = \SC {a \otimes b} {\chi_\rA \otimes \chi_\rB}  =  \SC {a} {\chi_\rA}  \SC {b} {\chi_\rB} = 1/(d_\rA d_\rB)$. \qed

The second consequence is the relation between the dimension and the maximum probability of a pure
state in the convex decomposition of the invariant state $\K\chi_ \rA$:
 
 \begin{lemma}\label{lem:pmax} For every system $\rA$, the maximum probability of a pure state in the convex decomposition of the invariant state is $p_{\max} = 1/d_\rA$. 
 \end{lemma}
 \Proof Let $\Phi \in \Stset_1(\rA\rB)$ be a purification of the
 invariant state $\K{\chi_\rA}$, chosen in such a way that the marginal on system $\rB$ is completely mixed. Let $a\in\Cntset(\rA)$ be an atomic
 effect with $\|a\|=1$. Then, equation \eqref{eqdaprovare} becomes
 \begin{equation*}
  \begin{aligned}
   \Qcircuit @C=1em @R=.7em @! R
    {& \multiprepareC{1}{\Phi}   & \qw\poloFantasmaCn{\rA}& \measureD a \\
    & \pureghost{\Phi}   & \qw \poloFantasmaCn{\rB} & \qw }
  \end{aligned}= \begin{aligned}
    \Qcircuit @C=1em @R=.7em @! R {&&  {p_{\max} } &&\prepareC {\psi} &\qw \poloFantasmaCn{\rB}&\qw  } \end{aligned}
 \end{equation*}
 where $\psi$ is some normalized pure state of system $\rB$.  Applying the deterministic effect $e$
 on system $\rB$ on both sides we obtain $\SC a {\chi_{\rA}} = p_{\max}$.  Finally, corollary \ref{cor:uno degli ultimi corollari}
 states $\SC a {\chi_\rA} = 1/d_\rA$. By comparison, we obtain $p_{\max} = 1/d_\rA$.  \qed

 Thanks to the compression axiom \ref{compression}, the notion of dimension can be applied not only
 to the whole state space $\Stset_1 (\rA)$ but also to its faces.  With \emph{face $F$ of the convex
   set} $\Stset_1 (\rA)$ we always mean the face $F_\rho$ identified by some state
 $\rho\in\Stset_1(\rA)$.
  
\begin{lemma}
  Let $F$ be a face of the convex set $\Stset_1 (\rA)$.  Every maximal set $\{\varphi_i\}_{i=1}^k$
  of perfectly distinguishable pure states in $F$ has the same cardinality $k$.  Precisely, if $F$
  is the face identified by $\rho\in\Stset_1 (\rA)$ and $\tE\in\Trnset(\rA, \rC)$ is the encoding in
  the ideal compression for $\rho$, then we have $k = d_\rC$
  \label{lem:cardisuprho}
\end{lemma}

\Proof The set $\{\tE\varphi_i\}_{i=1}^k\subset\Stset_1(\rC)$ is perfectly distinguishable by lemma
\ref{lem:distienc}, and it is maximal by corollary \ref{cor:maxienc}. Moreover, the states
$\{\tE\varphi_i\}_{i=1}^k$ are pure by lemma \ref{lem:pureenc}.  Hence, the cardinality $k$ of the
set $\{\varphi_i\}_{i=1}^k$ must be $k=d_\rC$.\qed
 
From now on the maximum number of perfectly distinguishable states in the face $F$ will be called
the \emph{dimension of the face $F$} and will be denoted by $|F|$.


\section{Decomposition into perfectly distinguishable pure states}\label{sec:decomposition}
In this section we show that in a theory satisfying our principles any state can be written as a
convex combination of perfectly distinguishable pure states.  In quantum theory, this corresponds to
the diagonalization of the density matrix.

To prove this result we need first a sufficient condition for the distinguishability of states, given in the following
\begin{lemma}
  Let $\{\rho_i\}_{i=1}^N \subset \Stset_1 (\rA)$ be a set of states.  If
  there exists a set of effects $\{b_i\}_{i=1}^N\subset \Cntset(\rA)$ (not necessarily an observation-test) such that
  $(b_i|\rho_j)=\delta_{ij}$, then the states $\{\rho_i\}_{i=1}^N$ are
  perfectly distinguishable.
  \label{lem:setofeffects}
\end{lemma}

\Proof For each $i=1, \dots,N$ consider the binary test $\{b_i, e-b_i\}$.  Since by hypothesis $\SC
{b_i} {\rho_j} = \delta_{ij}$, the test $\{b_i, e-b_i\}$ can perfectly distinguish $\rho_i$ from any
mixture of the states $\{\rho_j\}_{j \not = i}$.  In particular, this means that, for every $M< N$,
$\rho_{M+1}$ can be perfectly distinguished from the mixture $\omega_{M} = \sum_{j=1}^M \rho_j/M$.
Note that, by definition, the states $\{\rho_i\}_{i=1}^M$ belong to the face $F_{\omega_M}$.  We now
prove by induction on $M$ that the states $\{\rho_i\}_{i=1}^{M}$ are perfectly distinguishable. This
is true for $M=1$.  Now, suppose that the states $\{\rho_i\}_{i=1}^{M}$ are perfectly
distinguishable.  Since the state $\rho_{M+1}$ is perfectly distinguishable from $\omega_M$, by
lemma \ref{lem:refinement} we have that the states $\{\rho_i\}_{i=1}^{M+1}$ are perfectly distinguishable. Taking $M=N-1$ the
thesis follows.\qed

We now show that the invariant state $\chi$  is a mixture of perfectly
distinguishable pure states.
\begin{theorem}
  For every maximal set of perfectly distinguishable pure states
  $\{\varphi_i\}_{i=1}^{d_\rA}  \subset \Stset_1 (\rA)$ one has
  \begin{equation*}
    \chi  =   \frac 1 {d_\rA}\sum_{i=1}^{d_\rA}  {\varphi_i}.
  \end{equation*}
  \label{theo:maxdiscrinvar}
\end{theorem}

\Proof Let $\{a_i\}_{i=1}^{d_\rA}$ be the observation test such that $\SC {a_i} {\varphi_j}
=\delta_{ij}$, and $\Phi \in\Stset_1 (\rA\rB)$ be a purification of $\chi$, chosen in such a way
that the marginal on system $\rB$ is completely mixed (theorem \ref{theo:conjusys}).  Let
$\{\psi_i\}_{i=1}^{d_\rA} \subset \Stset_1 (\rB)$ be the pure states defined by
\begin{equation*}
  \begin{aligned}
    \Qcircuit @C=1em @R=.7em @! R
    {& \multiprepareC{1}{\Phi}   & \qw\poloFantasmaCn{\rA}& \measureD {a_i} \\
      & \pureghost{\Phi} & \qw \poloFantasmaCn{\rB} & \qw }
  \end{aligned}= \begin{aligned} \frac 1{d_\rA} \Qcircuit @C=1em
    @R=.7em @! R { \prepareC {\psi_i} &\qw \poloFantasmaCn{\rB}&\qw }
  \end{aligned}
\end{equation*}  
and, for each $i$, let $b_i$ be the atomic effect such that
\begin{equation} \label{dacitar}
  \begin{aligned}
    \Qcircuit @C=1em @R=.7em @! R
    {& \multiprepareC{1}{\Phi}   & \qw\poloFantasmaCn{\rA}& \qw \\
      & \pureghost{\Phi} & \qw \poloFantasmaCn{\rB} & \measureD {b_i}}
  \end{aligned}=
  \begin{aligned}
    \frac 1 {d_\rA} \Qcircuit @C=1em @R=.7em @! R { \prepareC
      {\varphi_i} &\qw \poloFantasmaCn{\rA}&\qw } \end{aligned}
\end{equation}
(here we used lemma \ref{lem:prepver} and the fact that $p_{\max} =
1/d_\rA$).  Then we have
\begin{align}\nonumber  \begin{aligned}
    \Qcircuit @C=1em @R=.7em @! R { \prepareC {\psi_i} &\qw
      \poloFantasmaCn{\rB}&\measureD {b_j} } \end{aligned} &= d_\rA
  \begin{aligned} \Qcircuit @C=1em @R=.7em @! R
    { \multiprepareC{1}{\Phi}   & \qw\poloFantasmaCn{\rA}& \measureD{a_i} \\
      \pureghost{\Phi} & \qw \poloFantasmaCn{\rB} & \measureD {b_j}}
  \end{aligned}\\ &=
  \begin{aligned}
    \Qcircuit @C=1em @R=.7em @! R { \prepareC {\varphi_j} &\qw
      \poloFantasmaCn{\rA}&\measureD{a_i} } \end{aligned} =
  \delta_{ij}. \label{eq:una delle ultime eq}
\end{align}
By lemma \ref{lem:setofeffects}, this implies that the states $\{\psi_i\}_{i=1}^{d_\rA}$ are
perfectly distinguishable. Now, since the marginal of $\K\Phi_{\rA\rB}$ on system $\rB$ is
completely mixed, theorem \ref{theo:interndiscr} states that the set $\{\psi_i\}_{i=1}^{d_\rA}$ is
maximal.  Let $\{b'_i\}_{i=1}^{d_\rA}$ the observation test such that $\SC {b_i'} {\psi_j} =
\delta_{ij}$.  By lemma \ref{lem:atomicobstest}, each $b_i'$ must be atomic.  On the other hand,
there is a unique atomic effect $b_i$ such that $\SC {b_i}{\psi_i} =1$ (theorem
\ref{theo:dualestato}). Therefore, $b_i' = b_i$. This means that the effects
$\{b_i\}_{i=1}^{d_\rA}$ form an observation test.  Once this fact has been proved, using Eq.
\eqref{dacitar} we obtain 
\begin{align*} 
\K \chi_\rA & =  \B {e_\rB}  \K{\Phi}_{\rA\rB} \\
  & =    \sum_i  \B {b_i}  \K  \Phi_{\rA\rB} \\
&= 1/d_\rA  \sum_i  \K{\varphi_i}.
\end{align*}  \qed

As a consequence, we have the following

\begin{corollary}[Existence of conjugate systems]
  For every system $\rA$ there exists a system $\tilde \rA$, called the \emph{conjugate system}, and
  a purification $\Phi\in\Stset_1(\rA\tilde \rA)$ of the invariant state $\chi_\rA$ such that
  $d_{\tilde \rA}= d_\rA$ and the marginal on $\tilde \rA$ is the invariant state
  $\chi_{\tilde\rA}$. The conjugate system $\tilde \rA$ is unique up to operational equivalence. 
  \label{cor:conjusys}
\end{corollary}

\Proof We first prove that $\tilde \rA$ is unique up to operational equivalence. The defining property of the conjugate system $\tilde \rA$ is that the marginal of $\Phi$ on 
 $\tilde \rA$ is the invariant state $\chi_{\tilde \rA}$, which is completely mixed.  Theorem \ref{theo:conjusys} then implies that $\tilde \rA$ is unique up to operational equivalence.   Let us now show the existence of $\tilde\rA$.  Take a purification of $\chi_\rA$, with purifying system $\tilde \rA$ chosen so that the marginal of $\Phi$ on $\tilde \rA$ is completely mixed
(this is possible thanks to theorem \ref{theo:conjusys}).  Now,  the states $\{\psi_i\}_{i=1}^{d_\rA}\subseteq\Stset(\rB)$,
defined by $\frac1 {d_\rA}|\psi_i):=[(a_i|\otimes\tI_{\tilde\rA}] |\Phi)$, are perfectly
distinguishable [see Eq. (\ref{eq:una delle ultime eq}) in the proof of theorem \ref{theo:maxdiscrinvar}]. Hence, by theorem \ref{theo:interndiscr} they are a maximal set of perfectly
distinguishable pure states. This implies $d_{\tilde \rA} = d_\rA$.  Finally, by theorem
\ref{theo:maxdiscrinvar} one has $1/d_{\tilde \rA}\sum_{i=1}^{d_{\tilde \rA}} \psi_i = \chi_{\tilde \rA}$.
\qed

\begin{corollary}
  The distance between the invariant state $\chi_\rA$ and an arbitrary
  pure state $\varphi \in \Stset_1 (\rA)$ is
\begin{equation*}
\|  \chi- \varphi \|  = \frac{2(d_\rA -1)}{d_\rA}.  
\end{equation*}
\end{corollary}
\Proof Take a maximal set of perfectly distinguishable pure states
$\{\varphi_i\}_{i=1}^{d_\rA}$ such that $\varphi_1 = \varphi$
(corollary \ref{cor:purebelongs}). Since $\chi= \sum_{i=1}^{d_\rA}
\varphi_i/d_\rA$ one has $\chi-\varphi = \frac{(d_\rA-1)}{d_\rA}
(\sigma-\varphi_1)$, where $\sigma= \sum_{i=2}^{d_\rA}
\varphi_i/(d_\rA-1)$. Hence, one has $\|\chi -\varphi \| =
\frac{(d_\rA-1)}{d_\rA} \| \sigma-\varphi_1\| =
\frac{2(d_\rA-1)}{d_\rA}$, having used that $\sigma$ and $\varphi_1$
are perfectly distinguishable and therefore $\| \sigma - \varphi_1 \|
= 2$ (see subsection II-I in Ref. \cite{purification}). \qed

We can now prove the following strong result:
\begin{theorem}[Spectral decomposition]
  For every system $\rA$, every mixed state can be written as a convex combination of perfectly
  distinguishable pure states.
  \label{theo:diag}
\end{theorem}
\Proof The proof is by induction on the dimension of the system. If $d_\rA =1$, the thesis trivially
holds. Now suppose that the thesis holds for any system $\rB$ with dimension $d_\rB\le N$, and take
a mixed state $\rho\in\Stset_1 (\rA)$ where $d_\rA = N+1$. There are two possibilities: either (1)
$\rho$ is not completely mixed  or  (2) $\rho$ is completely mixed.  Suppose that (1) $\rho$ is
not completely mixed. Then by the compression axiom \ref{compression} one can encode it in a system
$\rC$, using an encoding operation $\tE \in\Trnset(\rA, \rC)$.  Now, the maximum number of perfectly
distinguishable states in $\rC$ is equal to the maximum number of perfectly distinguishable states
in the face $F_\rho$ (corollary \ref{cor:maxienc}).  Since $\rho$ is not completely mixed, we must
have $d_\rC\le N$.  Using the induction hypothesis we then obtain that the state $\tE \rho
\in\Stset_1 (\rC)$ is a mixture of perfectly distinguishable pure states, say $\tE \rho = \sum_{i}
p_i \psi_i$.  Applying the decoding operation $\tD\in\Trnset(\rC,\rA)$ we get $\rho= \tD\tE \rho =
\sum_i p_i \tD \psi_i$.  Since by lemmas \ref{lem:pureenc} and \ref{lem:distienc} we know that the
states $\{\tD \psi_i\}_{i=1}^{d_\rC}$ are pure and perfectly distinguishable, this is the desired decomposition
for $\rho$.  Now suppose that $\rho$ is completely mixed (2).  Consider the half-line in
$\Stset_\Reals (\rA)$ defined by $\sigma_t = (1+t) \rho - t\chi$, $t \ge0$. Since the set of
normalized states $\Stset_1(\rA)$ is compact, the line will cross its border at some point $t_0$.
Therefore, one will have
\begin{equation*}
  \rho = \frac{1}{1+t_0} \sigma_{t_0} + \frac{t_0}{1+t_0}  \chi.   
\end{equation*}
for some state $\sigma_{t_0}$ on the border of $\Stset_1 (\rA)$, that is, for some state that is not
completely mixed.  But we know from the discussion of point (1) that the state $\sigma_{t_0}$ is a
mixture of perfectly distinguishable pure states, say $\sigma_{t_0} = \sum_{i=1}^k p_i \varphi_i$.
By lemma \ref{lem:extensiontomaximal}, this set can be extended to a maximal set of perfectly
distinguishable pure states $\{\varphi_i\}_{i=1}^{d_\rA}$.  On the other hand, theorem
\ref{theo:maxdiscrinvar} states that $\chi=\sum_{i=1}^{d_\rA} \varphi_i/d_{\rA}$. This implies the
desired decomposition
\begin{equation*}
  \rho = \sum_{i=1}^{d_\rA}  \left(\frac{q_i}{1+t_0}  + \frac{t_0}{d_\rA (1+t_0)} \right) \varphi_i ,
\end{equation*}
where $q_i=p_i$ for $1\leq i\leq k$, and $q_i=0$ otherwise.
\qed

It is easy to  show that the marginals of a pure bipartite state have the same spectral decomposition:  
 
 \begin{corollary}\label{cor:samespec}
Let  $\Psi \in\Stset_1 (\rA\rB)$ be a pure state, and let  $\rho $   and $\tilde \rho $ be the marginals of $\Psi$ on systems $\rA$ and $\rB$, respectively.   If $\rho$ has spectral decomposition $\rho = \sum_{i=1}^{d_\rA}   p_i  \varphi_i$,  with $p_i >0$ for every $i=1, \dots, r$, $r\le d_\rA$, then 
  $\tilde \rho$ has  spectral decomposition $\tilde \rho =  \sum_{i=1}^r  p_i  \psi_i$. 
\end{corollary} 
\Proof  Let $\{a_i\}_{i=1}^{d_\rA}$ be the observation-test such that $\SC {a_i}{\varphi_j} = \delta_{ij}$,  $\{b_i\}_{i=1}^{r}$ be the observation test such that  $\B {b_i}_\rB  \K{\Psi}_{\rA\rB}  =  p_i  \K {\varphi_i}_\rA$ for every $i \le r$.  For $i\le r$, define  the pure state $\psi_i\in\Stset_1 (\rB)$ and the probability $q_i$ via the relation  
\begin{equation*}
q_i \K {\psi_i}_{\rB}  :=  \B{a_i}_{\rA}  \K{\Psi}_{\rA\rB}. 
\end{equation*}
[Note that $\psi_i$ is pure due to the pure conditioning axiom]   By definition, we have
\begin{align*} 
q_i  \SC{b_j}{\psi_i}  & = \SC{a_i \otimes b_j}  {\Psi}  \\
 & =   \SC{a_i} {\varphi_j} \\
& =   p_i  \delta_{ij}  \qquad \forall i \le r, \forall j \le r.
\end{align*}
The above relation implies $q_i  =  \sum_{j=1}^{d_\rA}  q_i  \SC {b_j}{\psi_i}  = \sum_j  p_i \delta_{ij}  = p_i$ and $\SC{b_j}{\psi_i}  = \delta_{ij}$. Hence, the states $\{\psi_i\}_{i=1}^r$ are perfectly distinguishable.   On the other hand, we have  
 $\SC{a_i \otimes e_\rB}  {\Psi}    = \SC  {a_i}{\rho}  = 0  \quad \forall i >r,$
which implies $\B {a_i}_{\rA}  \K{\Psi}_{\rA\rB}   = 0$, $\forall i>r$ .  
Therefore, we obtained  
\begin{align*}
\K{\tilde \rho}_\rB & =  \B e_\rA  \K{\Psi}_{\rA\rB} \\
 &= \sum_{i=1}^{d_\rA}  \B {a_i}_{\rA} \K{\Psi}_{\rA\rB} \\
 &= \sum_{i=1}^{r}  \B {a_i}_{\rA} \K{\Psi}_{\rA\rB}\\
  &= \sum_{i=1}^r  p_i \K{\psi_i}_{\rA},
\end{align*} 
which is the desired spectral decomposition. \qed

The spectral decomposition of states has many consequences.  Here we just discuss the simplest ones, which are needed for the purpose of the derivation of quantum theory.

A first  consequence is the following lemma:
\begin{lemma}\label{lem:fazero}
  Let $\varphi\in\Stset_1(\rA)$ be a pure state and let $a \in \Cntset (\rA)$ be the unique atomic effect such that
  $\SC a \varphi=1$. If $\varphi $ is perfectly distinguishable from $\rho$, then $\SC a \rho =0$.
\end{lemma}

\Proof Let us write $\rho = \sum_{i=1}^k p_i \varphi_i$, with
$\{\varphi_i\}_{i=1}^k$ perfectly distinguishable pure states and
$p_i>0$ for each $i$.  Now, by lemma \ref{lem:refinement} the states
$\{ \varphi_1, \dots, \varphi_k,\varphi\}$ are perfectly
distinguishable, and by lemma \ref{lem:extensiontomaximal} this set
can be extended to a maximal set of perfectly distinguishable pure
states $\{\gamma_m\}_{m=1}^{d_\rA}$, with $\gamma_i=\varphi_i$ for
$i\leq k$ and $\gamma_{k+1} = \varphi$.  Denote by
$\{c_m\}_{m=1}^{d_\rA}$ the observation test that perfectly
distinguishes between the states $\{\gamma_m\}$.  Note that, by
definition, $\SC {c_{k+1}}{\varphi}=1$ and $\SC {c_{k+1}}{\varphi_j} =
0$ for every $j\not = k+1$. Also, recall that $c_{k+1}$ is atomic
(lemma \ref{lem:atomicobstest}).  By the duality of theorem
\ref{theo:dualestato} we have $a= c_{k+1}$, and, therefore, $\SC {a}
\rho= \sum_{i =1 }^k p_i \SC {c_{k+1}} {\psi_i} = 0$.\qed


Another consequence of theorem \ref{theo:diag} is the following characterization of the completely
mixed states as \emph{full rank} states:

\begin{corollary}{\bf (Characterization of completely mixed states)}
  A state $\rho\in \Stset_1(\rA)$, written as a mixture $\rho
  =\sum_{i=1}^{d_\rA} p_i \varphi_i$ of a maximal set of perfectly
  distinguishable pure states $\{\varphi_i\}_{i=1}^{d_\rA}$, is
  completely mixed if and only if $p_i >0$ for every $i=1, \dots,
  d_\rA$.
  \label{cor:complemi}
\end{corollary}
\Proof Necessity: If $p_i = 0$ for some $i$, then $\rho$ is perfectly distinguishable from $\varphi_i$.
Hence, it cannot be completely mixed.  Sufficiency: let $p_{\min} = \min\{ p_i, i=1, \dots,
d_\rA\}$.  Then we have $\rho = p_{\min} \chi + (1- p_{\min}) \sigma$, where $\sigma$ is the state
defined by $\sigma = 1/ (1-p_{\min}) \sum_{i=1}^{d_\rA} (p_i - p_{\min}/d_\rA) \varphi_i$.
Since $\rho$ contains $\chi$ in its convex decomposition, and since $\chi$ is completely mixed, we conclude that $\rho$ is completely mixed. \qed In particular, for two-dimensional systems we have the result:

\begin{corollary}\label{cor:pureborder}
  For $d_\rA=2$ any state on the border of $\Stset_1 (\rA)$ is pure.
\end{corollary}

Another consequence of theorem \ref{theo:diag} is that every element in the vector space
$\Stset_\Reals (\rA)$ can be written as a linear combination of perfectly distinguishable states:
\begin{corollary}\label{cor:spectraldecomp}
  For every $\xi \in\Stset_\Reals (\rA)$ there exists a maximal set of perfectly distinguishable
  pure states $\{\varphi_i\}_{i=1}^{d_\rA}$ and a set of real numbers $\{c_i\}_{i=1}^{d_\rA}$ such
  that $\K{\xi} = \sum_i c_i \K{\varphi_i}$.
\end{corollary}  

\Proof Write $\xi$ as $\xi =c_+ \rho -c_- \sigma$, where $c_+, c_-\ge 0$ and $\rho$ and $\sigma$ are
normalized states.  If $c_- =0$ there is nothing to prove, because $\xi$ is proportional to a state.
Then, suppose that $c_- >0$.  Write $\sigma$ as $\sigma =\sum_i p_i \psi_i$ where $\{\psi_i\}$ are
perfectly distinguishable and define $k = \max\{p_i\}$. Then one has $\chi + 1/(c_- k d_\rA) \xi =
(\chi - 1/(kd_\rA) \sigma) + c_+/(c_-kd_\rA) \rho$. Now, by definition $\chi - 1/(kd_\rA) \sigma$ is
proportional to a state: indeed we have $(\chi - 1/(kd_\rA) \sigma) = 1/d_\rA \sum_i (1- p_i/k)
\psi_i$, and, by definition $1-p_i/k\ge 0$.  Therefore $\chi + 1/(c_- k d_\rA) \xi$ is proportional
to a state, say $\chi + 1/(c_- k d_\rA) \xi = t \tau$, with $t>0$. Writing $\tau$ as $\tau = \sum_i
q_i \varphi_i$, where $\{\varphi_i\}_{i=1}^{d_\rA}$ is a maximal set of perfectly distinguishable
pure states, we then obtain $\xi = (c_- k d_\rA) (t \tau - \chi) =(c_- k d_\rA) \sum_i (t q_i -
1/d_\rA) \varphi_i$, which is the desired decomposition. \qed

In quantum theory, corollary \ref{cor:spectraldecomp} is equivalent to the fact that every Hermitian matrix is diagonal in a suitable orthonormal basis.  
A simple consequence of corollary \ref{cor:spectraldecomp} is the following
\begin{corollary}\label{cor:continuouspurestates}
For every system $\rA$ with $d_\rA=2$ there is a continuous set of pure states.
\end{corollary}
\Proof Let $\xi\in\Stset_\Reals (\rA)$ be an arbitrary vector such that $\SC e \xi =0$.  Note that
since the convex set $\Stset_1 (\rA)$ cannot be a segment (corollary \ref{cor:nosegment}), we must
have $D_\rA = {\rm dim}  [\Stset_\Reals (\rA)] >2$ and, therefore, the space of vectors $\xi$ such that $\SC e \xi = 0$ is at least
two-dimensional.  By corollary \ref{cor:spectraldecomp}, we have $\xi = c (\varphi_1
-\varphi_2) = 2c( \varphi_1 - \chi)$, where $c\ge 0$,  $\{\varphi_1, \varphi_2\}$ are two perfectly
distinguishable pure states and we used the fact that $\chi = \frac 12 (\varphi_1 + \varphi_2)$.
Let us define $\varphi_\xi := \varphi_1$.  With this definition, if $\varphi_{\xi_1} =
\varphi_{\xi_2}$ then one has $\xi_2 = t \xi_1$ for some $t\ge 0$.  Now, since there is a continuous
infinity of vectors $\xi$ (up to scaling), there must be a continuous set of pure states.  \qed

We conclude this section with the dual result to the ``spectral decomposition" of corollary \ref{cor:spectraldecomp}:  

\begin{corollary}\label{cor:spectraldecompeff} 
  For every $x \in\Cntset_\Reals (\rA)$ there exists a perfectly distinguishing observation-test $\{a_i\}_{i=1}^{d_\rA}$ and a set of real numbers $\{d_i\}_{i=1}^{d_\rA}$ such
  that $\B{x} = \sum_i d_i \B{a_i}$.
\end{corollary}

\Proof  Let $\Phi \in \Stset_1(\rA\tilde \rA)$ be a purification of the invariant state $\chi_\rA$, where $\tilde \rA$ is the conjugate system defined in  corollary \ref{cor:conjusys}.  
 Take the Choi vector $\K{R_x}_{\tilde \rA} : = \B {x}_\rA \K{\Phi}_{\rA\tilde \rA}$. By corollary \ref{cor:spectraldecomp}, there exists a maximal set of perfectly distinguishable pure states $\{\psi_i\}_{i=1}^{d_\rA}$ and a set of real numbers $\{c_i\}_{i=1}^{d_\rA}$ such that  $\K{R_x} = \sum_i c_i \K{\psi_i}$.  Let $\{a_i\}_{i=1}^{d_\rA}  \subset \Cntset(\rA)$ be the observation-test such that $ 1/d_\rA \K{\psi_i}_{\tilde \rA} =    \B{a_i}_\rA  \K{\Phi}_{\rA\tilde\rA}$ for every $i=1, \dots, d_\rA$ (recall that by corollary \ref{cor:conjusys} the marginal of $\Phi$ on system $\tilde \rA$ is the invariant state $\chi_{\tilde \rA}$ and $d_{\tilde \rA} = d_{\rA}$).  The test $\{a_i\}_{i=1}^{d_\rA}$ is perfectly distinguishing:  if $\{b_i\}_{i=1}^{d_\rA}$ is the observation-test such that  $\SC {b_i}{\psi_j} = \delta_{ij}$ and $\varphi_i\in\Stset_1 (\rA)$ is the state defined by $\K{\varphi_i}_\rA :=d_\rA     \B{b_i}_{\tilde \rA}  \K{\Phi}_{\rA\tilde\rA}$, then we have
\begin{align*}
\SC {a_i}{\varphi_j}  & =  d_\rA  \SC {a_i \otimes b_j}{\Phi}\\
& =  \SC {b_j}{\psi_i} \\
&= \delta_{ij}.   
 \end{align*}
 Moreover, we have 
 \begin{align*}
\B{x}_\rA  \K{\Phi}_{\rA\tilde\rA} & = \K{R_x}_{\tilde \rA} \\
& =  \sum_i  c_i \K{\psi_i}_{\tilde \rA}  \\
& =  \sum_i c_i d_\rA   \B{a_i}_{\rA}  \K{\Phi}_{\rA\tilde\rA}.  
 \end{align*}
 Since $\Phi$ is dynamically faithful, this implies $\B x =  \sum_i d_i  \B{a_i}$, where $d_i: = c_i  d_\rA$.  \qed

\section{Teleportation revisited}\label{sec:teleportation}
In this section we revisit probabilistic teleportation using the results about informational dimension. The key point is the section will be the proof the equality $D_\rA  = d_\rA^2$, which relates the dimension of the vector space $\Stset_\Reals (\rA)$ with the informational dimension $d_\rA$. 

\subsection{Probability of teleportation}
We start by showing a probabilistic teleportation scheme that achieves success probability $p_\rA = 1/d_\rA$ for every system $\rA$:

\begin{theorem}[Probability of teleportation]
  For every system $\rA$, probabilistic teleportation can be achieved
  with probability $p_\rA = 1/{d_\rA^2}$.
  \label{theo:probatele}
\end{theorem}

\Proof Let $\tilde \rA$ and $\K{\Phi }_{\rA\tilde\rA}$ be the conjugate system and the pure state defined in
corollary \ref{cor:conjusys}.  Then, the state $\K\Phi_{\rA\tilde\rA} \K{\Phi}_{\rA\tilde\rA}$
satisfies the identity
\begin{equation*}
  \begin{aligned}
    \Qcircuit @C=1em @R=.7em @! R {&\multiprepareC{1}{\Phi}&\qw\poloFantasmaCn{\rA}&\qw\\
      &\pureghost{\Phi}&\qw\poloFantasmaCn{\tilde\rA}&\measureD{e}\\
      &\multiprepareC{1}{\Phi}&\qw\poloFantasmaCn{\rA}&\measureD{e}\\
      &\pureghost{\Phi}&\qw\poloFantasmaCn{\tilde\rA}&\qw}
  \end{aligned}=
  \begin{aligned}
    \Qcircuit @C=1em @R=.7em @! R {&\prepareC{\chi}&\qw\poloFantasmaCn{\rA}&\qw\\
      &\prepareC{\chi}&\qw\poloFantasmaCn{\tilde\rA}&\qw}
  \end{aligned}=  \begin{aligned}
    \Qcircuit @C=1em @R=.7em @! R {&\prepareC{\chi}&\qw\poloFantasmaCn{\rA\tilde\rA}&\qw}
  \end{aligned}
\end{equation*}
On the other hand, by lemma \ref{lem:pmax} the maximum probability of a pure state in the convex
decomposition of $\chi_{\rA\tilde\rA}$ is $p_{\max} =1/d_{\rA\tilde\rA}$, and by corollaries
\ref{cor:dimprod} and \ref{cor:conjusys} one has $p_{\max} =1/(d_\rA d_{\tilde \rA})= 1/d_\rA^2$.
Therefore, by lemma \ref{lem:purificami-ens} there exists an atomic effect $E$ such that
\begin{equation}
  \begin{aligned}
    \Qcircuit @C=1em @R=.7em @! R {&\multiprepareC{1}{\Phi}&\qw\poloFantasmaCn{\rA}&\qw\\
      &\pureghost{\Phi}&\qw\poloFantasmaCn{\tilde\rA}&\multimeasureD{1}{E}\\
      &\multiprepareC{1}{\Phi}&\qw\poloFantasmaCn{\rA}&\ghost{E}\\
      &\pureghost{\Phi}&\qw\poloFantasmaCn{\tilde\rA}&\qw}
  \end{aligned}=\ \frac1{d_\rA^2}
  \begin{aligned}
    \Qcircuit @C=1em @R=.7em @! R {&\multiprepareC{1}{\Phi}&\qw\poloFantasmaCn{\rA}&\qw\\
      &\pureghost{\Phi}&\qw\poloFantasmaCn{\tilde\rA}&\qw}
  \end{aligned}
  \label{entaswa}
\end{equation}
and, since $\Phi$ is dynamically faithful,
\begin{align}
  \begin{aligned}
    \Qcircuit @C=1em @R=.7em @! R {&\multiprepareC{1}{\Phi}&\qw\poloFantasmaCn{\rA}&\qw\\
      &\pureghost{\Phi}&\qw\poloFantasmaCn{\tilde\rA}&\multimeasureD{1}{E}\\
      &&\qw\poloFantasmaCn{\rA}&\ghost{E}}
  \end{aligned}=\ \frac1{d_\rA^2}\ 
  \begin{aligned}
    \Qcircuit @C=1em @R=.7em @! R {&\qw\poloFantasmaCn{\rA}&\gate{\tI}&\qw\poloFantasmaCn{\rA}&\qw}
  \end{aligned}
  \label{probatelecirc}
\end{align}
as can be verified applying both members of Eq.~\eqref{probatelecirc}
to $\Phi$, thus obtaining Eq.~\eqref{entaswa}.  \qed

\subsection{Isotropic states and effects}
Here we define two maps that send reversible transformations of $\rA$ to reversible transformations
of $\tilde \rA$: the transpose and the conjugate.  Using these maps we will also define the notions
of isotropic states and effects and we will prove some properties of them.
 
Let us start from the definition of the transpose:
\begin{lemma}{\bf (Transpose of a reversible transformation)}
  Let $\Phi\in\Stset(\rA\tilde\rA)$ be a purification of the invariant state $\chi_\rA$. The
  reversible transformations of system $\tilde\rA$ are in one-to-one correspondence with the
  reversible transformations of system $\rA$ via the transposition $\tau$ defined as follows
  \begin{equation}
    \begin{aligned}
      \Qcircuit @C=1em @R=.7em @! R {&\multiprepareC{1}{\Phi}&\qw\poloFantasmaCn{\rA}&\gate{\tU}&\qw\poloFantasmaCn{\rA}&\qw\\
        &\pureghost{\Phi}&\qw\poloFantasmaCn{\tilde\rA}&\qw&\qw&\qw}
    \end{aligned}=
    \begin{aligned}
      \Qcircuit @C=1em @R=.7em @! R {&\multiprepareC{1}{\Phi}&\qw\poloFantasmaCn{\rA}&\qw&\qw&\qw\\
        &\pureghost{\Phi}&\qw\poloFantasmaCn{\tilde\rA}&\gate{\tU^{\tau}}&\qw\poloFantasmaCn{\tilde\rA}&\qw}
    \end{aligned}
    \label{transpo}
  \end{equation}
  \label{lem:transpose}
  [note that the transposition is defined with respect to the given state $\Phi$]
\end{lemma}

\Proof Since $(\tU\otimes\tI_{\tilde\rA})|\Phi)$ and $|\Phi)$ are purifications of the same state
$\chi_\rA$, there exists a reversible transformation $\tU^{\tau}\in \grp G_{\tilde\rA}$ such that
Eq.~\eqref{transpo} holds.  Since $\Phi$ is dynamically faithful on $\rA$, the map $\tU \mapsto
\tU^\tau$ is injective.  Furthermore, the map is surjective: for every reversible
$\tV\in\grp G_{\tilde\rA}$ the states $(\tI_\rA \otimes \tV)\K\Phi$ and $|\Phi)$ are two
purifications of the same state $\chi_{\tilde \rA}$, and, by the uniqueness of purification stated
in postulate \ref{purification}, there exists a reversible $\tU\in\grp G_{\rA}$ such that
\begin{equation}
  \begin{aligned}
    \Qcircuit @C=1em @R=.7em @! R {&\multiprepareC{1}{\Phi}&\qw\poloFantasmaCn{\rA}&\qw&\qw&\qw\\
      &\pureghost{\Phi}&\qw\poloFantasmaCn{\tilde\rA}&\gate{\tV}&\qw\poloFantasmaCn{\tilde\rA}&\qw}
  \end{aligned}=
  \begin{aligned}
    \Qcircuit @C=1em @R=.7em @! R {&\multiprepareC{1}{\Phi}&\qw\poloFantasmaCn{\rA}&\gate{\tU}&\qw\poloFantasmaCn{\rA}&\qw\\
      &\pureghost{\Phi}&\qw\poloFantasmaCn{\tilde\rA}&\qw&\qw&\qw}
  \end{aligned}
\end{equation}
namely $\tV=\tU^\tau$.\qed

The conjugate is just defined as the inverse of the transpose:
\begin{definition}\label{def:conjugaterep}
  Let $\tau$ be the transpose defined with respect to the state $\Phi\in\Stset_1(\rA\tilde\rA)$.  The
  conjugate of the reversible channel $\tU\in\grp G_\rA$ is the reversible channel
  $\tU^*\in\grp G_{\tilde\rA}$ defined by $\tU^*:=(\tU^\tau)^{-1}$.
\end{definition}

We can now give the definition of isotropic pure state (isotropic atomic effect):
\begin{definition}
  A pure state $\Psi\in\Stset(\rA\tilde\rA)$  (an atomic effect
  $F\in\Cntset(\tilde \rA\rA)$) is isotropic if it is invariant under
  the $\tU \otimes \tU^*$ (under $\tU^* \otimes \tU$).  Diagrammatically
  \begin{equation} \label{isotropic}
    \begin{split}
      \begin{aligned}
        \Qcircuit @C=1em @R=.7em @! R {&\multiprepareC{1}{\Psi}&\qw\poloFantasmaCn{\rA}&\gate{\tU}&\qw\poloFantasmaCn{\rA}&\qw\\
          &\pureghost{\Psi}&\qw\poloFantasmaCn{\tilde\rA}&\gate{\tU^*}&\qw\poloFantasmaCn{\tilde\rA}&\qw}
      \end{aligned} &=
    \begin{aligned}
      \Qcircuit @C=1em @R=.7em @! R
      {&\multiprepareC{1}{\Psi}&\qw\poloFantasmaCn{\rA}&\qw\\
        &\pureghost{\Psi}&\qw\poloFantasmaCn{\tilde\rA}&\qw}
    \end{aligned} \qquad \forall \tU \in \grp G_\rA \\
    \left(\begin{aligned}
        \Qcircuit @C=1em @R=.7em @! R {&\qw\poloFantasmaCn{\tilde \rA}&\gate{\tU^*}&\qw\poloFantasmaCn{\tilde \rA}&\multimeasureD{1}{F}\\
          &\qw\poloFantasmaCn{\rA}&\gate{\tU}&\qw\poloFantasmaCn{\rA}&\ghost{F}}
      \end{aligned}\right.
    &= \left. \begin{aligned} \Qcircuit @C=1em @R=.7em @! R
        {&\qw\poloFantasmaCn{\tilde \rA}&\multimeasureD{1}{F}\\
          &\qw\poloFantasmaCn{\rA}&\ghost{F}\\}
      \end{aligned} \qquad \forall \tU \in \grp G_\rA \right)
  \end{split}
\end{equation}
\end{definition}

An example of isotropic state is $\Phi$: indeed, by definition of conjugate we have, for every
$\tU\in\grp G_\rA$,
\begin{align*}
(\tU \otimes \tU^*) \K \Phi &= (\tU \otimes (\tU^\tau)^{-1})  \K \Phi \\
&= (\tI_\rA \otimes  (\tU^\tau)^{-1} \tU^\tau) \K \Phi = \K \Phi.
\end{align*} 
As a consequence, the teleportation effect $E$ is isotropic: indeed one has
\begin{align*}
  \begin{aligned}
    \Qcircuit @C=1em @R=.7em @! R {&\multiprepareC{1}{\Phi}&\qw\poloFantasmaCn{\rA}&\qw&\qw\\
      &\pureghost{\Phi}&\qw\poloFantasmaCn{\tilde \rA}& \gate {\tU^*}  &\multimeasureD{1}{E}\\
      &\multiprepareC{1}{\Phi}&\qw\poloFantasmaCn{\rA}&\gate{\tU} &  \ghost{E}\\
      &\pureghost{\Phi}&\qw\poloFantasmaCn{\tilde\rA}&\qw&\qw}
  \end{aligned} &= \begin{aligned}
    \Qcircuit @C=1em @R=.7em @! R {&\multiprepareC{1}{\Phi}&\qw\poloFantasmaCn{\rA}&\gate{\tU^{-1}}&\qw\poloFantasmaCn{\rA}&\qw \\
      &\pureghost{\Phi}&\qw\poloFantasmaCn{\tilde\rA} &\multimeasureD{1}{E} &&\\
      &\multiprepareC{1}{\Phi}&\qw\poloFantasmaCn{\rA}&  \ghost{E}&&\\
      &\pureghost{\Phi}&\qw\poloFantasmaCn{\tilde\rA}&\gate{\tU^\tau} &\qw\poloFantasmaCn{\tilde\rA}&\qw}
  \end{aligned}\\
   &=\frac1{d_\rA^2}
  \begin{aligned}
   \Qcircuit @C=1em @R=.7em @! R {&\multiprepareC{1}{\Phi}&\qw\poloFantasmaCn{\rA}&\gate{\tU^{-1}} & \qw\poloFantasmaCn{\rA}&\qw\\
      &\pureghost{\Phi}&\qw\poloFantasmaCn{\tilde\rA}&\gate{\tU^\tau} &\qw\poloFantasmaCn{\tilde\rA} &\qw}
      \end{aligned}\\
      &=\frac1{d_\rA^2}
  \begin{aligned}
    \Qcircuit @C=1em @R=.7em @! R {&\multiprepareC{1}{\Phi}&\qw\poloFantasmaCn{\rA}&\qw\\
      &\pureghost{\Phi}&\qw\poloFantasmaCn{\tilde\rA}&\qw}
  \end{aligned}\\
  &=  \begin{aligned}
    \Qcircuit @C=1em @R=.7em @! R {&\multiprepareC{1}{\Phi}&\qw\poloFantasmaCn{\rA}&\qw\\
      &\pureghost{\Phi}&\qw\poloFantasmaCn{\tilde \rA}&\multimeasureD{1}{E}\\
      &\multiprepareC{1}{\Phi}&\qw\poloFantasmaCn{\rA}&  \ghost{E}\\
      &\pureghost{\Phi}&\qw\poloFantasmaCn{\tilde\rA}&\qw}
  \end{aligned}
\end{align*}
which implies $ \B E (\tU^*\otimes \tU) = \B E$, since the state $\Phi \otimes \Phi$ is dynamically faithful.  

We now show that all isotropic pure states (isotropic atomic effects) are connected to the state
$\Phi$ (to the effect $E$) through a local reversible transformation.
\begin{lemma}
  If a pure state $\Psi\in\Stset_1(\rA\tilde\rA)$ is isotropic then
  $|\Psi) =(\tV\otimes\tI_{\tilde\rA})|\Phi)$
  for some reversible transformation $\tV\in\grp G_\rA$ such that $\tV
  \tU = \tU \tV$ for every $\tU \in \grp G_\rA$.
\label{lem:isostat}
\end{lemma}

\Proof Since
$\Psi$ satisfies Eq. (\ref{isotropic}), its marginal on system $\tilde \rA$ is the invariant
state $\K{\chi_{\tilde \rA}}$. Since $\Psi$ and $\Phi$ are
purifications of the same state, there must exist a reversible channel
$\tV \in \grp G_{ \rA}$ such that $\K{\Psi} = (\tV\otimes\tI_{\tilde \rA})
\K{\Phi}$.  Moreover, we have  for every $\tU \in \grp G_\rA$
\begin{align*}
 (\tU \tV  \tU^{-1} \otimes \tI_{\tilde \rA}) \K{\Phi} &= 
 (\tU \tV \otimes \tU^*) \K{\Phi} \\
 & =  (\tU  \otimes \tU^*) \K{\Psi} \\
 & =  \K{\Psi} \\
 & =  (\tV \otimes \tI_{\tilde \rA}) \K{\Phi}. 
\end{align*}
Since $\Phi$ is dynamically faithful, the above equation implies $\tU \tV \tU^{-1} = \tV$ for every $\tU \in \grp G_\rA$.  
\qed
By the duality between states and effects, it is easy to obtain the following:
\begin{lemma}
  Let $A\in \Cntset (\rA\tilde\rA)$ be the atomic effect such that $\SC A \Phi =1$.  If an atomic
  effect $F\in\Cntset(\tilde \rA\rA)$ is isotropic then $(F|_{\tilde \rA\rA}=(A|_{\tilde
    \rA\rA}(\tI_{\tilde\rA}\otimes \tV)$ for some reversible transformation $\tV\in\grp G_\rA$ such
  such that $\tV \tU = \tU \tV$ for every $\tU \in \grp G_\rA$.
  \label{lem:isoeff}
\end{lemma}
\Proof Let $\Psi$ be the pure state such that $\SC F \Psi =1$.  Clearly $\Psi$ is isotropic: one has
$\B F (\tU\otimes \tU^*) \K \Psi = \SC F \Psi =1$, and, therefore, $(\tU \otimes \tU^*) \K \Psi = \K
\Psi$.  By lemma \ref{lem:isostat}, there exists a reversible transformation $\tV$ such that $\K \Psi =
(\tV^{-1} \otimes \tI_{\tilde\rA}) \K \Phi$  and $\tV ^{-1}\tU = \tU \tV^{-1}$ for every $\tU \in \grp G_\rA$. 
Now, this implies $\B F (\tV^{-1} \otimes \tI_{\tilde \rA}) \K\Phi = \SC F \Psi = 1$, which by theorem 
\ref{theo:dualestato} implies $\B F = \B A (\tV \otimes \tI_{\tilde \rA})$.\qed

As a consequence, every isotropic effect is connected to the teleportation effect by a local
reversible transformation:
\begin{corollary}
  If an atomic effect $F\in\Cntset (\tilde \rA\rA)$ is isotropic
  then $(F|_{\tilde \rA\rA}=(E|_{\tilde \rA\rA}(\tI_{\tilde\rA}\otimes
  \tV)$ for some reversible transformation $\tV\in\grp G_\rA$ such
  that $\tV \tU = \tU \tV$ for every $\tU \in \grp G_\rA$.
  \label{cor:isoeff2}
\end{corollary}
\Proof Since $\B E$ and $\B F$ are both isotropic, lemma \ref{lem:isoeff} implies that they are both connected to $\B A$ through a local reversible transformation, say $\tV$ and $\tW$, respectively. Therefore, they are connected to each other through the transformation $\tW\tV^{-1}$. \qed

\subsection{Dimension of the state space}
In this subsection we use the local distinguishability axiom to prove the equality $D_\rA =
d_\rA^2$ (see theorem \ref{theo:squareconv}). As a consequence,
we will be able to represent the states of a system $\rA$ as square $d_\rA\times d_\rA$ hermitian complex
matrices, that is, hermitian operators on the complex Hilbert space $\Cmplx^{d_\rA}$.  Theorem \ref{theo:squareconv}
is thus the point where the complex field (as opposed to the real field)  enters in our derivation.  Notice that, even if the local distinguishability excludes quantum theory on real Hilbert spaces since the very beginning,  to prove the emergence of complex Hilbert spaces we need to use all the six principles.  

Due to local distinguishability, any bipartite state $\Psi \in\Stset(\rA\rB)$ can be written as
\begin{equation*}
  \K \Psi = \sum_{i=1}^{D_\rA} \sum_{j=1}^{D_\rB}  \Psi_{ij} ~ \K{\alpha_i} \K{\beta_j},
\end{equation*} 
where $\{\alpha_i\}$ ($\{\beta_j\}$) is a basis for the vector space
$\Stset_\Reals(\rA)$ ($\Stset_\Reals (\rB)$).  Similarly, a bipartite effect $F
\in\Cntset(\rB\rA)$ can be written as
\begin{equation*}
  \B F=\sum_{k=1}^{D_\rB}\sum_{l=1}^{D_\rA} F_{kl} ~  \B{\beta_k^*} \B{\alpha_l^*}
\end{equation*} 
with $\SC {\alpha_l^*}{\alpha_i} = \delta_{i l}$ and
$\SC{\beta^*_k}{\beta_j} = \delta_{jk}$.  Finally, a transformation
$\tC$ from $\rA$ to $\rB$ can be written as
\begin{equation*}
  \tC  = \sum_{j=1}^{D_\rB} \sum_{i=1}^{D_\rA} C_{ji}   ~\K {\beta_j} \B {\alpha_i^*}
\end{equation*} 
In this matrix representation, the teleportation diagram of Eq.
(\ref{probatelecirc}) becomes
\begin{equation} \label{matrixteleportation}
  \Phi E  =  \frac{I_{D_\rA}}{d_\rA^2} ,
\end{equation}
where $I_{D_\rA}$ is the identity matrix in dimension $D_\rA$.  On the
other hand, we also have
\begin{equation*}
  1\ge \SC E  \Phi = \Tr[\Phi E] = \frac{D_\rA}{d_\rA^2}
\end{equation*} 
and, therefore,
\begin{equation*}
  D_\rA \le d_\rA^2.
\end{equation*} 
We now show that one has the equality, using the following standard
lemma:

\begin{lemma}
  With a suitable choice of basis for the vector space $\Stset_\Reals
  (\rA)$, every reversible transformation $\tU \in \grp G_\rA$ is represented by
  a matrix $M_\tU$ of the form
  \begin{equation}
    M_\tU=\left(\begin{array}{c|c}
        1&0\\
        \hline 0&O_\tU
      \end{array}\right),
    \label{block}
  \end{equation}
  where $O_\tU$ is an orthogonal $(D_\rA-1) \times (D_\rA-1)$ matrix.
  \label{lem:orthogo}
\end{lemma}

\Proof Let $\{\xi_i\}$ be a basis for $\Stset_\Reals (\rA)$, chosen in such a way that the first
basis vector is $\chi$, while the remaining vectors satisfy $\SC e {\xi_i} = 0, \forall i = 2,
\dots, D_\rA$.  Such a choice is always possible since every vector $v \in\Stset_\Reals (\rA)$ can
be written as $v = \SC e v ~ \chi + \xi $, where $\xi$ satisfies $\SC e \xi = 0$.  Now, since $\tU
\chi = \chi$, the first column of $M_\tU$ must be $(1, 0 ,\dots ,0)^T$. Moreover, since for every
normalized state $\rho$, $\tU \rho$ is a normalized state, one must have $\B e \tU \K \xi =0$ for
every $\xi$ such that $\SC e \xi =0$. Hence, the first row of $M_\tU$ must be $(1, 0,\dots, 0 )$,
namely $M_\tU$ has the block form of Eq. (\ref{block}). It remains to show that, with a suitable
choice of basis, the matrix $O_\tU$ in the second block can be chosen to be orthogonal.  Observe that, by definition the matrices $\{M_\tU\}_{\tU \in \grp G_\rA}$ form a representation of the group $\grp G_\rA$: indeed, one has $M_{\tI}  =  I_{D_\rA}$ and $M_{\tU \tV} =  M_{\tU}  M_{\tV}$ for every $\tU, \tV \in \grp G$. 
Consider the
positive definite matrix $P$ defined by the integral
\begin{equation*}
  P:=\int{\rm d} \tU O_\tU^T O_\tU,
\end{equation*}
where ${\rm d} \tU$ is the Haar measure on the compact group $\grp G_\rA$ (see  corollary 30 of Ref. \cite{purification} for the proof of compactness)  and $A^T$ denotes the transpose of
$A$.  By definition, one has $P^T = P$ and $O^T_\tU P O_\tU=P$ for every $\tU \in \grp G_\rA$.  Let
us now define the new representation
\begin{equation*}
  O'_\tU:=P^{\frac12}O_\tU P^{-\frac12},
\end{equation*}
obtained from $O_\tU$ by a change of basis in the subspace spanned by $\{\xi_i\}_{i=2}^{D_\rA}$.
With this choice, each matrix $O'_\tU$ is orthogonal:
\begin{equation*}
\begin{split}
  {O'}_\tU^T {O'}_\tU&=\left(P^{\frac12} O_\tU P^{-\frac12}\right)^T \left (P^{\frac12}O_\tU P^{-\frac12}\right)\\
  &=P^{-\frac12} \left (O^T_\tU P  O_\tU\right) P^{-\frac12}= I_{D_\rA-1}.
\end{split}
\end{equation*}
\qed

As a consequence, we have the following:
\begin{corollary}
  For every system $\rA$, the group of reversible transformations
  $\grp G_\rA$ is (isomorphic to) a compact subgroup of ${\mathbb
    O}(D_\rA-1)$.
  \label{cor:orthogo2}
\end{corollary}  

\begin{lemma}\label{lem:EPhi=1}
  Let $E\in\Cntset(\rA \tilde \rA)$ be the teleportation effect of Eq.
  (\ref{probatelecirc}). Then, one has $\SC E \Phi =1$.
\end{lemma}
\Proof Let $A\in\Cntset (\rA\tilde \rA)$ be the atomic effect such that $\SC A \Phi =1$. We now
prove that $A = E$.  Indeed, by corollary \ref{cor:isoeff2} there exists a reversible transformation
$\tV \in\grp G_\rA$ such that $\B A = \B E (\tV \otimes \tI_{\tilde \rA})$.  Using a basis for
$\Stset_\Reals (\rA)$ such that the transformations in $\grp G_\rA$ are represented by orthogonal
matrices as in Eq. (\ref{block}), one has
\begin{align}
\nonumber1 &= \SC A \Phi\\
\nonumber& = \B  E  (\tV\otimes \tI_{\tilde \rA})  \K \Phi\\
\nonumber & =\Tr [E M_\tV  \Phi] \\
\nonumber &= \Tr[\Phi E M_\tV]\\
\nonumber &= \frac{\Tr[M_\tV]}{d_\rA^2} , 
\end{align}
having used Eq. (\ref{matrixteleportation}) for the last equality.    
Using the inequality $\Tr[M_\tV] \le \Tr[I_{D_\rA}]$, that holds for every orthogonal $D_\rA \times D_\rA$ matrix, we then obtain 
\begin{align}
\nonumber 1 & = \frac{\Tr[M_\tV]}{d_\rA^2} \\
\nonumber &\le \frac{\Tr[I_{D_\rA}]}{d_\rA^2}\\
\nonumber & = \Tr[E\Phi]\\
\nonumber & = \SC E \Phi\\
\nonumber & \le 1  ,
\end{align}
ans, therefore $\SC  E \Phi = 1$ 
 \qed 

\begin{theorem}[Dimension of the state space]\label{theo:squareconv}
The dimension $D_\rA$ of the vector space generated by the states in $\Stset(\rA)$ is $D_\rA=d_\rA^2$.  
\end{theorem}
\Proof  Using lemma \ref{lem:EPhi=1} and Eq. (\ref{matrixteleportation}) we obtain $1 = \SC E \Phi  = \Tr[E\Phi]  =  \Tr [I_{D_\rA} ]  /d_\rA^2  = D_\rA/d_\rA^2$.  Hence, $D_\rA = d_\rA^2$\qed 

An interesting consequence of the relation $\SC E \Phi =1$ is the following
\begin{corollary}[No inversion]
  Let us write an arbitrary state $\rho\in\Stset_1(\rA)$ as $\rho =
  \chi_\rA + \xi$, with $\SC e \xi = 0$.  Then, the linear map $\tN$
  defined by $\tN (\rho) = \chi_\rA - \xi$ is not a physical
  transformation.
  \label{cor:noinversion}
\end{corollary}
\Proof Write the state $\Phi$ as $\Phi = \chi_\rA \otimes \chi_{\tilde \rA} + \Xi$.  Since $\B e_\rA
\K \Phi_{\rA\tilde\rA} = \K \chi_{\tilde \rA}$ one must have $\B e_\rA \K \Xi_{\rA\tilde \rA} =0$.
Therefore, $\Xi$  must be of the form $\Xi =  \sum_i  \alpha_i \otimes \beta_i$ with $\SC e {\alpha_i} = 0$ for all $i$.   Applying the transformation $\tN$ one then obtains $(\tN \otimes \tI_{\tilde\rA}) \Phi =
\chi_\rA \otimes \chi_{\tilde\rA} - \Xi$. We now prove that this is not a state, and therefore,
$\tN$ cannot be a physical transformation.  Let $E$ be the teleportation effect. Since $\SC E
\Phi=1$, we have $1 = \SC E {\chi_\rA \otimes \chi_{\tilde \rA}} + \SC E \Xi = 1/d^2_{\rA} + \SC E
\Xi$.  Now, we have
\begin{align*}
  (E|(\tN \otimes \tI_{\tilde \rA}) |\Phi)&=\frac1{d_\rA^2}-(E |\Xi)=\frac2{d_\rA^2}-1,
\end{align*}
Since this quantity is negative for every $d_\rA >1$, the map $\tN$ cannot be a physical
transformation. \qed

\begin{corollary}\label{cor:noinversion2}
  The matrix $M_\tN$ defined as 
\begin{equation}
    M_\tN=\left(\begin{array}{c|c}
        1&0\\
        \hline 0&-I_{D_\rA-1}
      \end{array}\right),
  \end{equation}
  cannot represent a physical transformation of system $\rA$.
\end{corollary}


\section{Derivation of the qubit}\label{sec:qubits}

In this section we show that every two-dimensional system in our theory is a \emph{qubit}. With this
expression we mean that the normalized states in $\Stset_1(\rA)$ can be represented as density matrices
for a quantum system with two-dimensional Hilbert space.  With this choice of representation we also
show that the effects in $\Cntset(\rA)$ are all the positive Hermitian matrices bounded by the
identity, and that the reversible transformations $\grp G_\rA$ act on the states by conjugation with
unitary matrices in $\mathbb{SU} (2)$.

The first step is to prove that the set of normalized states $\Stset_1 (\rA)$ is a sphere.  The idea
of the proof is a simple geometric observation: in the ordinary three-dimensional space the sphere
is the only compact convex set that has an infinite number of pure states connected by orthogonal
transformations. The complete proof is given in the following
\begin{theorem}[The Bloch sphere]
  The normalized states of a system $\rA$ with $d_\rA=2$ form a sphere and the group $\grp G_\rA$ is
  $\mathbb {SO} (3)$.
  \label{theo:qubit}
\end{theorem}

\Proof According to corollary \ref{cor:orthogo2}, the group of reversible transformations $\grp
G_\rA$ is a compact subgroup of the orthogonal group ${\mathbb O}(3)$. It cannot be the whole
${\mathbb O} (3)$ because, as we saw in corollary \ref{cor:noinversion2}, the inversion $-I$ cannot
represent a physical transformation.  We now show that $\grp G_\rA$ must be $\mathbb{SO}(3)$ by
excluding all the other possibilities.  From corollary \ref{cor:continuouspurestates} we know that
the system $\rA$ has a continuum of pure states.  Therefore, the group $\grp G_\rA$ must contain a continuous
set of transformations.  Now, from the classification of the closed subgroups of $\mathbb O (3)$ we
know that there are only two possibilities: \emph{i)} $\grp G_\rA$ is $\mathbb {SO} (3)$ and
\emph{ii)} $\grp G_\rA$ is the subgroup generated by $\mathbb {SO} (2)$, the group of rotations
around a fixed axis, say the $z$-axis, and possibly by the reflections with respect to planes
containing the $z$-axis.  Note that the reflection in the $xy$-plane is forbidden, because the
composition of this reflection with the rotation of $\pi$ around the $z$-axis would give the
inversion, which is forbidden by corollary \ref{cor:noinversion}.  The case \emph{ii)} is excluded because in this case the action of the group $\grp
G_\rA$ cannot be transitive.  The detailed proof is as follows: because of the $\mathbb{SO} (2)$
symmetry, the set of pure states must contain at least a circle in the $xy$-plane.  This circle will
be necessarily invariant under all operations in the group.  However, since the convex set of states
is three dimensional, there is at least a pure state outside the circle.  Clearly, there is
no way to transform a state on the circle into  a state outside the circle by means of an operation in $\grp
G_\rA$. This is in contradiction with the fact that every two pure states are connected by a
reversible transformation.  Hence, the case $\emph{ii)}$ is ruled out.  The only remaining
alternative is \emph{i)}, namely that $\grp G_\rA = \mathbb{SO} (3)$ and, hence, the set of pure
states generated by its action on a fixed pure state is a sphere.  \qed

Since the convex set of density matrices on a two-dimensional Hilbert space is a sphere, we can
represent the states in $\Stset_1(\rA)$ as density matrices.  Precisely, we can choose three
orthogonal axes passing through the center of the sphere and call them $x,y,z$ axes, take
$\varphi_{+, k}, \varphi_{-,k}$, $k=x,y,z$ to be the two perfectly distinguishable pure states in
the direction of the $k$-axis and define $\sigma_k : = \varphi_{k,+} - \varphi_{k,-}$.  From the
geometry of the sphere we know that any state $\rho \in\Stset_1(\rA)$ can be written as
\begin{equation}\label{operationalbloch}
\K\rho = \K \chi  +  \frac 12  \sum_{k=x,y,z}  n_k  \K {\sigma_k}  \qquad \sum_{k=x,y,z}  n_k^2 \le 1,
\end{equation}
where the pure states are those for which $\sum_{k=x,y,z} n_k^2 =1$.  The Bloch representation $S_\rho$  of
quantum state $\rho$  is then obtained by associating the basis vectors $\chi, \sigma_x, \sigma_y,
\sigma_z$ to the matrices
\begin{align*}
\nonumber &S_\chi = \frac 12 \begin{pmatrix}   1 & 0 \\0&1  \end{pmatrix} \quad S_{\sigma_x} =  \begin{pmatrix}   0 & 1 \\1&0  \end{pmatrix}  \\  
 \label{mappalinearestates}
 &S_{\sigma_y} =  \begin{pmatrix}   0 & -i \\i &0  \end{pmatrix} \quad S_{\sigma_z} =  \begin{pmatrix}   1 & 0 \\0&-1  \end{pmatrix}
 \end{align*}
 and by defining $S_\rho$ by linearity from Eq. (\ref{operationalbloch}).  Clearly, in this way we
 obtain $S_\rho = \frac 12 \begin{pmatrix} 1 + n_z & n_x-in_y \\ n_x + i n_y & 1-n_z \end{pmatrix}$,
 which is the expression of a generic density matrix. Denoting by $M_2 (\Cmplx)$ the set of complex
 two-by-two matrices we have the following
\begin{corollary}[Qubit density matrices]
For  $d_\rA = 2$ the set of states $\Stset_1 (\rA)$ is isomorphic to the set of density matrices in $M_{2} (\Cmplx)$ through the isomorphism $\rho  \mapsto S_\rho$.   
\end{corollary}

Once we decide to represent the states in $\Stset_1 (\rA)$ as matrices, the effects in
$\Cntset(\rA)$ are necessarily represented by matrices too.  The matrix representation of an effect,
given by the map $a \in\Cntset (\rA) \mapsto E_a \in M_2(\Cmplx)$ is defined uniquely by the
relation
\begin{equation*}
\Tr[  E_a  S_\rho] =  \SC a \rho \qquad \forall \rho \in\Stset(\rA).
\end{equation*} 
We then have the following
\begin{corollary}
  For $d_\rA = 2$ the set of effects $\Cntset (\rA)$ is isomorphic to the set of positive Hermitian
  matrices $P \in M_{2} (\Cmplx)$ such that $P\le I$.
\end{corollary}
\Proof Clearly the matrix $E_a$ must be positive for every effect $a$, since we have $ \Tr[E_a
S_\rho] = \SC a \rho \ge 0$ for every density matrix $S_\rho$.  Moreover, since we have $\Tr[E_a
S_\rho] =\SC a \rho\le1$ for every density matrix $S_\rho$, we must have $E_a \le I$.  Finally, we
know that for every couple of perfectly distinguishable pure states $\varphi, \varphi_\perp$ there
exists an atomic effect $a$ such that $\SC a \varphi = 1$ and $\SC a {\varphi_\perp} = 0$. Since the
two pure states $\varphi, \varphi_\perp$ are represented by orthogonal rank-one projectors
$S_\varphi$ and $S_{\varphi_\perp}$, we must have $E_a = S_\varphi$.  This proves that the atomic
effects are the whole set of positive rank-one projectors.  As a consequence, also every positive
matrix $P$ with $P\le I$ must represent some effect $a$. \qed

Finally, the reversible transformations are represented as conjugations by unitary matrices in
$\mathbb{SU}(2)$:
\begin{corollary}\label{cor:unitaryconjugation}
  For every reversible transformation $\tU \in \grp G_\rA$ with $d_\rA =2$ there exists a unitary
  matrix $U \in\mathbb {SU} (2)$ such that
\begin{equation}\label{unitaryconjugation} 
S_{\tU \rho}  = U S_\rho U^\dag \qquad \rho \in\Stset (\rA).
\end{equation} 
Conversely,  for every $U\in\mathbb {SU} (2)$ there exists a reversible transformation $\tU \in \grp G_\rA$ such that Eq. (\ref{unitaryconjugation}) holds.  
\end{corollary}
\Proof Every rotation of the Bloch sphere is represented by conjugation by some $\mathbb {SU }(2)$
matrix.  Conversely, every conjugation by an $\mathbb {SU }(2)$ matrix represents some rotation on
the Bloch sphere.  On the other hand, we know that $\grp G_\rA$ is the group of all rotations on the
Bloch sphere (theorem \ref{theo:qubit}). \qed

Note that we proved that all two-dimensional systems $\rA$ and $\rB$ in our theory have the same
states ($\Stset_1 (\rA) \simeq \Stset_1 (\rB)$), the same effects ($\Cntset(\rA) \simeq \Cntset(\rB)$),
and the same reversible transformations ($\grp G_\rA \simeq \grp G_\rB$), but we did not show that $\rA$ and
$\rB$ are operationally equivalent.  For example, $\rA$ and $\rB$ could be different when we compose
them with a third system $\rC$: the set of states $\Stset_1(\rA\rC)$ and $\Stset_1 (\rB\rC)$ could
be non-isomorphic.  The fact that every couple of two-dimensional systems $\rA$ and $\rB$ are operationally equivalent will be proved later (cf. corollary \ref{cor:equidim}).

We conclude this section with a simple fact that will be very useful later:

\begin{corollary}{\bf (Superposition principle for qubits)}\label{cor:superpositionqubit}
  Let $\{\varphi_1, \varphi_2\} \subset \Stset_1(\rA)$ be two perfectly distinguishable pure states
  of a system $\rA$ with $d_\rA =2$. Let $\{a_1, a_2\}$ be the observation-test such that $\SC
  {a_i}{\varphi_j}= \delta_{ij}$.  Then, for every probability $0\le p \le 1$ there exists a pure state
  $\psi_p \in\Stset_1(\rA)$ such that
\begin{align}\label{superpositionqubit}
\SC {a_1}{\psi_p}  = p  \qquad \SC {a_2 }{\psi_p}  = 1-p.
\end{align}
Precisely, the set of pure states $\psi_p \in\Stset_1 (\rA)$ satisfying Eq. (\ref{superpositionqubit}) is
a circle in the Bloch sphere.
\end{corollary}
\Proof Elementary property of density matrices.  \qed

\section{Projections}\label{sec:projections}

In this section we define the projection on a face $F$ of the convex set $\Stset_1 (\rA)$ and we
prove several properties of projections.  The projection on the face $F$ will be defined as an
atomic operation $\Pi_F\in\Trnset(\rA)$ that acts as the identity on states in the face $F$ and that
annihilates the states on the \emph{orthogonal face} $F^\perp$.  In the following we first introduce
the concept of orthogonal face, then prove the existence and uniqueness of projections, and finally give some
useful results on the projection of a pure state on two orthogonal faces.

\subsection{Orthogonal faces and orthogonal complements} 
 
In order to introduce the notion of orthogonal face we need first a few elementary results.  
We start by showing that there is a canonical way to associate a state $\omega_F$ to a face $F$: 
\begin{lemma}[State associated to a face]
Let $F$ be a face of the convex set $\Stset_1 (\rA)$ and let $\{\varphi_i\}_{i=1}^{|F|}$  be a maximal set of perfectly
  distinguishable pure states in $F$.   Then  the state
  $\omega_F:=\frac1{|F|}\sum_{i=1}^{|F|}\varphi_i$ depends only on the face $F$ and not on the particular set  $\{\varphi_i\}_{i=1}^{|F|}$.  Morever, $F$ is the face identified by $\omega_F$
  \label{lem:indepchi}
\end{lemma}

\Proof Suppose that $F$ is the face identified by $\rho$ and let $\tE \in\Trnset (\rA,\rC)$ (resp.
$\tD \in\Trnset (\rC, \rA)$) be the encoding (resp. decoding) in the ideal compression for $\rho$.
By lemma \ref{lem:pureenc} and corollary \ref{cor:maxienc}, $\{\tE\varphi_i\}_{i=1}^{|F|}$ is a maximal set
of perfectly distinguishable pure states of $\rC$ and by theorem \ref{theo:maxdiscrinvar} one has
$\chi_\rC = \frac 1 {|F|} \sum_{i =1}^{|F|} \tE \varphi_i$.  Hence, $\omega_F=\frac 1 {|F|}
\sum_{i=1}^{|F|} \varphi_i = \frac1{|F|} \sum_{i=1}^{|F|}\tD \tE\varphi_i=\tD\chi_\rC$.  Since the
right-hand side of the equality is independent of the particular set $\{\varphi_i\}_{i=1}^{|F|}$,
the state $\omega_F$ in the left-hand side is independent too.   To prove that $F$ is the face identified by $\omega_F$ it is enough to observe that  $\omega_F$ is completely mixed relative to $F$:  this fact follows from the relation $\omega_F  =  \tD  \chi_{\rC}$ and from lemma \ref{lem:encrhointernal} \qed

We now define the \emph{orthogonal complement} of the state
$\omega_F$:
\begin{definition}\label{def:stateorthocomp}
The \emph{orthogonal complement of the state} $\omega_F$  is the state $\omega_F^\perp \in\Stset_1 (\rA)  \cup \{0\}$ defined as follows: 
\begin{enumerate}
\item if $|F| = d_\rA$, then $\omega_F^\perp = 0$
\item if $F < d_\rA$, then $\omega_F^\perp$ is defined by the relation
\begin{equation}\label{definingrelation}
\chi_\rA   =  \frac{|F|}{d_\rA}  \omega_F  +  \frac{d_\rA - |F|}{d_\rA}  \omega_{F}^\perp
\end{equation}
\end{enumerate}
\end{definition}

An easy way to write the orthogonal complement is 
\begin{lemma}\label{lem:concretecomplement}
  Take a maximal set $\{\varphi_i\}_{i=1}^{|F|}$ of perfectly distinguishable pure states in $F$ and
  extend it to a maximal set $\{\varphi_i\}_{i=1}^{d_\rA}$ of perfectly distinguishable pure states
  in $\Stset_1 (\rA)$, then for $|F| < d_\rA$ we have
\begin{equation*}
\omega_F^\perp  = \frac 1 {d_\rA - |F|}  \sum_{i=|F| +1}^{d_\rA}  \varphi_i.
\end{equation*}  
\end{lemma}
\Proof By definition, for $|F| <d_\rA$ we have $\omega_F^\perp = \frac 1 {d_\rA -|F|} ( d_\rA
\chi_\rA - |F| \omega_F ) $.  Substituting the expressions $\chi_\rA = \frac 1 {d_\rA}
\sum_{i=1}^{d_\rA} \varphi_i$ and $\omega_F = \frac 1 {|F|} \sum_{i=1}^{|F| }\varphi_i$ we then
obtain the thesis. \qed Note, however, that by definition the orthogonal complement
$\omega_F^{\perp}$ depends only on the face $F$ and not on the choice of the maximal set in lemma
\ref{lem:concretecomplement}.

An obvious consequence of lemma \ref{lem:concretecomplement} is 
\begin{corollary}\label{cor:obvious}
The states $\omega_F$ and  $\omega_F^\perp$ are perfectly distinguishable. 
\end{corollary}
\Proof Take a maximal set $\{\varphi_i\}_{i=1}^{|F|}$ of perfectly distinguishable pure states in
$F$, extend it to a maximal set $\{\varphi_i\}_{i=1}^{d_\rA}$, and take the observation-test such
that $\SC {a_i} {\varphi_j} = \delta_{ij}$.  Then the binary test $\{a_F, e-a_F\}$, defined by
$a_F:= \sum_{i=1}^{|F|} a_i$ distinguishes perfectly between $\omega_F$ and $\omega_F^\perp$. \qed

We say that a state $\tau\in\Stset_1 (\rA)$ is \emph{perfectly distinguishable from the face $F$} if
$\tau$ is perfectly distinguishable from every state $\sigma$ in the face $F$.  With this definition
we have the following
\begin{lemma}\label{lem:trecondizioni}
The following are equivalent: 
\begin{enumerate}
\item $\tau$ is perfectly distinguishable from the face $F$
\item $\tau$  is perfectly distinguishable from $\omega_F$
\item $\tau$ belongs to the face identified by $\omega_F^\perp$, i.e. $\tau \in F_{\omega_F^\perp}$. 
\end{enumerate}
\end{lemma}
\Proof ($1 \Leftrightarrow 2$) $\tau$ is perfectly distinguishable from $\omega_F$ if and only if
then there exists a binary test $\{a,e-a\}$ such that $\SC a \tau =1$ and $\SC a {\omega_F}=0$.  By
lemma \ref{lem:detuponrho} this is equivalent to the condition $\SC a \tau =1$ and $a=_{\omega_F}
0$, that is, $\tau$ is distinguishable from any state $\sigma$ in the face identified by $\omega_F$,
which by definition is $F$. ($2 \Rightarrow 3$)  Let $\{\varphi_i\}_{i=1}^{|F|}$ be a maximal set of
perfectly distinguishable states in $F$, $\omega_F  =  \frac 1 {|F|}  \sum_{i =1}^{|F|} \varphi_i$, and let $\{\varphi_i\}_{i=|F| +1}^k$ be the maximal set of
perfectly distinguishable pure states in the spectral decomposition $\tau = \sum_{i=|F| +1}^k p_i  \varphi_i$,  with $p_i>0$ for every $i = |F| +1, \dots, k$. Since $\tau $ is perfectly
distinguishable from $\omega_F$, by lemma \ref{lem:refinement} we have that the states
$\{\varphi_i\}_{i=1}^{k}$ are all perfectly distinguishable. Let us extend this set to a maximal set
$\{\varphi_i\}_{i=1}^{d_\rA}$.  By lemma \ref{lem:concretecomplement} have $\omega_F^\perp = \frac 1
{d_\rA -|F|}   \sum_{i=|F|+1}^{d_\rA} \varphi_i$. Hence, all the states $\{\varphi_i\}_{i =  |F| +1}^{d_\rA}$ are in the face $F_{\omega_F^\perp}$. Since $\tau$ is a mixture of these states, it also belongs to the face  $F_{\omega_F^\perp}$.  
  ($3 \Rightarrow 2$) Since $\omega_F$ and $\omega_F^\perp$ are
perfectly distinguishable, if $\tau$ belongs to the face identified by $\omega_F^\perp$, then by
lemma \ref{lem:inclusiondiscrimination} $\tau$ is perfectly distinguishable from $\omega_F$.
\qed

\begin{corollary}\label{cor:distinguishablityfromconvexcomb}
  If $\rho$ is perfectly distinguishable from $\sigma$ and from $\tau$, then $\rho$ is perfectly
  distinguishable from any convex mixture of $\sigma$ and $\tau$.
\end{corollary}
\Proof Let $F$ be the face identified by $\rho$. Then by lemma \ref{lem:trecondizioni} we have
$\sigma, \tau\in F_{\omega_F^\perp}$. Since $F_{\omega_F^\perp}$ is a convex set, any mixture of
$\sigma$ and $\tau$ belongs to it.  By lemma \ref{lem:trecondizioni}, this means that any mixture of
$\sigma$ and $\tau$ is perfectly distinguishable from $\rho$. \qed
 
We are now ready to give the definition of orthogonal face:  
\begin{definition}[Orthogonal face]
The \emph{orthogonal face}  $F^\perp$  is the set  of all  states that are perfectly distinguishable from the face $F$.
\end{definition}
By lemma \ref{lem:trecondizioni} it is clear that $F^\perp$ is the face identified by
$\omega_F^\perp$, that is $F^\perp = F_{\omega_F^\perp}$.

In the following we list few elementary facts about orthogonal faces:  
\begin{lemma}\label{lem:list}  
The following properties hold
\begin{enumerate}
\item $|F^\perp|  = d_\rA - |F|$
\item  $\chi_\rA=\frac{|F|}{d_\rA}\omega_F+\frac{|F^\perp|}{d_\rA}\omega_{F^\perp}$
\item $\omega_{F^\perp}  = \omega_F^\perp$
\item $\omega_{F^\perp}^\perp  = \omega_F$
\item  $\left(F^\perp\right)^\perp = F$.
\end{enumerate}
\end{lemma}
\Proof \emph{Item 1.}  If $|F| = d_\rA $ the thesis is obvious.  If $|F|< d_\rA$, take a maximal set
$\{ \varphi_i\}_{i=1}^{|F|}$ (resp.  $\{ \varphi_j\}_{j=|F|+1}^{|F| + |F^\perp|}$) of perfectly
distinguishable pure states in $F$ (resp. $F^{\perp}$).  Hence we have
\begin{equation*}
  \omega_F = \frac 1 {|F|}  \sum_{i=1}^{|F|} \varphi_i    \qquad \left ( {\rm resp.} \quad \omega_{F^\perp}  = \frac 1 {|F^\perp|}  \sum_{j=|F|+1}^{|F| +  |F^\perp|}   \varphi_j \right).
\end{equation*}
By corollary \ref{cor:obvious}  the states $\omega_F$ and $\omega_{F^\perp}$ are perfectly distinguishable.  Hence, the states
$\{\varphi_i\}_{i=1}^{|F| + |F^\perp|}$ are perfectly distinguishable jointly (lemma
\ref{lem:refinement}).  Now, we must have $|F| + |F^\perp| = d_\rA$, otherwise there would be a pure
state $\psi$ that is perfectly distinguishable from the states $\{\varphi_i\}_{i=1}^{|F| +
  |F^\perp|}$.  This implies that $\psi$ belongs to $F^\perp$ and that states $\{\psi\} \cup \{
\varphi_j\}_{j=|F|+1}^{|F| + |F^\perp|}$ are perfectly distinguishable in $F^\perp$, in
contradiction with the hypotheses that the set $\{ \varphi_j\}_{j=|F|+1}^{|F| + |F^\perp|}$ is maximal in
$F^\perp$.  \emph{Item 2} Immediate from item 1 and definition \ref{def:stateorthocomp}. \emph{Item 3 and 4} Both items follow by comparison
of item 2 with Eq.  \ref{definingrelation}.  \emph{Item 5} By condition 3 of lemma
\ref{lem:trecondizioni}, $\left(F^\perp\right)^\perp$ is the face identified by the state
$\omega_{F^\perp}^\perp$, which, by item 4, is $\omega_F$.  Since the face identified by $\omega_F$
is $F$, we have $\left(F^\perp\right)^\perp = F$.  \qed

We now show that there is a canonical way to associate an effect $a_F$ to a face $F$:

\begin{definition}[Effect associated to a face]
 We say that $a_F \in\Cntset(\rA)$ is the \emph{effect associated to the face}  $F \subseteq \Stset_1 (\rA)$ if and only if $a_F  =_{\omega_F}  e  $  and $a_F =_{\omega_F^\perp}  0$.
 \label{def:effface}
 \end{definition}
In other words, the definition imposes that $\SC {a_F} \rho = 1$ for every $\rho \in F$  and $\SC {a_F} \sigma = 0 $ for every $\sigma \in F^\perp$.

\begin{lemma}
  A state $\rho\in\Stset_1(\rA)$ belongs to the face $F$ if and only
  if $(a_F|\rho)=1$.
  \label{lem:identifytest}
\end{lemma}

\Proof By definition, if $\rho$ belongs to $F$, then $(a_F|\rho)=1$. Conversely, if
$(a_F|\rho)=1$, then $\rho$ is perfectly distinguishable from $\omega^\perp_{F}$, because $\SC {a_F}{\omega_F^\perp} =0$.  Now, we know that $\omega_F^\perp$ is equal to $\omega_{F^\perp}$ (item 4 of lemma
\ref{lem:list}). By item 2 of lemma \ref{lem:trecondizioni} the fact that $\rho$ is perfectly
distinguishable from $\omega_{F^\perp}$ implies that $\rho$ belongs to $\left( F^\perp
\right)^\perp$, which is just $F$ (item 5 of lemma \ref{lem:list}). \qed

We now show that the effect $a_F$ associated to the face $F$ exists and is unique.
A preliminary result needed to this purpose is the following:  

\begin{lemma}\label{lem:concreteaF}
  The effect $a_F$ must have the form $a_F = \sum_{i=1}^{|F|} a_i$, where
  $a_i$ is the atomic effect such that $\SC {a_i}{\varphi_i} =1$ and
  $\{\varphi_i\}_{i=1}^{|F|}$ is a maximal set of perfectly
  distinguishable pure states in $F$.
\end{lemma}

\Proof By corollary \ref{cor:spectraldecompeff} we have that $a_F$ can
be written as $\B{a_F} = \sum_i d_i \B{ a_i}$ where
$\{a_i\}_{i=1}^{d_\rA}$ is a perfectly distinguishing test.  Moreover,
since $a_F$ is an effect, we must have $d_i\ge 0 $ forall $i=1, \dots,
d_\rA$.  Now, by definition we have $\SC {a_F}{\omega_F^\perp} =0$,
which implies $d_i \SC {a_i} {\omega_F^\perp} = 0$ for every $i=1,
\dots, d_\rA$, that is, $\SC {a_i}{\omega_F^\perp} = 0$ whenever $d_i
\not = 0$.  Let us focus on the values of $i$ for which $d_i \not =
0$. Let $\varphi_i$ be the pure state such that $\SC {a_i}{\varphi_i}
= 1$.  The condition $\SC {a_i}{\omega_F^\perp} = 0$ implies that
$\varphi_i$ is perfectly distinguishable from $\omega_F^\perp$.
Therefore, $\varphi_i$ belongs to $(F^\perp)^\perp$, which is $F$.
Since by definition we must have $\SC {a_F}{\varphi_i} = 1$, this also
implies that $d_i = 1$.  In summary, we proved that $a_F = \sum'_i
a_i$ where the prime means that the sum is restricted to those values
of $i$ such that $\varphi_i \in F$.  The condition $a_F =_{\omega_F}
e$ also implies that the number of terms in the sum must be exactly
$|F|$.  The thesis is then proved by suitably relabelling the effects
$\{a_i\}_{i=1}^{d_\rA}$, in such a way that $\varphi_i $ belongs to
$F$ for every $i=1, \dots, |F|$.  \qed

\begin{lemma}\label{lem:uniqueaF}
  The effect $a_F$ associated to the face $F$ is unique.
\end{lemma}

\Proof Suppose that $a_F= \sum_{i=1}^{|F|} a_i$ and $a_F' =
\sum_{i=1}^{|F|} a_i'$ are two effects associated to the face $F$,
both written as in lemma \ref{lem:concreteaF}.  Let
$\{\varphi_i\}_{i=1}^{|F|}$ (resp. $\{\varphi_i'\}_{i=1}^{|F|}$) be
the maximal set of perfectly distinguishable pure states in $F$ such
that $\SC {a_i}{\varphi_i} = 1$ for every $i=1, \dots, |F|$ (resp.
$\SC {a'_i}{\varphi'_i} = 1$ for every $i=1, \dots, |F|$), and let
$\{\psi_j\}_{j=1}^{|F^\perp|}$ be a maximal set of perfectly
distinguishable pure states in $F^\perp$. Since $\omega_F$ and
$\omega_{F}^\perp$ are perfectly distinguishable, the states
$\{\varphi_i\}_{i=1}^{|F| } \cup \{\psi_j\}_{j=1}^{|F^\perp|} $ (resp.
$\{\varphi'_i\}_{i=1}^{|F| } \cup \{\psi_j\}_{j=1}^{|F^\perp|} $ are
perfectly distinguishable (lemma \ref{lem:refinement}).  Moreover, the
set is maximal since $|F| + |F^\perp| = d_\rA$.  Let $b_j$ be the
atomic effect such that $\SC {b_j}{\psi_j} =1$.  Then, the test that
distinguishes the states $\{\varphi_i\}_{i=1}^{|F|} \cup
\{\psi_j\}_{j=1}^{|F^\perp|} $ (resp.  $\{\varphi'_i\}_{i=1}^{|F|}
\cup \{\psi_j\}_{j=1}^{|F^\perp|} $ is given by $\{a_i\}_{i=1}^{|F|}
\cup \{b_j\}_{j=1}^{|F^\perp|} $ (resp.  $\{a'_i\}_{i=1}^{|F|} \cup
\{b_j\}_{j=1}^{|F^\perp|}$ and its normalization reads
\begin{align*}
  e &=  \sum_{i=1}^{|F|}  a_i  + \sum_{j=1}^{|F^\perp|}  b_j  = a_F  +     \sum_{j=1}^{|F^\perp|}  b_j ~,\\
  e & = \sum_{i=1}^{|F|} a'_i + \sum_{j=1}^{|F^\perp|} b_j = a'_F +
  \sum_{j=1}^{|F^\perp|} b_j ~.
\end{align*}
By comparison we obtain $a_F = a_F'$.  \qed

\subsection{Projections}
We are now in position to define the projection on a face:
\begin{definition}[Projection]
  Let $F$ be a face of $\Stset_1(\rA)$. A {\em projection} on the face
  $F$ is an atomic transformation $\Pi_F$ such that
  \begin{enumerate}
  \item $\Pi_F=_{\omega_F} \tI_\rA$
  \item $\Pi_{F}=_{\omega^\perp_{F}}0$
  \end{enumerate}
  \label{def:project}
  When $F$ is the  face identified by a pure state $\varphi\in\Stset_1 (\rA)$, we have $F= \{\varphi\}$ and call $\Pi_{\{\varphi\}}$ a \emph{projection on the pure state $\varphi$}.   
\end{definition}
The first condition in definition \ref{def:project} means that the projection $\Pi_F$ does not disturb the states in the face $F$.
The second condition means that $\Pi_{F}$ annihilates all states in the orthogonal face $F^\perp$. As a notation, we will indicate with $\Pi_F^\perp$ the projection on the face $F^\perp$, that is, we will use the definition $\Pi_F^\perp := \Pi_{F^\perp}$.

An equivalent condition for $\Pi_F$ to be a projection on the face $F$ is the following:
\begin{lemma}
  Let $\{\varphi_i\}_{i=1}^{d_\rA}$ be a maximal set of perfectly distinguishable pure states for
  system $\rA$. The transformation $\Pi_F$ in $\Trnset(\rA)$ is a projection on the face generated
  by the subset $\{\varphi_{i}\}_{i=1}^{|F|}$ if and only if
  \begin{enumerate}
  \item $\Pi_F=_{\omega_F} \tI_\rA$
  \item $\Pi_F|\varphi_l)=0$ for all $l>|F|$
  \end{enumerate}
  \label{lem:project}
\end{lemma}

\Proof The condition is clearly necessary, since by Definition \ref{def:project}
$\Pi_F|\varphi_l)=0$ for $l>|F|$. On the other hand, if $\Pi_F|\varphi_l)=0$ for $l> |F|$ then by
definition of $\omega^\perp_F$ we have $\Pi_F|\omega^\perp_F)=0$, and, therefore $\Pi_F
=_{\omega_F^\perp} 0$.\qed

A result that will be useful later is:  
\begin{lemma}\label{lem:pitensorid}
The transformation $\Pi_{F} \otimes \tI_\rB$  is a projection on the face $\tilde F$ identified the state $\omega_F \otimes \chi_\rB$.
\end{lemma}

\Proof  $\Pi_{F} \otimes \tI_\rB$ is atomic, being the product of two atomic transformations.  We now show that $\Pi_{F}  \otimes  \tI_\rB  =_{\omega_F \otimes \chi_\rB}   \tI_\rA \otimes \tI_\rB$:  Indeed, by the local tomography axiom it is easy to see that every state $\sigma \in F_{\omega_F \otimes \chi_\rB}$ can be written as  $\K \sigma =  \sum_{i=1}^r  \sum_{j=1}^{d_\rB}  \sigma_{ij}     \K {\alpha_i}  \K{\beta_j} $, where $\{  \alpha_i\}_{i=1}^r$ is a basis for $\Span(F)$ and $\{\beta_j\}_{j=1}^{d_\rB}$ is a basis for $\Stset_1 (\rB)$.   Since $\Pi_F  =_{\omega_F}   \tI_\rA$,  we have 
\begin{align*}
\K \sigma &= (\Pi_F \otimes \tI_\rB) \K \sigma \\
&=   \sum_{i=1}^r  \sum_{j=1}^{d_\rB}  \sigma_{ij}    \Pi_F \K {\alpha_i}  \K{\beta_j}  \\
&=   \sum_{i=1}^r  \sum_{j=1}^{d_\rB}  \sigma_{ij}     \K {\alpha_i}  \K{\beta_j}  \\
& =\K  \sigma ,      
\end{align*}
which implies $\Pi_{F}  \otimes  \tI_\rB  =_{\omega_F \otimes \chi_\rB}   \tI_\rA \otimes \tI_\rB$. 
Finally, note that $\omega_{\tilde F}  =  \omega_F \otimes \chi_\rB$, while $\omega_{\tilde F}^\perp  =  \omega_F^\perp  \otimes \chi_\rB$.  Since we have $(\Pi_F  \otimes \tI_\rB)  \K {\omega_{\tilde F}^\perp}  =  \Pi_F   \K  {\omega_F^\perp}  \otimes \K { \chi_\rB} =0$, we can conclude $\Pi_F\otimes \tI_\rB  =_{\omega_{\tilde F}^\perp}  0$. Hence $\Pi_F \otimes \tI_\rB$ is a projection on $\tilde F$. \qed

In the following we will show that for every face $F$ there exists a unique projection $\Pi_F$ and we will prove several properties of projections.  Let us start from an elementary observation:

\begin{lemma}\label{lem:aA=a}
Let $\varphi$ be a pure state in the face $F \subseteq \Stset_1 (\rA)$ and let $a\in\Cntset(\rA)$ be the atomic effect such that $\SC a \varphi =1$. If $\tA \in \Trnset(\rA)$ is an atomic transformation such that $\tA =_{\omega_F}  \tI_\rA$, then $\B a \tA = \B a$.   Moreover, if $a_F$ is the effect associated to the face $F$, then we have $\B {a_F}  \tA = \B {a_F} $.  
\end{lemma}

\Proof By lemma \ref{lem:atomicity}, the effect $\B {a}\tA $ is atomic.  Now, since
$\tA|\varphi)=|\varphi)$, we have $(a|\tA|\varphi)=\SC {a} \varphi=1$.  However, by theorem
\ref{theo:dualestato} $\B a$ is the unique atomic effect such that $\SC {a}
{\varphi}=1$.  Hence, $(a|\tA=(a|$.  Moreover, writing $a_F$ as $a_F = \sum_{i=1}^{|F|}  a_i$ with $\SC {a_i}{\varphi_i} =1$, $\varphi_i \in F$  (lemma \ref{lem:concreteaF}), we    obtain $\B{a_F} \tA = \sum_{i=1}^{|F|}  \B {a_i}  \tA = \sum_{i=1}^{|F|} \B{ a_i} =\B{ a_F}$.\qed

When applied to the case of projections, the above lemma gives the following
\begin{corollary}
  Let $\varphi$ be a pure state in the face $F\subseteq \Stset_1(\rA)$ and let $a \in\Cntset(\rA)$ be the atomic effect such that
  $\SC {a}\varphi =1$. Then, we have $(a|\Pi_F=(a|$.  Moreover, if $a_F$ is the effect associated to the face $F$, then we have $\B {a_F}  = \B {a_F} \Pi_F$.  
  \label{cor:projeff}
\end{corollary}

The counterpart of corollary \ref{cor:projeff} is given as follows: 

\begin{lemma}\label{lem:zeroeff}
Let $\psi$ be a pure state in the face $F^\perp$ and let $b$ be the atomic effect such
  that $\SC {b}\psi =1$. Then, we have $(b|\Pi_F=0$.  Moreover, if $a^\perp_F$ is the effect associated to the face $F^\perp$, then we have $\B  {a_F^\perp}  \Pi_F = 0$.
\end{lemma}
\Proof   By lemma \ref{lem:atomicity}, the effect $\B {b}\Pi_F $ is atomic.  Hence, $\B b \Pi_F$ must be proportional to an atomic effect $b'$ with $|\!|  b'|\!| = 1$, for some proportionality constant $\lambda \in [0,1]$, that is $\B b \Pi_F  =  \lambda \B{b'}$.  We want to prove that $\lambda$ is zero. By contradiction, suppose  that $\lambda \not = 0$.  Let $\psi'$ be the pure state such that $\SC {b'}{\psi'} =1$.  Now, since
$\Pi_F \K{\omega_F^\perp}=0$, we have $0= ( b| \Pi_F |\omega^\perp_F)=\lambda  (b' |\omega_F^\perp)$, which implies $(b'|\omega_F^\perp) =0$.   Hence, $\psi'$ is perfectly distinguishable from $\omega_F^\perp$, which in turn implies that $\psi'$  belongs to $\left(F^\perp\right)^\perp =F$. We then have $\lambda = \B b \Pi_F  \K{\psi'} =  \SC b {\psi'} = 0$ (the last equality follows from the fact that $\psi$ and $\psi'$ belong to $F^\perp$ and $F$, respectively, and hence are perfectly distinguishable).  This is in contradiction with the assumption $\lambda \not =0$, thus concluding the proof that  $\B b \Pi_F  =0$.    Moreover, writing  $a_F^\perp$ as  $a_F^\perp  = \sum_{i=1}^{|F^\perp|}  b_i$ with $\SC{b_i}{\psi_i} = 1$, $\psi_i \in F^\perp$, we obtain $\B {a_F^\perp}  \Pi_F  =\sum_{i=1}^{|F^\perp| }   \B {b_i}  \Pi_F = 0$. \qed

Combining corollary \ref{cor:projeff}  and lemma \ref{lem:zeroeff} we obtain an important property of projections, expressed by the following:
\begin{corollary}\label{cor:effettodipi} 
If $\Pi_F$ is a projection on the face $F$, then one has $\B {e_\rA} \Pi_F  = \B {a_F}$. 
\end{corollary}
\Proof The thesis follows from  corollary \ref{cor:projeff}  and lemma \ref{lem:zeroeff}  and from the fact that $a_F + a_F^\perp = e$. \qed

In the following we will see that for every face $F$ there exists a unique projection.  To prove
that, let us start from the existence:
\begin{lemma}[Existence of projections]\label{lem:existence}
  For every face $F$ of $\Stset_1(\rA)$ there exists a projection
  $\Pi_F$.  
\end{lemma}

\Proof 
By lemma \ref{lem:atomrealiz}, there exists a system $\rB$ and an atomic transformation $\tA\in\Trnset(\rA,\rB)$ with
$(e|_\rB\tA=(a_F|$. Then, if $\Psi_{\omega_F}\in\Stset (\rA \rC)$ is a purification of $\omega_F$,
we can define the state $|\Sigma)_{\rB\rC}:=(\tA\otimes\tI_\rC)|\Psi_{\omega_F})_{\rA\rC}$.  By
lemma \ref{lem:atomicity}, $\Sigma$ is a pure state.  Moreover, the pure states $\Sigma$ and
$\Psi_{\omega_F}$ have the same marginal on system $\rC$: indeed, we have $\B {e_\rB} \K \Sigma =
[\B {e_\rB} \tA] \K{\Psi_{\omega_F}} = \B {a_F} \K{\Psi_{\omega_F}}$ and, by definition, $a_F =_{\omega_F} e_\rA$, which by theorem \ref{theo:uponinput} implies
$\B{a_F} \K{\Psi_{\omega_F}} = \B{e_\rA}\K{\Psi_{\omega_F}}$.  If $\varphi_0$ and $\psi_0$ are two
arbitrary pure states of $\rA$ and $\rB$, respectively, the uniqueness of purification stated by
Postulate \ref{purification} implies that there exists a reversible channel $\tU \in\grp G_{\rA\rB}$
such that
\begin{equation}\label{questa}
\begin{split}   
 \begin{aligned}
    \Qcircuit @C=1em @R=.7em @! R {&\prepareC{\psi_0}&\qw\poloFantasmaCn{\rB}&\qw\\
      &\multiprepareC{1}{\Psi_{\omega_F}}&\qw\poloFantasmaCn{\rA}&\qw\\
      &\pureghost{\Psi_{\omega_F}}&\qw\poloFantasmaCn{\rC}&\qw}
  \end{aligned} &=
   \begin{aligned}
    \Qcircuit @C=1em @R=.7em @! R {&\prepareC{\varphi_0}&\qw\poloFantasmaCn{\rA}&\ghost{\tU} & \qw\poloFantasmaCn{\rB}&\qw\\
      &\multiprepareC{1}{\Sigma}&\qw\poloFantasmaCn{\rB}&\multigate{-1}{\tU} &\qw\poloFantasmaCn{\rA}&\qw\\
      &\pureghost{\Sigma}&\qw\poloFantasmaCn{\rC}&\qw &\qw&\qw}
  \end{aligned}\\
  &= \begin{aligned}
    \Qcircuit @C=1em @R=.7em @! R {&&&\prepareC {\varphi_0}& \qw \poloFantasmaCn{\rA} &\ghost{\tU}&\qw\poloFantasmaCn{\rB}&\qw\\
      &\multiprepareC{1}{\Psi_{\omega_F}}&\qw\poloFantasmaCn{\rA}&\gate{\tA}&\qw\poloFantasmaCn{\rB}&\multigate{-1}{\tU}&\qw\poloFantasmaCn{\rA}&\qw\\
        &\pureghost{\Psi_{\omega_F}}&\qw\poloFantasmaCn{\rC}&\qw&\qw&\qw&\qw&\qw}
  \end{aligned}
\end{split}
\end{equation}
Now, take the atomic effect $b \in\Cntset(\rB)$ such that $(b|\psi_0)=1$, and define the
transformation 
$\Pi_F \in\Trnset(\rA)$ as 
\begin{equation*}
\begin{aligned}
    \Qcircuit @C=1em @R=.7em @! R {&  \qw \poloFantasmaCn{\rA}  & \gate{\Pi_F}  &  \qw \poloFantasmaCn{\rA}  &\qw}
\end{aligned}
= \begin{aligned}
    \Qcircuit @C=1em @R=.7em @! R {&&\prepareC {\varphi_0}& \qw \poloFantasmaCn{\rA} &\ghost{\tU}&\qw\poloFantasmaCn{\rB}&\measureD b\\
      &\qw\poloFantasmaCn{\rA}&\gate{\tA}&\qw\poloFantasmaCn{\rB}&\multigate{-1}{\tU}&\qw\poloFantasmaCn{\rA}&\qw}
  \end{aligned}
\end{equation*}  Applying
$b$ on both sides of Eq. (\ref{questa}) we then obtain
\begin{equation*}
  (\Pi_F \otimes\tI_\rC)|\Psi_{\omega_F})=|\Psi_{\omega_F}),
\end{equation*} 
and, therefore, $\Pi_F=_{\omega_F}\tI_\rA$. Moreover, the transformation $\Pi_F$ is atomic, being
the composition of atomic transformations (lemma \ref{lem:atomicity}).  Finally, we have $\Pi_F
=_{\omega_F^\perp} 0$: indeed, by construction of $\Pi_F$ we have
\begin{align*}
\B {e_\rA}  \Pi_F  \K \rho  &=    \B {e_\rA \otimes b} \tU  ( \tA \otimes \tI_\rA) \K  {\rho\otimes \varphi_0}\\
 &\le    \B {e_\rA \otimes e_\rB} \tU  ( \tA \otimes \tI_\rA) \K  {\rho\otimes \varphi_0}\\
 & =  \B {e_\rA}  \tA  \K \rho \\
 & =  \SC {a_F} \rho.
\end{align*} 
This implies $\B {e_\rA} \Pi_F\K{\omega_{F}^\perp} =\SC {a_F}{\omega_F^\perp} 0$, and, therefore, $
\Pi_F=_{\omega_{F}^\perp}0$.  In conclusion, $\Pi_F$ is the desired projection. \qed

To prove the uniqueness of the projection $\Pi_F$ we need two auxiliary lemmas, given in the
following.
\begin{lemma} \label{lem:projchoi} Let $\Phi\in\Stset_1(\rA\tilde\rA)$ be
  a purification of the invariant state $\chi_\rA$, and
  let $\Pi_F\in\Trnset(\rA)$ be a projection on the face $F \subseteq \Stset_1 (\rA)$. Then, the pure state
  $\Phi_F \in \Stset_1(\rA\tilde\rA)$ defined by
  \begin{equation}
    |\Phi_F) := \frac{d_\rA} {|F|}
    (\Pi_F\otimes\tI_{\tilde A} )|\Phi)
  \end{equation}
  is a purification of $\omega_F$.
\end{lemma}

\Proof The state $\Phi_F$ is pure by lemma \ref{lem:atomicity}. Let us choose a maximal set of
perfectly distinguishable pure states $\{\varphi_i\}_{i=1}^{d_\rA}$ such that
$\{\varphi_i\}_{i=1}^{|F|}$ is maximal in $F$. Now, we have
\begin{align*}
  (e_{\tilde\rA}| \,  |\Phi_F)_{\rA\tilde\rA}&=\frac{d_\rA} {|F|} \left[  \Pi_F \otimes (e_{\tilde\rA}|\right]  |\Phi)_{\rA\tilde\rA}\\
  &=\frac {d_\rA} {|F|}\Pi_F|\chi_\rA),
  \end{align*}
  having used the relation $  \B{e_{\tilde \rA}}  \K{\Phi}_{\rA\tilde\rA}  =  \K{\chi_\rA}$ (corollary \ref{cor:conjusys}).  We then obtain  
\begin{align*}
(e_{\tilde\rA}| \,  |\Phi_F)_{\rA\tilde\rA}  &=\frac {d_\rA} {|F|}\Pi_F|\chi_\rA) \\
&  =  \frac 1 {|F|}  \sum_{i=1}^{d_\rA} \Pi_F \K{\varphi_i}\\
&= \frac1 {|F|}\sum_{i=1}^{|F|} |\phi_i)\\
  &=|\omega_F),
\end{align*}
having used that $\chi_\rA = \sum_{i=1}^{d_\rA} \varphi_i/d_\rA$ (theorem \ref{theo:maxdiscrinvar}),
and the definition of $\Pi_F$. \qed

\begin{lemma}
  Let $\Pi_F \in\Trnset(\rA)$ be a projection. A transformation
  $\tC\in\Trnset (\rA)$ satisfies $\tC=_{\omega_F} \tI_\rA$ if and
  only if
  \begin{equation}
    \tC\Pi_F=\Pi_F.
    \label{upep}
  \end{equation}
  \label{lem:idupthe}
\end{lemma}

\Proof Let $\Phi_F$ be the purification of $\omega_F$ defined in lemma \ref{lem:projchoi}.  Since
$\tC =_{\omega_F} \tI_\rA$, we have $(\tC\otimes \tI) \K{\Phi_F} = \K{\Phi_F}$.  In other words, we
have $ (\tC \Pi_F \otimes \tI)\K\Phi = (\Pi_F \otimes \tI) \K\Phi$.  Since $\Phi$ is dynamically
faithful, this implies that $\tC\Pi_F=\Pi_F$.  Conversely, Eq.~\eqref{upep} implies that for
$\sigma\in F_{\omega_F}$, $\tC|\sigma)=\tC\Pi_F|\sigma)=\Pi_F|\sigma)=|\sigma)$, namely
$\tC=_{\omega_F}\tI_{\rA}$.\qed

\begin{theorem}[Uniqueness of projections]
   The projection $\Pi_F$ satisfying Definition
  \ref{def:project} is unique.
  \label{theo:uniproj}
\end{theorem}

\Proof Let $\Pi_F$ and $\Pi'_F$ be two projections on the same face
$F$, and define the pure states $\Phi_F$ and $\Phi_F'$ as in lemma
\ref{lem:projchoi}. 
Now,  $\Phi_F$ and
$\Phi_F'$ are both purifications of the same state $\tilde \omega_F
\in\tilde\rA$ : indeed, one has
 \begin{align*}
 \B {e_\rA} \K{\Phi_F}_{\rA\tilde\rA} &= \frac {d_\rA}{|F|}   [\B {e_\rA}  \Pi_F] \K{\Phi}_{\rA\tilde\rA} \\
 & =\frac {d_\rA}{|F|}   \B {e_F}  \K{\Phi}_{\rA\tilde\rA} \\
 & =  \frac {d_\rA}{|F|}   [\B {e_\rA}  \Pi'_F] \K{\Phi}_{\rA\tilde\rA}\\
 &= \B {e_\rA} \K{\Phi'_F}_{\rA\tilde\rA},
 \end{align*}
 having used the relation $\B {e_\rA}  \Pi_F  =  \B{a_F}  =  \B{e_\rA}  \Pi'_F$, which comes from corollary  \ref{cor:projeff} and from the uniqueness of the effect $a_F$   (lemma \ref{lem:uniqueaF}).  
By the uniqueness of purification, we have $\K{\Phi_F'} =
(\tU \otimes \tI_{\tilde \rA}) \K{\Phi_F}$ for some reversible transformation $\tU\in\grp G_\rA$.
This implies $(\Pi'_F \otimes \tI_{\tilde \rA}) \K{\Phi}=(\tU\Pi_F\otimes \tI_{\tilde \rA})
\K{\Phi}$, and, since $\Phi$ is dynamically faithful, $\Pi'_F=\tU\Pi_F$. Since by Definition
\ref{def:project} we have $\Pi'_F=_{\omega_F}\tI_\rA$ and $\Pi_F=_{\omega_F}\tI_\rA$, we can
conclude that $\tU =_{\omega_F} \tI_\rA$. Finally, using lemma \ref{lem:idupthe} with $\tC  = \tU$ we obtain  
$\Pi'_F=\tU\Pi_F=\Pi_F$.\qed

We now show a few simple properties of projections. In the following, given a maximal set of
perfectly distinguishable pure states $\{\varphi_i\}_{i=1}^{d_\rA}$ and any subset
$V\subseteq\{1,\dots,d_\rA\}$ we define (with a slight abuse of notation) $\omega_V:=\sum_{i\in
  V}\varphi_i/|V|$, and $\Pi_V$ as the projection on the face $F_V:= F_{\omega_V}$. We will refer to
$F_V$ as \emph{the face generated by $V$}.

\begin{lemma}
  For two arbitrary subsets $V, W\subseteq\{1,\dots,d_\rA\}$ one has
  \begin{equation*}
     \Pi_V\Pi_W=\Pi_{V \cap W}.
  \end{equation*}
  In particular, if $V \cap W = \emptyset$ one has $\Pi_V \Pi_W = 0$.
  \label{lem:intersection}
\end{lemma}
\Proof First of all, $\Pi_V \Pi_W$ is atomic, being the product of two atomic transformations.
Moreover, since the face $F_{V \cap W}$ is contained in the faces $F_V$ and $F_W$, we have $\Pi_V
\Pi_W \K \rho = \Pi_V \K \rho = \K \rho$ for every $\rho \in F_{V \cap W}$.  In other words, $\Pi_V
\Pi_W=_{\omega_{V \cap W} } \tI_\rA $.  Moreover, if $l \not \in V \cap W$ we have $\Pi_V \Pi_W
\K{\varphi_l} =0$. By lemma \ref{lem:project}  and and by the uniqueness of projections (theorem \ref{theo:uniproj}) we then obtain that $\Pi_V \Pi_W$ is the projection on the face generated by $V\cap W$.  \qed

\begin{corollary}[Idempotence]
  Every projection $\Pi_F$ satisfies the identity
  $\Pi_F^2=\Pi_F$.
  \label{cor:squareproj}
\end{corollary}

\Proof Consider a maximal set of perfectly distinguishable pure states $\{\varphi_i\}_{i=1}^{d_\rA}$
such that $\{\varphi_i\}_{i\in V}$ is maximal in $F$. In this way $F$ is the face generated by $V$,
and, therefore $\Pi_F=\Pi_V$. The thesis follows by taking $V = W$ in lemma \ref{lem:intersection}.  \qed

\begin{corollary}
  For every state $\rho \in \Stset_1 (\rA)$ such that $\rho\not\in F^\perp$, the normalized state
  $\rho'$ defined by
  \begin{equation}
    \K{\rho'} = \frac{\Pi_F  \K \rho}{\B e \Pi_F  \K \rho} 
    \label{eq:tric}
  \end{equation}  belongs to the face $F$.  
  \label{cor:projrho}
\end{corollary}
\Proof By corollary \ref{cor:effettodipi}, we have $\B e \Pi_F =
(a_F|$.  Since $\rho\not\in F^\perp$, we must have
$(e|\Pi_F|\rho)=(a_F|\rho)>0$, and, therefore, the state $\rho'$ in Eq.~\eqref{eq:tric} is well
defined. Moreover, using the definition of $\rho'$ we obtain 
\begin{align*}
\SC {a_F} {\rho'} &= \frac{\B {a_F} \Pi_F \K \rho}{ \B e \Pi_F \K{\rho}}\\
&=1,  
\end{align*}
having used corollaries  \ref{cor:projeff} and \ref{cor:effettodipi} for the last equality.
Finally, lemma
\ref{lem:identifytest} implies that $\rho' $ belongs to the face $F$.
\qed

\begin{corollary}\label{cor:postproj}
  Let $\Pi_{\{\varphi\}}$ be the projection on the pure state $\varphi \in\Stset_1(\rA)$ and $a$ be the atomic effect such that $\SC {a}{\varphi}=1$. Then for
  every state $\rho \in\Stset_1(\rA)$ one has $\Pi_{\{\varphi\}}
  \K{\rho} =  p \K{\varphi} $ where $p =\SC {a} \rho$.
\end{corollary}
\Proof Recall that, by corollary \ref{cor:effettodipi}, we have $\B {a}  =  \B e  \Pi_{\{\varphi\}}$.     If $\SC{a} \rho=0$ then clearly $\Pi_{\{\varphi\}}|\rho)=0$.  Otherwise, the proof is a straightforward application of corollary \ref{cor:projrho}. \qed

We conclude the present subsection with a result that will be useful in the next subsection.
\begin{lemma}
  An  atomic transformation $\tA\in\Trnset (\rA)$ satisfies $\tA=_{\omega_F} \tI_\rA$ if and only
  if
  \begin{equation}
    \Pi_F\tA=\Pi_F.
    \label{revidupthe}
  \end{equation}
  \label{lem:revidupthe}
\end{lemma}

\Proof Suppose that $\tA=_{\omega_F} \tI_\rA$.    Let $\Phi \in\Stset_1 (\rA\tilde\rA)$ be a purification of the invariant state $\chi_\rA$ and define the two pure states
\begin{align*}
|\Phi_F) &:= \frac{d_{\rA}}{|F|} ~  (\Pi_F\otimes\tI_{\tilde \rA})|\Phi)\\
|\Phi_F') &=: \frac{d_\rA}{|F|} ~ (\Pi_F\tA\otimes\tI_{\tilde \rA})   |\Phi).
\end{align*} 
Then we have
\begin{align*}   
\B{e_{\rA}}  | \Phi_F') & = [\B {a_F}  \tA]  \K{\Phi} \\
& =  \B {a_F}  \K{\Phi}\\
& = \B {e_\rA}  \K{\Phi_F},      
\end{align*} 
having used the condition $\B {a_F} \tA = \B {a_F}$ (lemma \ref{lem:aA=a}).   Now, we proved that $\Phi_F$ and $\Phi_F'$ have the same marginal on system $\tilde \rA$.
By the uniqueness of purification, there exists a reversible transformation $\tV\in\grp G_\rA$ such
that $\K {\Phi_F'} = (\tV \otimes \tI_{\tilde \rA}) \K{\Phi_F}$. Since $\Phi$ is dynamically
faithful, this implies
  $\Pi_F\tA=\tV\Pi_F$.
  Now, for every $\rho$ in $F$ one has $ \tV|\rho)=\tV\Pi_F|\rho)=\Pi_F\tA|\rho)=|\rho)$, namely
  $\tV=_{\omega_F}\tI_\rA$.  
  Applying lemma \ref{lem:idupthe} with $\tC  = \tV \Pi_F$ and using the idempotence of projections we then obtain 
  \begin{align*}
  \Pi_F\tA&=\tV\Pi_F\\
  & =  (\tV \Pi_F)  \Pi_F   \\
  &=\Pi_F\Pi_F\\
  &=\Pi_F.
  \end{align*}
  Conversely, suppose that Eq.~\eqref{revidupthe} is satisfied. Let $\varphi\in F$ be a pure state
  in $F$ and $a$ be the atomic effect such that $\SC {a} \varphi =1$. Then, we have
\begin{equation*}
 \B {a}  \tA \K\varphi =  (a|\Pi_F\tA|\varphi)=(a| \Pi_F |\varphi)=\SC {a} \varphi =1,
\end{equation*}
having used the relation $\B a \Pi_F = \B a$  (corollary \ref{cor:projeff}).  Then, by theorem \ref{theo:dualeeffetto},
$\tA\varphi=\varphi$. Since $\varphi \in F$ is arbitrary, this implies $\tA =_{\omega_F} \tI_\rA$.
\qed

\subsection{Projection of a pure state on two orthogonal faces}

In Section \ref{sec:qubits} we proved a number of results concerning two-dimensional systems.  Some
properties of two-dimensional systems will be extended to the case of generic systems using the
following lemma:

\begin{lemma}
  Consider a pure state $\varphi\in\Stset_1(\rA)$ and two complementary projections $\Pi_F$ and
  $\Pi^\perp_{F}$. Then, $\varphi$ belongs to the face identified by the state
  $\K\theta:=(\Pi_F+\Pi^\perp_{F})\K \varphi$.
  \label{lem:prorefin}
\end{lemma}

\Proof If $\Pi_F \K\varphi = 0$ (resp. $\Pi_{F}^\perp \K \varphi = 0$), then there is nothing to
prove: this means that $\Pi_{F}^\perp \K \varphi =\K \varphi$ (resp.  $\Pi_{F} \K \varphi =\K
\varphi$) and the thesis is trivially true.  Suppose now that $\Pi_F\K \varphi \not= 0$ and
$\Pi_{F}^\perp \K \varphi \not = 0$.  Using the notation $\Pi_1:= \Pi_F$, $\Pi_2 : = \Pi_F^\perp$, we can
define the two pure states $|\varphi_i):=\Pi_i|\varphi)/ \B e \Pi_i|\varphi)$, $i=1,2$, and the
probabilities $p_i = \B e \Pi_i \K \varphi$. In this way we have $\Pi_i\K{\varphi}=p_i\K{\varphi_i}$
for $i=1,2$ and $\theta = p_1 \varphi_1 + p_2 \varphi_2$.  Taking the atomic effect $\B {a_i}$ such
that $\SC {a_i}{\varphi_i}=1$ we have $a_{F_\theta} = a_1 + a_2$, where $a_{F_\theta}$ is the effect
associated to the face $F_\theta$.  Recalling that $\B {a_i} \Pi_i = \B{a_i}$ for $i=1,2$ (corollary
\ref{cor:projeff}), we then conclude the following
\begin{equation*}
  \begin{split}
    (a_{F_\theta}|\varphi) &=[(a_1|+(a_2|]|\varphi)\\
    &= (a_1|   \Pi_1 |\varphi) +  (a_2| \Pi_2|\varphi)\\
    & = \sum_{i=1,2}  p_i  \SC {a_i}  {\varphi_i} =1.
  \end{split}
\end{equation*}
Finally, lemma \ref{lem:identifytest} yields $\varphi\in F_\theta$.\qed

A  consequence of lemma \ref{lem:prorefin}  is the following
\begin{lemma}\label{lem:senzanome}
  Let $\varphi \in \Stset_1 (\rA)$ be a pure state, $a \in \Cntset(\rA)$ be the unique atomic
  effect such that $\SC {a} \varphi=1$, and $F$ be a face in $\Stset_1(\rA)$. If $\rho$ is
  perfectly distinguishable from $\Pi_F \K \varphi$ and from $\Pi^\perp_{F} \K \varphi $ then $\rho$
  is perfectly distinguishable from $\K \varphi$.  In particular, one has $\SC {a} \rho =0$.
\end{lemma}

\Proof Since $\rho$ is perfectly distinguishable from $\Pi_F\K \varphi$ and $\Pi^\perp_{F} \K
\varphi$, it is also perfectly distinguishable from any convex combination of them (corollary
\ref{cor:distinguishablityfromconvexcomb}).  Equivalently, $\rho$ is perfectly distinguishable from the face
$F_\theta$ identified by $\K\theta:= \Pi_F \K\varphi + \Pi_{F}^\perp \K \varphi$. In particular, it
must be perfectly distinguishable from $\varphi$, which belongs to $F_\theta$ by virtue of lemma
\ref{lem:prorefin}.  If $a$ is the atomic effect such that $\SC {a} \varphi =1$,
then by lemma \ref{lem:fazero} we have $\SC {a} \rho = 0$. \qed

A technical result that will be useful in the following is:  
\begin{lemma}\label{lem:ultimo,speriamo}
Let $\varphi \in \Stset_1 (\rA)$ be a pure state such that $\Pi_F \K \varphi \not = 0$ and $\Pi_F^\perp \K{\varphi} \not = 0$.   Define the pure states $\K{\varphi_1} :=  \Pi_F  \K \varphi  / \B e  \Pi_F  \K  \varphi $ and  $\K{\varphi_2}: =  \Pi_F^\perp  \K \varphi  /\B e  \Pi^\perp_F \K  \varphi $ and the mixed state $\K \theta : = (\Pi_F  +  \Pi_F^\perp) \K \varphi $. Then, we have 
\begin{align*}
 \Pi_F   \Pi_{F_\theta}  &  =  \Pi_{\{\varphi_1\}}\\
 \Pi_F^\perp   \Pi_{F_\theta} & =  \Pi_{\{\varphi_2\}}.
\end{align*}
\end{lemma}
\Proof Let $\{\psi_i\}_{i=1}^{|F|}$ be a maximal set of perfectly distinguishable pure states in $F$, chosen in such a way that $\psi_1 = \varphi_1$, and let $\{\psi_i\}_{i=|F| +1}^{d_\rA}$ be a maximal set of perfectly distinguishable pure states in $F^\perp$, chosen in such a way that $\psi_{|F|+1} =  \varphi_2$.   Defining the sets $V:=\{  1, \dots, |F|\}$, $W : = \{  |F| +1, \dots, d_\rA\}$, and $U:= \{1, |F| +1\}$ we then have $\Pi_V  =   \Pi_F  $,  $\Pi_W  = \Pi_{F}^\perp$  and $\Pi_U  =  \Pi_{F_\theta}$.  Using lemma \ref{lem:intersection} we obtain 
\begin{align*}
\Pi_F \Pi_\theta  & = \Pi_V \Pi_U  \\
& = \Pi_{V \cap U}  \\
& = \Pi_{\{\psi_1\}}\\
& = \Pi_{\{\varphi_1\}}  
\end{align*} 
and  
\begin{align*}
\Pi_F^\perp \Pi_\theta  & = \Pi_W \Pi_U  \\
& = \Pi_{W \cap U}  \\
& = \Pi_{\{\psi_{|F|  +1}\}}\\
& = \Pi_{\{\varphi_2\}}  
\end{align*} 
\qed

We conclude this subsection with an important observation about the group of reversible transformations
that act as the identity on two orthogonal faces $F$ and $F^\perp$. If $F$ is a face of $\Stset_1
(\rA)$, let us define $\grp G_{F, F^\perp}$ as the group of all reversible transformations $\tU
\in \grp G_\rA$ such that
\begin{equation*}
 \tU =_{\omega_F} \tI_\rA, \qquad \tU =_{\omega_F^\perp}  \tI_\rA.
\end{equation*}
Then we have the following
\begin{theorem}\label{theo:circ}
For every face $F \subset \Stset_1 (\rA)$ such that $F \not = \{0\}$ and $F \not = \Stset_1 (\rA)$, the group   $\grp G_{F, F^\perp}$ is topologically equivalent to a circle.
\end{theorem}
\Proof Let $\tU$ be a transformation in $\grp G_{F, F^\perp}$, $\Phi
\in\Stset(\rA\tilde\rA)$ be a purification of the invariant state
$\chi_\rA$ and $\K{\Phi_\tU}:= (\tU \otimes \tI_{\tilde\rA}) \K\Phi$
be the Choi state of $\tU$.  Define the orthogonal faces $\tilde F: = F_{\omega_F \otimes \chi_{\tilde \rA}} $ and $\tilde F^\perp =F_{\omega_{\tilde F}  \otimes \chi_{\tilde \rA}}$, and the projections $\Pi_{\tilde F}  : =  \Pi_F  \otimes \tI_{\tilde \rA} $  and    $\Pi^\perp_{\tilde F}  : =  \Pi^\perp_F  \otimes \tI_{\tilde \rA} $ (see lemma \ref{lem:pitensorid}).
Using lemma
\ref{lem:revidupthe} we then obtain
\begin{align*}
\Pi_{\tilde F} \K {\Phi_\tU} & =  (\Pi_ F \otimes \tI_{\tilde \rA})  \K {\Phi_\tU} \\ 
 &=  (\Pi_F\tU \otimes \tI_{\tilde \rA})  \K \Phi \\
 & = (\Pi_F \otimes \tI_{\tilde \rA}) \K \Phi\\
& =  \frac{|F|} {d_\rA}  \K{\Phi_F},  
\end{align*} 
and, similarly, 
\begin{align*}
\Pi_{\tilde F}^\perp  \K {\Phi_\tU}  & = (\Pi_F^\perp \otimes \tI_{\tilde \rA})  \K {\Phi_\tU}\\
 &=  (\Pi^\perp_F\tU \otimes \tI_{\tilde \rA})  \K \Phi \\
 & = (\Pi^\perp_F \otimes \tI_{\tilde \rA}) \K \Phi\\
 & =  \frac{|F^\perp|} {d_\rA} \K {\Phi_{F^\perp}}. 
 \end{align*}  
This means that the projections of $\Phi_\tU$ on the faces $\tilde F$ and
$\tilde F^\perp$ are independent of $\tU$.  Also, it means that $\Phi_\tU$
belongs to the face $F_\theta$ identified by the state $\K\theta :=
\frac{|F| } {d_\rA} \K {\Phi_F} + \frac{|F^\perp|} {d_\rA}
\K{\Phi_{F^\perp}}$ (lemma \ref{lem:prorefin}).  Now, by the compression axiom, $F_\theta$ is
isomorphic to the state space of a qubit, say with $\Phi_F$ and
$\Phi_{F^\perp}$ indicating the north and south poles of the Bloch
sphere, respectively, and we know that all the Choi
states $\{\Phi_\tU\}_{\tU \in \grp G_{F, F^\perp}}$ are at the same
latitude (precisely, the latitude is the angle $\zeta$ given by
$\cos \zeta = (|F|-|F^\perp|)/d_\rA$).  This implies that the
states $\{\Phi_\tU\}_{\tU \in \grp G_{F, F^\perp}}$ are a subset of a
circle $\mathsf C_\zeta$ in the Bloch sphere describing the face $F_\theta$.  Precisely, the circle $\mathsf C_\zeta$ is given by 
\begin{align*}
\mathsf C_\zeta : =  \left \{  \Psi  \in  F^{\phantom{\perp}}_\theta ~|~   \right.   \Pi_{\{\Phi_F\}}   \K\Psi  &=  \frac{|F|}{d_\rA}  \K{ \Phi_F},  \\
    \Pi_{\{\Phi_{F^\perp}\}}   \K\Psi  &=  \frac{|F^\perp|}{d_\rA}  \left. \K{ \Phi_{F^\perp}} \right\}
\end{align*}
We now prove that in fact they are the whole circle. Let  $\Psi$ be a state in $\mathsf
C_\zeta$.  Since $\K\Psi$ belongs to the face $F_\theta$, we obtain  
\begin{align*}
(\Pi_F \otimes \tI_{\tilde \rA})  \K \Psi  & =  \Pi_{\tilde F}  \K  \Psi  \\
& = \Pi_{\tilde F} \Pi_{F_\theta} \K  \Psi \\
& =  \Pi_{ \{\Phi_F\}}    \K\Psi\\
& =   \frac{|F|} {d_\rA}   \K{\Phi_F}     
\end{align*} 
[the third equality comes from lemma \ref{lem:ultimo,speriamo} with the substitutions $F \to \tilde F$,   $\varphi  \to  \Psi $, $\varphi_1 \to \Phi_F$, and $\varphi_2 \to  \Phi_{F^\perp}$], 
and, similarly,
\begin{align*}
(\Pi_F^\perp \otimes \tI_{\tilde \rA})  \K \Psi  & =  \Pi_{\tilde F}^\perp  \K  \Psi  \\
& = \Pi_{\tilde F}^\perp \Pi_{F_\theta} \K  \Psi \\
& =  \Pi_{ \{\Phi_{F^\perp}\}}    \K\Psi\\
& =   \frac{|F^\perp|} {d_\rA}   \K{\Phi_{F^\perp}}.     
\end{align*}
Therefore, we have
\begin{align*}
\B {e_\rA}  \K \Psi &= [ \B {a_F}  + \B {a_F^\perp} ] \K \Psi\\
& = [\B {e_\rA} \Pi_F  \otimes \tI_{\tilde \rA}  ] \K \Psi + [  \B {e_\rA}  \Pi_F^\perp \otimes \tI_{\tilde \rA}  ] \K \Psi \\
& =  \frac{|F| } {d_\rA} \B {e_\rA}  {\K \Phi_F}  + \frac {|F^\perp|} {d_\rA} \B {e_\rA}  \K{\Phi_{F^\perp}} \\
& =   [\B {e_\rA} \Pi_F  \otimes \tI_{\tilde \rA}  ] \K \Phi + [  \B {e_\rA}  \Pi_F^\perp \otimes \tI_{\tilde \rA}  ] \K \Phi  \\
&= [ \B {a_F}  + \B {a_F^\perp} ] \K \Phi\\
& = \B {e_\rA}  \K \Phi\\ 
&= \K{\chi_{\tilde \rA}}.
\end{align*} 
Since $\Psi$ and $\Phi$ are both purifications of the invariant state $\chi_{\tilde \rA}$, by the
uniqueness of purification there must be a reversible transformation $\tU \in \grp G_\rA$ such that
$\K \Psi = (\tU \otimes \tI_{\tilde \rA}) \K\Phi$.  Finally, it is easy to check that $\Pi_F \tU =
\Pi_F$ and $\Pi_F^\perp \tU = \Pi_{F}^\perp$, which, by lemma \ref{lem:revidupthe} implies
$\tU=_{\omega_F} \tI_{\rA}$ and $\tU=_{\omega_F^\perp} \tI_\rA$.  This proves that the Choi states $\{\Phi_\tU\}_{\tU \in \grp G_{F, F^\perp}}$ are the whole circle $\mathsf
C_\zeta$. Since the Choi isomorphism is continuous in the operational norm (see theorem 14 of \cite{purification}),  the group $\grp G_{F, F^\perp}$ is topologically equivalent to a circle.  \qed

\section{The superposition principle}\label{sec:superposition}
The validity of the superposition principle,
proved for two-dimensional systems using the geometry of the Bloch sphere (corollary \ref{cor:superpositionqubit}), can be now extended to
arbitrary systems thanks to lemma \ref{lem:prorefin}.

\begin{theorem} \label{theo:superpositiongeneral}{\bf (Superposition principle for general systems)}
  Let $\{\varphi_i\}_{i=1}^{d_\rA} \subseteq \Stset_1 (\rA)$ be a
  maximal set of perfectly distinguishable pure states and
  $\{a_i\}_{i=1}^{d_\rA}$ be the observation-test such that $\SC {a_i} {\varphi_j}
  = \delta_{ij}$. Then, for every choice of probabilities
  $\{p_i\}_{i=1}^{d_\rA} $, $ p_i\ge 0, \sum_{i=1}^{d_\rA} p_i =1 $
  there exists at least one pure state $\varphi_{\mathbf
    p}\in\Stset_1(\rA)$ such that
  \begin{equation}\label{superpuno}
  p_i = \SC
  {a_i}{\varphi_{\mathbf p}} \qquad \forall i=1, \dots, d_\rA.
  \end{equation}
  \label{lem:allprobN}
  or, equivalently, 
  \begin{equation}\label{superpdue}
  \Pi_{\{\varphi_i\}}  \K{\varphi_{\mathbf p}}  = p_i  \K{\varphi_i}  \qquad \forall i  =1, \dots, d_\rA,  
  \end{equation}
  where $\Pi_{\{\varphi_i\}}$ is the projection on $\varphi_i$.
\end{theorem}

\Proof Let us first prove the equivalence between Eqs. (\ref{superpuno}) and (\ref{superpdue}).    From Eq. (\ref{superpdue}) we obtain Eq. (\ref{superpuno}) using the relation $\B e \Pi_{\{i\}} = \B{a_i}$, which follows from corollary \ref{cor:effettodipi}.  Conversely, from Eq. (\ref{superpuno}) we obtain Eq. (\ref{superpdue})  using corollary  \ref{cor:postproj} . Now, we will prove Eq. (\ref{superpuno})
by induction. The statement for $N=2$ is proved by corollary \ref{cor:superpositionqubit}.  Assume
that the statement holds for every system $\rB$ of dimension $d_\rB=N$ and suppose that $d_\rA=N+1$.
Let $F$ be the face identified by $\omega_F = 1/N \sum_{i=1}^N \varphi_i$ and $F^\perp$ be the
orthogonal face, identified by the state $\varphi_{N+1}$.  Now there are two cases:  either $p_{N+1}  =1$  or $p_{N+1}  \not = 1$.  If $p_{N+1}=1$, then there is nothing to prove:  the desired state is $\varphi_{N+1}$.   Then, suppose that $p_{N+1}  \not = 1$.  Using the induction hypothesis and the
compression axiom \ref{compression} we can find a state $\psi_{\mathbf q} \in F$ such that $\SC
{a_i} {\psi_{\mathbf q}} = q_i$, with $q_i=p_i/(1-p_{N+1})$, $i=1, \dots, N$.  Let us then define a
new maximal set of perfectly distinguishable pure states $\{\varphi'_i\}_{i=1}^{N+1}$ 
, with $\varphi'_1=\psi_{\mathbf q}$ and $\varphi'_{N+1}=\varphi_{N+1}$.  Note that one has $\omega_F = 1/N \sum_{i=1}^N \varphi'_i$, that is, $F$ is the face generated by the states $\{\varphi_i'\}_{i=1}^N$.    Now
consider the two-dimensional face $F'$ identified by $\theta = 1/2 (\varphi_1' + \varphi_{N+1}')$. By corollary
\ref{cor:superpositionqubit} (superposition principle for qubits) we know that there exists a pure state
$\varphi \in F'$ with $(a'_1|\varphi)=1-p_{N+1}$ and $(a'_{N+1}|\varphi)=p_{N+1}$. 
Let us define $V :=\{1, \dots, N\}$ and $W: = \{ 1, N+1\}$. Then,  we have $\Pi_F = \Pi_V$  and $\Pi_{F'} = \Pi_W$, and  by lemma
\ref{lem:intersection},
\begin{align*}
\Pi_F |\varphi) &=\Pi_F\Pi_{F'} |\varphi)\\
&=\Pi_{V\cap W}|\varphi)\\
  & =\Pi_{ \{\varphi_1'\} } |\varphi)\\
  & =\Pi_{\{\psi_{\mathbf q}\}}|\varphi)\\
  &=(1-p_{N+1})|\psi_{\mathbf q}),
  \end{align*} 
having used corollary \ref{cor:postproj}  for the last equality. 
Finally, for $i=1,\dots,N$ we have
\begin{align*}
(a_i|\varphi)&=(a_i|\Pi_F|\varphi)\\
&=(1-p_{N+1}) \SC {a_i}  {\psi_{\mathbf q}}\\
&= (1-p_{N+1}) q_i\\
 &=p_i.
 \end{align*}
 On the other hand we have $(a_{N+1}|\varphi)=(a'_{N+1}|\varphi)=p_{N+1}$.
 \qed

\subsection{Completeness for purification}
Using the superposition principle and the spectral decomposition of theorem \ref{theo:diag} we can now show that every state of system $\rA$ has a purification in $\rA\rB$ provided $d_\rB \ge d_\rA$:
\begin{lemma}\label{lem:completenessforpurification}
For every state $\rho \in \Stset_1 (\rA)$ and for every system $\rB$ with $d_\rB \ge d_\rA$  there exists a purification of $\rho$  in  $\Stset_1(\rA\rB)$.
\end{lemma}  

\Proof Take the spectral decomposition of $\rho$, given by $\rho =
\sum_{i=1}^{d_\rA} p_i \varphi_i $, where $\{p_i\}$ are probabilities
and $\{\varphi_i\}_{i=1}^{d_\rA} \subset \Stset_1 (\rA)$ is a maximal
set of perfectly distinguishable pure states.  Let $\{\psi_i\}_{i=
  1}^{d_\rB}$ be a maximal set of perfectly distinguishable pure
states and $\{a_i\}_{i=1}^{d_\rA} \subset \Cntset (\rA)$ (resp.
$\{b_i\}_{i=1}^{d_\rB} \subset \Cntset(\rB)$) be the test such that
$\SC {a_i}{\varphi_j} = \delta_{ij}$ (resp. $\SC {b_i}{\psi_j} =
\delta_{ij}$).  Clearly, $\{\varphi_i \otimes \psi_j\}$ is a maximal
set of perfectly distinguishable pure states for $\rA\rB$.  Then, by
the superposition principle (theorem \ref{theo:superpositiongeneral})
there exists a pure state $\Psi_{\rho}$ such that $\SC {a_i \otimes b_j} {\Psi_\rho} = p_i \delta_{ij} $.  
Equivalently, we have $\B {b_i}_\rB \K {\Psi_\rho}_{\rA\rB} = p_i ~ \K {\varphi_i}_\rA$  for every $i=1, \dots, d_\rA$ and $\B {b_i}_\rB \K {\Psi_\rho}_{\rA\rB}=0$ for $i>d_\rA$.
Summing over $i$ we then obtain $\B e_\rB \K{\Psi_\rho}_{\rA\rB} =
\sum_{i=1}^{d_\rB} \B{b_i}_\rB \K{ \Psi_\rho}_{\rA\rB} = \sum_{i=1}^{d_\rA}
p_i\K{\varphi_i}_\rA = \K \rho_\rA$. \qed

In the terminology of Ref. \cite{purification}, lemma \ref{lem:completenessforpurification} states that a system $\rB$ with $d_\rB \ge d_\rA$  is complete for the purification of system $\rA$.

As a consequence of lemma \ref{lem:completenessforpurification} we have the following:
\begin{corollary}\label{cor:tutticoniugati}
Every system $\rB$ with $d_\rB = d_\rA $ is operationally equivalent to the conjugate system $\tilde \rA$.  
\end{corollary}
\Proof By corollary \ref{lem:completenessforpurification}, the invariant state $\chi_\rA \in\Stset_1 (\rA)$ has a purification $\Psi$ in $\Stset_1 (\rA\rB)$. By corollary \ref{cor:samespec},  the marginal of $\Psi$ on $\rB$ is the invariant state $\chi_\rB$.   By definition, this means that $\rB$  is a conjugate system of $\rA$.  Since the conjugate system $\tilde \rA$ is unique up to operational equivalence (corollary \ref{cor:conjusys}),  this implies the thesis. \qed  
  
\subsection{Equivalence of systems with equal
  dimension}\label{subsec:equivalence}
We are now in position to prove that two systems $\rA$ and $\rB$ with
the same dimension are operationally equivalent, namely that there is
a reversible transformation from $\rA$ to $\rB$.  In other words, we
prove that the informational dimension classifies the systems of our theory up to
operational equivalence.  The fact that this property is derived from
the principles, rather than being assumed from the start, is one of
the important differences of our work with respect to Refs.
\cite{Har01,DakBru09,Mas10}. Another difference is that here the equivalence of systems with the same dimension is proved after the derivation of the qubit, whereas in Refs.  \cite{Har01,DakBru09,Mas10} the derivation of the qubit requires the equivalence of  systems with the same dimension.

\begin{corollary}{\bf (Operational equivalence of systems with equal dimension)}
Every two systems $\rA$ an $\rB$ with $d_\rA = d_\rB$ are operationally equivalent.  \label{cor:equidim}
\end{corollary}

\Proof  By corollary \ref{cor:tutticoniugati}, $\rA$ and $\rB$ are both operationally equivalent to the conjugate system $\tilde \rA$. Hence, they are operationally equivalent to each other.  \qed

\subsection{Reversible operations of  perfectly distinguishable pure  states}

An important consequence of the superposition principle is the possibility of transforming an arbitrary maximal set of perfectly distinguishable pure states into another via a reversible transformation:  
\begin{corollary}\label{cor:reversibleclassicaloperations}
Let $\rA$ and $\rB$ be two systems with $d_\rA = d_\rB=:d$ and let $\{\varphi_i\}_{i=1}^d$  (resp.  $\{\psi_i\}_{i=1}^d$) be a maximal set of perfectly distinguishable pure states in $\rA$ (resp. $\rB$).   Then, there exists a reversible transformation $\tU \in \Trnset (\rA, \rB)$ such that $\tU \K{\varphi_i}  =\K{ \psi_i}$.
\end{corollary}

\Proof  Let $\Phi  \in \Stset (\rA \tilde\rA)$ be a purification of the invariant state $\chi_\rA$. Although we know that $\rA$  and $\tilde \rA$ are operationally equivalent  (corollary \ref{cor:tutticoniugati}) we use the notation $\rA$  and $\tilde \rA$ to distinguish between the two subsystems of $\rA\tilde\rA$.  Define the pure state $\tilde \varphi_i$ via the relation   $\B {a_i}_\rA    \K{\Phi}_{\rA\tilde\rA} = \frac 1 d \K {\tilde \varphi_i}_{\tilde \rA}$,  where $\{a_i\}_{i=1}^{d}$ is the observation-test such that $\SC {a_i }{\varphi_i}  = \delta_{ij}$.    
Let $\{\tilde a_i\}_{i=1}^d$ be the observation-test such that $\SC{\tilde a_i}  {\tilde \varphi_j} = \delta_{ij}$. Then, by lemma \ref{lem:prepver} we have 
\begin{equation}\label{bastaaa}
\B {\tilde a_i}_{\tilde \rA}  \K\Phi_{\rA\tilde\rA} =    \frac 1d  \K{\varphi_i}_{\rA}.
\end{equation}
On the other hand,  if $\{\tilde b_i\}_{i=1}^d$ is the observation-test such that $(\tilde b_i|\tilde \psi_j) = \delta_{ij}$, then
using the superposition principle (theorem \ref{theo:superpositiongeneral}) we can construct  a state $\Psi \in\Stset_1(\rB\tilde \rA)$ such that  $\SC {b_i \otimes \tilde a_j} {\Psi}  =  \delta_{ij}/d$, or, equivalently, 
\begin{equation}\label{bastaaaa}
\B  {\tilde a_i}_{\tilde \rA} \K {\Psi}_{\rB \tilde \rA}  =  \frac 1 d  \K  {\psi_i}_\rB. 
\end{equation}  
Now, $\Phi$ and $\Psi$ have the same marginal on system $\tilde \rA$:  they are both purifications of the invariant state $\chi_{\tilde \rA}$.  Moreover, $\rA$ and $\rB$ are operationally equivalent because they have the same dimension (corollary \ref{cor:equidim}).    Hence, by the uniqueness of purification, there must be a  reversible transformation $\tU \in  \Trnset(\rA, \rB)$ such that  
\begin{equation}\label{bastaaaaa}
\K\Psi_{\rB \tilde \rA} = (\tU \otimes \tI_{\tilde \rA})  \K \Phi_{\rA\tilde\rA}.
\end{equation}   
Combining Eqs. (\ref{bastaaa}), (\ref{bastaaaa}), (\ref{bastaaaaa}) we finally obtain  
\begin{align*}
  \frac 1 d   \tU \K  {\varphi_i}_\rA  &=    \left[\tU \otimes \B  {\tilde a_i}_{\tilde \rA}  \right] \K {\Phi}_{\rB \tilde \rA}\\
& = \B  {\tilde a_i}_{\tilde \rA}   \K {\Psi}_{\rB \tilde \rA}  \\ 
& =  \frac 1 d  \K  {\psi_i}_\rB ,
\end{align*}
that is, $\tU \K {\varphi_i} = \K {\psi_i}$ for every $i=1, \dots, d$. \qed

\section{Derivation of the density matrix formalism}\label{sec:matrix}

The goal of this section is to show that our set of axioms implies that
\begin{itemize}
\item the set of states for a
system $\rA$ of dimension $d_{\rA}$ is the set of density matrices on the Hilbert space
$\Cmplx^{d_\rA}$
\item  the set of effects is the set of positive matrices bounded by the identity
\item the pairing between a state and an effect is given by the trace of the product of the
corresponding matrices.
\end{itemize}
Using the result of theorem \ref{theo:statesspecify}, we will then obtain that all the physical transformations in our theory are exactly the physical transformations allowed in quantum mechanics.   This will conclude our derivation of quantum theory.

\subsection{The basis}\label{subsec:basis}

In order to specify the correspondence between states and matrices we choose a particular basis for
the vector space $\Stset_\Reals (\rA)$.  For this purpose, we adopt the choice of basis used in Ref. \cite{Har01}. The basis is constructed as follows: Let us first choose a
maximal set of $d_\rA$ perfectly distinguishable states $\{\varphi_m\}_{m=1}^{d_\rA}$, and declare
that they are the first $d_\rA$ basis vectors.  Then, for every $m<n$ the face $F_{mn}$ generated
by $\{\varphi_m, \varphi_n\}$ defines a ``two-dimensional subsystem": precisely, the face $F_{mn}:=
F_{\omega_{mn}}$ with $\omega_{mn} :={\frac{\varphi_m+ \varphi_n} 2}$ can be ideally encoded in a
two-dimensional system.  Now, the convex set of states of a two-dimensional system is the Bloch
sphere, and we can choose the $z$-axis to be the line joining the two states $\{\varphi_m,
\varphi_n\}$, e.g. with  the positive direction of the $z$-axis being the direction from $\varphi_m$
to $\varphi_n$.  Once the direction of the $z$-axis has been specified, we can choose the $x$ and
$y$ axes.  Note that any couple of orthogonal directions in the plane orthogonal to $z$-axis is a valid choice for the 
$x$- and $y$-axes (here we do not restrict ourselves to the choice of a right-handed coordinate system).  At the moment there is no relation among the different choices of axes made for
different values of $m$ and $n$. However, to prove that the states are represented by positive
matrices, later we will have to find a suitable way of connecting all these choices of axes.

Let $\varphi_{x, +}^{mn}, \varphi_{x, -}^{mn} \in F_{mn}$ ($\varphi_{y, +}^{mn},\varphi_{y,
  -}^{mn} \in F_{mn}$) be the two perfectly distinguishable states in the direction of the
$x$-axis ($y$-axis) and define
\begin{equation}\label{def:sigmaxyz}
\sigma_{k}^{mn} := \varphi_{k,+}^{mn} - \varphi_{k,-}^{mn} \qquad k =x,y . 
\end{equation} 
An immediate observation is the following:  
\begin{lemma}\label{lem:trivialfrombloch}
The four vectors $\{ \varphi_m, \varphi_n, \sigma_x^{mn}, \sigma_y^{mn} \} \subseteq
\Stset_{\Reals} (\rA)$ are linearly independent. Moreover, denoting by $a_l \in \Cntset (\rA)$ the atomic effect such that $\SC {a_l}{\varphi_l}  = 1$ we have $\SC {a_l}{\sigma^{mn}_k}  =  0$ for every $l,m,n  \in \{1, \dots, d_\rA\}$ and for every $k = x,y$. 
\end{lemma}
\Proof  Linear independence is evident from the geometry of the Bloch sphere.  Moreover, for $ l \not \in  \{m,n\}$ the states $\varphi_{k , \pm}^{mn}$ are perfectly distinguishable from $\varphi_l$, and, therefore $\SC {a_l} {\sigma_k^{mn}}  =  0$.   If $l \in \{m,n\} $, since the states  $\varphi_{k , \pm}^{mn}$, $k = x,y$ lie on the equator of the  Bloch sphere, we know that $\SC {a_l}  {\varphi_{k,\pm}^{mn}}  =  1/2$  for $k = x,y$.  Hence, $\SC {a_l} {\sigma_k^{mn}}  =\frac  12 - \frac 12  0$.  \qed

We now show that the collection of all vectors obtained in this way is a basis for $\Stset_{\Reals}
(\rA)$.  To this purpose we use the following
\begin{lemma}Let $V\subset\{1,\dots,d_\rA\}$, and consider the
  projection $\Pi_V$. Then, for $m\in V$ and $n\not\in V$, one has
  $\Pi_V|\sigma_{k}^{mn})=0$ for $k=x,y$.
  \label{lem:proj}
\end{lemma}

\Proof Using lemma \ref{lem:intersection} and corollary
\ref{cor:postproj} we obtain
\begin{equation*}
  \begin{split}
    \Pi_{V}|\varphi_{k,\pm}^{mn})&=\Pi_{V}\Pi_{\{m,n\}}|\varphi_{k,\pm}^{mn})\\
    &=\Pi_{\{m\}}|\varphi_{k, \pm}^{mn})\\
    &= |\varphi_m) ~ (a_m|\varphi_{k,\pm}^{mn})
  \end{split}
\end{equation*}
Since the face $F_{mn}$  is isomorphic to the Bloch sphere and the state since $\varphi_{k\pm}^{mn}$, $k = x,y$ lie on the equator of the Bloch sphere, we know that
$(a_m|\varphi_{k\pm}^{mn})=\frac12$. This implies
\begin{align*} 
  \Pi_{V}|\sigma_{k}^{mn}) & =  \Pi_{V}     \left (  |\varphi_{k,+}^{mn}) -  |\varphi_{k,-}^{mn}) \right)\\
  & =\K{\varphi_m} \left (\frac 12 - \frac 12 \right) =0 .
 \end{align*} \qed

\begin{lemma}
  The vectors $\{\varphi_n\}_{m=1}^{d_\rA} \cup \{\sigma_k^{mn}\}_{n>m=1,\dots,d_\rA\
    k=x,y}$ form a  basis for  $\Stset_{\mathbb R}(\rA)$.
  \label{lem:matrep}
\end{lemma}

\Proof Since the number of vectors is exactly $d_\rA^2$, to prove that they form a basis it is
enough to show that they are linearly independent.  Suppose that there exists a vector of
coefficients $\{c_m\}\cup\{c_{k}^{mn}\}$ such that
\begin{equation*}
 \sum_m c_m\varphi_m+\sum_{n>m,\ k=x,y}c_{k}^{mn}\sigma_{k}^{mn}=0.
\end{equation*}
Applying the projection $\Pi_{\{m,n\}}$ on both sides and using lemma \ref{lem:proj} we obtain
\begin{equation*}
  c_m|\varphi_m)+c_n|\varphi_n)+c^{x}_{mn}|\sigma_x^{mn})+c_{y}^{mn}|\sigma_y^{mn})=0.
\end{equation*}
However, we know that the vectors $\{\varphi_m,\varphi_n,\sigma_x^{mn},\sigma_y^{mn}\}$ are linearly
independent. Consequently, $c_m=c_n=c^k_{mn}=0$ for all $m,n,k$.  \qed

\subsection{The matrices}

Since the state space $\Stset(\rA)$ for system $\rA$ spans a real vector space of dimension
$D_\rA=d_\rA^2$, we can decide to represent the vectors $\{\varphi_m\}_{m=1}^{d_\rA} \cup
\{\sigma_k^{mn}\}_{n>m=1,\dots,d_\rA\ k=x,y}$ as Hermitian $d_\rA \times d_\rA$ matrices.
Precisely, we associate the vector $\varphi_m$ to the matrix $S_{\varphi_m}$ defined by
\begin{equation}\label{matphii}
  \left[S_{\varphi_m}\right]_{rs} =  \delta_{rm} \delta_{sm},
\end{equation} 
the vector $\sigma^{mn}_{x}$ to the matrix 
\begin{equation}\label{matsx}
  \left[S_{\sigma^{mn}_x}\right]_{rs} =  \delta_{rm} \delta_{sn}  + \delta_{rn} \delta_{sm}
\end{equation}
and the vector $\sigma^{mn}_{y}$ to the matrix 
\begin{equation}
  \left[S_{\sigma^{mn}_y}\right]_{rs} = i \lambda \left( \delta_{rm} \delta_{sn}  - \delta_{rn} \delta_{sm}\right).\label{eq:itsimportant}
\end{equation}
where $\lambda$ can take the values $+1$ or $-1$. The freedom in the
choice of $\lambda$ will be useful in subsection
\ref{tensor}, where we will introduce the representation of composite
systems of two qubits.  However, this choice of sign plays no role in
the present subsection, and for simplicity we will take the positive
sign.

Recall that in principle any orthogonal direction in the plane orthogonal to the $z$-axis can be chosen to be
the $x$-axis.  In general, the other possible choices for the $x$-axis will lead to matrices of the form
\begin{equation}
  \left[S_{\sigma^{mn}_{x,\theta}}\right]_{rs} =  \delta_{rm} \delta_{sn}  e^{i\theta} + \delta_{rn} \delta_{sm} e^{-i\theta} \qquad \theta \in [0, 2\pi)\label{eq:rotax},
\end{equation}
and the corresponding choice for the $y$-axis will lead to a matrices of the form 
\begin{equation}
  \left[S_{\sigma^{mn}_{y,\theta}}\right]_{rs} =   i \lambda \left(  \delta_{rm} \delta_{sn}      e^{i\theta} - \delta_{rn} \delta_{sm} e^{-i\theta} \right) \qquad \theta \in [0, 2\pi)\label{eq:rotay},
\end{equation}

Since the vectors $\{\varphi_m\}_{m=1}^{d_\rA} \cup \{\sigma_k^{mn}\}_{n>m=1,\dots,d_\rA ;\ k=x,y}$
are a basis for the real vector space $\Stset_{\Reals} (\rA)$, we can expand any state $\rho \in
\Stset(\rA)$ on them:
\begin{equation}\label{eq:expandrho} 
  \K \rho = \sum_m \rho _m \K{\varphi_m}+\sum_{n>m,\ k=x,y} \rho_{k}^{mn} \K{\sigma_{k}^{mn}}
\end{equation}
and the expansion coefficients $\{\rho_m\}_{m=1}^{d_\rA } \cup \{\rho^{mn}_{k}\}_{n>m =1, \dots,
  d_\rA; \ k = x,y}$ are all real.  Hence, each state $\rho$ is in one-to-one correspondence with a
Hermitian matrix, given by
\begin{equation}\label{eq:expandsrho} 
  S_\rho = \sum_m  \rho_m S_{\varphi_m}+\sum_{n>m,\ k=x,y} \rho_{k}^{mn}  S_{\sigma_{k}^{mn}}.
\end{equation}

Since effects are linear functionals on states, they are also represented by Hermitian matrices. We
will indicate with $E_a$ the Hermitian matrix associated to the effect $a \in \Cntset (\rA)$.  The matrix $E_a$ is uniquely defined by the relation:
\begin{equation*}\SC a \rho = \Tr[E_a S_\rho]. 
\end{equation*}

In the rest of the section we show that the set of matrices $\{S_\rho ~|~ \rho \in \Stset_1 (\rA)\}$
is the whole set of positive Hermitian matrices with unit trace and that the set of matrices $\{E_a
~|~ a \in \Cntset(\rA)\}$ is the set of positive Hermitian matrices bounded by the identity.

 Let us start from some simple facts:

\begin{lemma}
  The invariant state $\chi_\rA$ has matrix representation $S_{\chi_\rA}=\frac {I_{d_\rA}}{d_\rA}$,
  where $I_{d_\rA}$ is the identity matrix in dimension $d_\rA$.
\end{lemma}

\Proof Obvious from the expression $\chi_\rA =\frac1d\sum_m\varphi_m$ and from the matrix
representation of the states $\{\varphi_m\}_{m=1}^{d_\rA}$ in Eq. (\ref{matphii}).\qed
 
 \begin{lemma}\label{lem:matai}
   Let $a_m \in\Cntset(\rA)$ be the atomic effect such that $\SC {a_m} {\varphi_m}  =1$.  Then, the effect $a_{m}$ has matrix representation $E_{a_m}$ such
   that $E_{a_{m}}=S_{\varphi_m}$.
\end{lemma}

\Proof Let $\rho\in\Stset_1 (\rA)$ be an arbitrary state.  Expanding $\rho$ as in Eq. (\ref{eq:expandrho})  and using lemma \ref{lem:trivialfrombloch} we obtain $\SC {a_m} {\rho}  = \rho_m$.  On the other hand, by Eq. (\ref{eq:expandsrho}) we have that $\rho_m$ is the $m$-th diagonal element of the matrix $S_\rho$: by definition of $S_{\varphi_m}$ [eq.  (\ref{matphii})], this implies  $\rho_m =  \Tr [  S_{\varphi_m}   S_\rho  ]$.     Now, by construction we have $\Tr[E_{a_m}  S_\rho]  =  \SC {a_m} \rho = rho_m =  \Tr[S_{\varphi_m}  S_\rho]$ for every $\rho \in \Stset_1 (\rA)$. Hence, $E_{a_m}  =  S_{\varphi_m}$. \qed

\begin{lemma}
  The deterministic effect $e\in\Cntset(\rA)$ has matrix representation $E_e=I_{d_\rA}$.
\end{lemma}

\Proof Obvious from the expression $e= \sum_m a_{m}$, combined with lemma \ref{lem:matai} and Eq.
(\ref{matphii}).\qed

\begin{corollary}\label{cor:unittrace}
  For every  state $\rho \in \Stset_1 (\rA)$ one has
  \begin{equation*}
    \Tr[S_\rho]=1.
  \end{equation*}
\end{corollary}
\Proof $\Tr[ S_{\rho}] = \Tr[E_e S_\rho] = \SC e \rho = 1$. \qed

\begin{theorem}\label{theo:sqrts}
  The matrix elements of $S_\varphi$ for a pure state $\varphi\in\Stset_1 (\rA)$ are
  $(S_{\varphi})_{mn}=\sqrt{p_m p_n}e^{i\theta_{mn}}$, with   $\sum_{m=1}^{d_\rA}   p_m  =1$,   $\theta_{mn}  \in [0, 2\pi)$, $\theta_{mn}=0$ and
  $\theta_{mn}=-\theta_{nm}$.
\end{theorem}

\Proof First of all, the diagonal elements of $S_\varphi$ are given by $[S_{\varphi}]_{mm}  =  \SC  {a_m}  \varphi$ [cf. Eqs. (\ref{eq:expandrho}) and (\ref{eq:expandsrho})].  Denoting the $m$-th element by $p_m$, we clearly have  $\sum_{m=1}^{d_\rA}  p_m  =  \SC e \varphi = 1$.  Now,  the projection $\Pi_{\{m,n\}}|\varphi)$ is a state in the face $F_{mn}$,  and, by our choice of representation, the corresponding matrix $S_{\Pi_{\{m,n\}}  \K \varphi}$ is proportional to a pure qubit state (non-negative rank-one matrix). 
On the other hand,  it is easy to see from Eqs. (\ref{eq:expandrho}) and (\ref{eq:expandsrho})  that $S_{\Pi_{\{m,n\}}|\varphi)}$ is the matrix with the same elements as $S_\varphi$ in the block corresponding to the qubit $(m,n)$ and 0 elsewhere.  In order to be positive and rank-one the corresponding $2\times2$
sub-matrix must have the off-diagonal elements $(S_{\varphi})_{mn}=\sqrt{p_m p_n}e^{i\theta_{mn}}$, for some $\theta_{mn} \in [0, 2\pi)$ with $\theta_{nm}  =  - \theta_{mn}$.  Repeating the same argument for all choices of indices $m,n$, the thesis follows.\qed

\begin{theorem}
  For a pure state $\varphi \in\Stset_1(\rA)$, the corresponding atomic effect $a_\varphi$ such that
  $(a_\varphi|\varphi)=1$ has a matrix representation $E_{\varphi}$ with the property that
  $E_{\varphi}=S_\varphi$.
  \label{theo:E=S}
\end{theorem}

\Proof We already know that the statement holds for $d_\rA=2$, where we proved the Bloch sphere
representation, equivalent to the fact that states and effects are represented as $2 \times 2$
positive complex matrices, with the set of pure states identified with the set of all rank-one
projectors.  Let us now consider a generic system $\rA$. For every $m<n$, the face $F_{mn}$
generated by $\{\varphi_m, \varphi_n\}$ can be encoded in a two-dimensional system. Therefore, the
matrices $S_{\Pi_{\{m,n\}} \K \varphi}$ and $E_{\B a \Pi_{\{m,n\}}}$ are positive (also, recall that
all matrix elements outside the $(m,n)$ block are zero).  Let $\varphi^{(mn)}_\perp$ be the pure
state in the face $F_{mn}$ that is perfectly distinguishable from $\Pi_{\{m,n\}} \K \varphi$. Note
that, since $\varphi^{(mn)}_\perp$ belongs to the face $F_{mn}$, it is also perfectly
distinguishable from $\Pi_{\{1, \dots, d_\rA\} \setminus \{m,n\}}\K \varphi$.  Hence,
$\varphi^{(mn)}_\perp$ is perfectly distinguishable from $\varphi$ and, in particular, $(a|
\varphi^{(mn)}_\perp) = 0$ (lemma \ref{lem:senzanome}).  This implies the relation
\begin{align*}
  \Tr\left[ E_{\B a\Pi_{\{m,n\}}}  S_{\K{\varphi^{(mn)}_\perp}} \right]  &=  \B {a} \Pi_{\{m,n\}}  | {\varphi^{(mn)}_\perp})\\
  & = ( a  | {\varphi_\perp^{(mn)}}  ) =0.
\end{align*}
Now, since the matrix $E_{\B a \Pi_{\{m,n\}}}$ is positive, the above relation implies $E_{\B a
  \Pi_{\{m,n\}}} = c_{mn} S_{\Pi_{\{m,n\}} \K \varphi}$, where $c_{mn} \ge 0$.  Finally, repeating
the argument for all possible values of $(m,n)$, we obtain that $c_{mn} = c$ for every $m,n$, that
is, $E_a = c S_\varphi$.  Taking the trace on both sides we obtain $\Tr[E_a] = c$. To prove that
$c=1$, we use the relation $\Tr [E_a] /d_{\rA} = \SC a {\chi_\rA} = 1/d_\rA$. \qed

We conclude with a simple corollary that will be used in the next subsection:
\begin{corollary}\label{cor:effettospannato}
  Let $\varphi \in\Stset_1 (\rA)$ be a pure state and let $\{\gamma_i\}_{i=1}^{r} \subset
  \Stset_1 (\rA)$ be a set of pure states.  If the state $\varphi$ can be written as
  \begin{align*} \K \varphi = \sum_i
  x_i \K{\gamma_i} 
  \end{align*} 
  for some real coefficients $\{x_i\}_{i=1}^r$, then the atomic effect $a$ such that $\SC a \varphi = 1$ is given by
  \begin{equation*}
    \B a = \sum_i x_i  \B {c_i},   
  \end{equation*}
  where $c_i$ is the atomic effect such that $\SC{c_i}{\gamma_i} =1$.
\end{corollary}
\Proof For every $\rho\in\Stset (\rA)$ by theorem \ref{theo:E=S} one has
\begin{align*}
\SC a \rho & =  \Tr[  E_a S_\rho] = \Tr[S_\varphi  S_\rho] =\sum_i x_i  \Tr[S_{\gamma_i}  S_\rho]\\
&= \sum_i x_i \Tr[E_{c_i}  S_\rho] =\sum_i x_i \SC {c_i} \rho ,   
\end{align*}
thus implying the thesis.\qed

\subsection{Choice of axes for a two-qubit system}\label{tensor}

If $\rA$ and $\rB$ are two systems with $d_\rA = d_\rB = 2$, then we can use two different
types of matrix representations for the states of the composite system$\rA\rB$:

The first type of representation is the representation $S_\varphi$ introduced through lemma
\ref{lem:matrep}: here we will refer to it as the \emph{standard representation}. Note that there are many different representations of this type because for every pair $(m,n)$ there is freedom in choice of the $x$- and $y$-axis [cf. Eqs. (\ref{eq:rotax} ) and (\ref{eq:rotay})] 

The second type of representation is the \emph{tensor
product representation} $T_\varphi$, defined by the tensor product of matrices representing states of
systems $\rA$ and $\rB$: 
for a state $\K\rho =\sum_{i,j} \rho_{ij} \K{ \alpha_i} \K{ \beta_j}$, with $\alpha_i \in \Stset(\rA), \beta_j \in\Stset(\rB)$,  we have
\begin{equation}\label{deftensorrep}
  T_\rho  := \sum_{i,j}  \rho_{ij}  S^{\rA}_{\alpha_i}  \otimes S^{\rB}_{\beta_j} , 
\end{equation}
where $S^{\rA}$ (resp. $S^{\rB}$) is the matrix representation for system $\rA$ (resp. $\rB$).   Here the freedom is in the choice of the axes for the Bloch spheres of qubits $\rA$ and $\rB$.   
Since $\rA$ and $\rB$ are operationally equivalent, we will indicate the elements of the bases for $\Stset_\Reals (\rA)$ and $\Stset_\Reals (\rB)$ with the same letters:  $\{\varphi_m\}_{m=1}^2$ for the two perfectly distinguishable pure states and $\{\sigma_k\}_{k=x,y}$ for the remaining basis vectors.  

We now show a few properties of the tensor representation.  Let $F_A$ denote the matrix corresponding to the effect $ A \in \Cntset(\rA\rB)$ in the tensor
representation, that is, the matrix defined by
\begin{equation}
\SC  A \rho : =\Tr[F_A  T_\rho] \qquad \forall \rho \in \Stset (\rA\rB).
\end{equation}
It is easy to show that the matrix representation for effects must satisfy the analogue of Eq. (\ref{deftensorrep}):
\begin{lemma}\label{lem:a corto di nomi}
Let  $A \in \Cntset(\rA\rB)$ be a bipartite effect, written as $\B A =\sum_{i,j}  A_{ij}  \B {a_i}  \B {b_j}$. Then one has
\begin{equation*}
F_A = \sum_{i,j}    A_{ij}  ~E^\rA_{a_i} \otimes E^\rB_{b_j},
\end{equation*} 
where $E^\rA_{a_i}$ (resp. $E^\rB_{b_j}$) is the matrix representing the single-qubit effect $a_i$ (resp.
$b_j$) in the standard representation for qubit $\rA$ (resp. $\rB$).
\end{lemma}
\Proof For every bipartite state $\K \rho = \sum_{k,l} \rho_{kl} \K {\alpha_k} \K{\beta_l}$ one has
\begin{align*}
\Tr[F_A  T_\rho]  &=\SC A \rho\\
&=\sum_{i,j,k,l}   A_{ij} \rho_{kl}  \SC  {a_i}{\alpha_k}  \SC {b_j}{\beta_l}\\
&= \sum_{i,j,k,l}   A_{ij}   \rho_{kl}  \Tr[ E^\rA_{a_i}  S^\rA_{\alpha_k}]  \Tr   [ E\rB_{b_j}  S^\rB_{\beta_l}]\\
&= \sum_{i,j,k,l}    A_{ij}  \rho_{kl}  \Tr[ (E^\rA_{a_i}  \otimes  E^\rB_{b_j})  T_{\alpha_k \otimes \beta_l}]\\
&= \sum_{i,j}  A_{ij}  \Tr [  (E^\rA_{a_i}  \otimes E^\rB_{b_j} )  T_\rho ]  
\end{align*}
which implies the thesis. \qed

\begin{corollary}
  Let $\Psi \in\Stset_1 (\rA\rB)$ be a pure state and let $A\in\Cntset (\rA\rB)$ be the atomic
  effect such that $\SC A \Psi =1$. Then one has $F_A = T_\Psi$.
  \label{cor:F=T}
\end{corollary}
\Proof Let $\{a_i\}_{i=1}^4$ (resp. $\{\beta_j\}_{j=1}^4$) be a set of pure states that span
$\Stset_\Reals (\rA)$ (resp. $\Stset_\Reals (\rB)$) and expand $\Psi$ as $\K\Psi = \sum_{i,j} c_{ij}
\K {\alpha_i} \K {\beta_j}$.  Then, corollary \ref{cor:effettospannato} yields $\B A = \sum_{i,j}
c_{ij} \B {a_i} \B{b_j}$ where $a_i$ and $b_j$ are the atomic effects such that $\SC {a_i}{\alpha_i}
= \SC{b_j} {\beta_j} = 1$.  Therefore, we have
\begin{align*}
  F_A = \sum_{i,j} c_{ij} E^\rA_{a_i} \otimes E^\rB_{b_j}= \sum_{i,j} c_{ij}
  S^\rA_{\alpha_i} \otimes S^\rB_{\beta_j} =T_\Psi.
\end{align*}
\qed

\begin{corollary}
  For every bipartite state $\rho \in\Stset_1 (\rA\rB)$, $d_\rA = d_\rB = 2$ one has $\Tr [T_\rho]
  =1$.
\end{corollary}

\Proof For each qubit we have
\begin{equation}
  E_{a_1} =
  \begin{pmatrix} 1 & 0 \\ 0& 0 \end{pmatrix},\ E_{a_2} =
  \begin{pmatrix} 0 & 0 \\ 0& 1 \end{pmatrix}.
\end{equation}
Hence, $E^\rA_{e_\rA}  =  E^\rB_{e_\rB} =   I$, where $I$ is the $2\times 2$ identity matrix.   
By lemma \ref{lem:a corto di nomi},  we then have $F_{e_\rA \otimes e_\rB} =   
I \otimes I$ and, therefore $\Tr [T_\rho] = \Tr [F_{e_\rA \otimes e_\rB}  T_\rho  ]= \SC {e_\rA \otimes e_\rB} {\rho} =1$.
\qed 

Finally, an immediate consequence of local distinguishability is the following:
\begin{lemma}\label{lem:tensoraction}
Suppose that  $\tU \in \grp G_\rA$ and $\tV\in \grp G_\rB$ are two reversible transformations for qubits $\rA$ and $\rB$, respectively, and that $U, V \in \mathbb{SU} (2)$ are such that  
\begin{align*}
S^{\rA}_{\tU \rho}  &=  U S^{\rA}_\rho U^\dag \qquad \forall \rho \in \Stset_1 (\rA)\\
S^{\rB} _{\tV \sigma}  &=  V S^{\rB}_\sigma  V^\dag  \qquad \forall \sigma \in \Stset_1 (\rB).
\end{align*}  
Then, we have $T_{(\tU \otimes \tV)  \tau }   =  (U \otimes V)  T_\tau  (U^\dag \otimes V^\dag)$ for every $\tau \in \Stset_1 (\rA\rB)$.
\end{lemma}
 \Proof  The thesis follows by linearity expanding $\tau$ as $\tau = \sum_{i,j=1}^4  \tau_{ij}  \alpha_i \otimes \beta_j$, where $\{\alpha_i\}_{i=1}^4$  and $\{\beta_j\}_{j =1}^4$ are bases for the $\Stset_\Reals (\rA)$ and $\Stset_\Reals (\rB)$. \qed

The rest of this subsection is aimed at showing that, with a suitable choice of matrix representation for system $\rB$, the
standard representation coincides with the tensor representation, that is, $S_\rho =T_\rho$ for
every $\rho\in\Stset(\rA\rB)$.  This technical result is important because some properties used in
our derivation are easily proved in the standard representation, while the property expressed by lemma \ref{lem:tensoraction} is easily  proved
in the tensor representation: it is then essential to show that we can construct a representation
that enjoys both properties. 

The four states $\{ \varphi_m \otimes \varphi_n\}_{m,n=1}^2$ are clearly a maximal set of perfectly distinguishable pure states in  $\rA\rB$.   In the following we will construct the standard representation starting from this set.  
\begin{lemma}
  For a composite system $\rA\rB$ with $d_\rA=d_\rB=2$ one can choose the standard
  representation in such a way that the following equalities hold 
    \begin{align}
   & S_{\varphi_m\otimes \varphi_n}=T_{\varphi_m\otimes \varphi_n},\label{diag}\\
    &S_{\varphi_m\otimes \sigma_k}=T_{\varphi_m\otimes \sigma_k}, \qquad k = x,y\label{quba},\\
   & S_{\sigma_k\otimes \varphi_m}=T_{\sigma_k\otimes \varphi_m}, \qquad k=x,y\label{qubb}.
  \end{align}
  \label{lem:almost}
\end{lemma}

\Proof   Let us choose single-qubit representations $S^\rA$ and $S^\rB$ that satisfy Eqs. (\ref{matphii}), (\ref{matsx}), and (\ref{eq:itsimportant}).   
On the other hand, choosing tho states $\{\varphi_n \otimes \varphi_n\}$ in lexicographic order as the four distinguishable states for the standard representation, we have  
\begin{align*}
&[S_{\varphi_1 \otimes \varphi_1}]_{rs} =  \delta_{1r} \delta_{1s} \qquad   [S_{\varphi_1 \otimes \varphi_2}]_{rs} =  \delta_{2r} \delta_{2s} \\
&[S_{\varphi_2 \otimes \varphi_1}]_{rs} =  \delta_{3r} \delta_{3s} \qquad   [S_{\varphi_2 \otimes \varphi_2}]_{rs} =  \delta_{4r} \delta_{4s} .\\
\end{align*}    
With this choice, we get $S_{\varphi_m \otimes \varphi_n} = S^\rA_{\varphi_m}  \otimes S^\rB_{\varphi_n}  =  T_{\varphi_m \otimes \varphi_n}$ for every $m,n=1,2$.  This  proves Eq. (\ref{diag}).    
Let us now prove Eqs.~\eqref{quba} and (\ref{qubb}).  Consider the two-dimensional face $F_{11,12}$, generated by the states $\varphi_1 \otimes \varphi_1$ and $\varphi_1\otimes \varphi_2$.  This face is the face identified by the state $\omega_{11,12}: = \varphi_1  \otimes \chi_\rB$, and we have $F_{11,12} \simeq \{\varphi_1\} \otimes  \Stset_1 (\rB)$.  Therefore we can choose the vectors  $\sigma_k^{11,12}$, $k=x,y$  to satisfy the relation  $\sigma_k^{11,12}: =  \varphi_1  \otimes \sigma_k$, $k =x,y$. 
Now, in the standard representation we have 
\begin{align*} 
[S_{\sigma_x}^{11,12}]_{rs}  &=  \delta_{r1}  \delta_{s2}  +  \delta_{r2}  \delta_{s1}\\
[S_{\sigma_y}^{11,12}]_{rs}  &= i \lambda  ( \delta_{r1}  \delta_{s2}  -  \delta_{r2}  \delta_{s1})
\end{align*} 
[cf. Eqs. (\ref{matsx}) and (\ref{eq:itsimportant})].    This implies $S_{\sigma^{11,12}_k}  =  S^\rA_{\varphi_{11}}  \otimes S^\rB_{\sigma_k} = T_{\varphi_{11}\otimes \sigma_k}$, for $k=x,y$.  Repeating the same argument for the face $F_{22,21}$, $F_{11,21}$, and $F_{21,22}$ we obtain the proof of Eqs.   ~\eqref{quba} and (\ref{qubb}).  \qed

In order to prove that, with a suitable choice of axes, the standard representation coincides with the tensor representation---i.e.  
$S_\rho = T_\rho$ for every $\rho \in \Stset (\rA\rB)$---it remains to find a choice of axes such that
$S_{\sigma_k \otimes \sigma_l} = T_{\sigma_k \otimes \sigma_l}$, $k=x,y$. This will be proved in the
following.

\begin{lemma}
Let $\Phi \in \Stset_1 (\rA\rB)$ be a pure state such that   $\SC {a_1 \otimes a_1} \Phi = \SC {a_2 \otimes a_2} \Phi = \frac 1/2$ [such a state exists due to the superposition principle].  
 With a suitable choice of the matrix representation $S^\rB$,  the state $\Phi $ is represented by the matrix  
    \begin{equation}\label{isolem}
    T_{\Phi}=\frac12
    \begin{pmatrix}
      1&0&0&1\\
      0&0&0&0\\
      0&0&0&0\\
      1&0&0&1\\
    \end{pmatrix}.
  \end{equation}
 Moreover, one has
  \begin{equation}\label{phispan}
    \Phi=\chi_\rA\otimes\chi_\rB+\frac14(\sigma_x\otimes\sigma_x-\sigma_y\otimes\sigma_y+\sigma_z\otimes\sigma_z).
  \end{equation}
\end{lemma}

\Proof  

Let us start with the proof of  Eq. (\ref{isolem}).    For every reversible transformation $\tU \in \grp G_\rA$, let $\tU^*\in\grp G_\rB$  be the conjugate of $\tU$, defined with respect to the state $\Phi$.  Since all $2\times2$ unitary (non-trivial) representations of $\mathbb{SU}(2)$ are unitarily equivalent,
by a suitable choice of the standard representation $S^{\rB}_\rho$ for system $\rB$, one has
\begin{align}\label{eq:complexconj}
S^{\rB}_{\tU^*\rho}=U^* S^{\rB}_\rho U^T,
\end{align} 
where $U^*$ and $U^T$ are the complex conjugate and the transpose of the matrix $U\in\mathbb{SU} (2)$ such that $S^{\rA} _{\tU \rho}  = U S_\rho^{\rA}  U^\dag$.   

Due to Eq. (\ref{eq:complexconj}) and to lemma \ref{lem:tensoraction},  the isotropic state $\Phi$ must satisfy the condition
$(U\otimes U^*) T_\Phi (U^\dag\otimes U^T)=T_\Phi, \forall U \in \mathbb{SU} (2)$. Now, the unitary representation $\{U \otimes
U^*\}$ has two irreducible subspaces and the projectors on them are given by the matrices
\begin{align*}
  P_0&=\frac12
  \begin{pmatrix}
    1&0&0&1\\
    0&0&0&0\\
    0&0&0&0\\
    1&0&0&1
  \end{pmatrix}\\
\quad P_1&=\frac12
  \begin{pmatrix}
    1&0&0&-1\\
    0&2&0&0\\
    0&0&2&0\\
    -1&0&0&1
  \end{pmatrix} =  I \otimes I - P_0,
\end{align*}
where $I$ is the $2\times 2$ identity matrix.
The most general form for $T_\Phi$ is then the following
\begin{align*}
  T_\Phi &=  x_0  P_0  +  x_1  P_1 \\
    & =  (x_0 - x_1)   P_0  + x_1  I \otimes I\\ 
  &=\begin{pmatrix}
    \alpha+\beta&0&0&\beta\\
    0&\alpha&0&0\\
    0&0&\alpha&0\\
    \beta&0&0&\alpha+\beta
  \end{pmatrix},
\end{align*}
having defined $\alpha: = x_1 $ and $\beta: =(x_0-x_1)/2$.  

Now, by construction the state $\Phi$ satisfies the condition 
\begin{align*}\B {a_m}_\rA  \K\Phi_{\rA\rB}=\frac12 \K{\varphi_m}_\rB \quad m=1,2.
\end{align*} 
By definition of the tensor representation, the conditional states $\B {a_m}_\rA  \K{\Phi}_{\rA\rB} $  are described by the diagonal blocks of the matrix $T_\Phi$:  
\begin{align}\label{eq:pureblocks}
S^\rB_{\B {a_1}  \K\Phi_{\rA\rB}}  =    \begin{pmatrix}\alpha+\beta&0\\0&\alpha\end{pmatrix} \qquad   S^\rB_{\B {a_2}  \K\Phi_{\rA\rB}}  =\begin{pmatrix}\alpha&0\\0&\alpha+\beta\end{pmatrix}.
\end{align} 
Since the states $\varphi_1$ and $\varphi_2$ are pure, the above matrices must be be rank-one. Moreover, their trace must be equal to $\SC {a_m \otimes e_\rB}  \Phi= 1/2 \SC{e_\rB} {\varphi_m}=\frac12$, $m=1,2$.  Then we have two possibilities.  Either \emph{i)} $\alpha=0$ and $\beta=\frac12$ or
\emph{ii)} $\alpha=-\beta=\frac12$. In the case \emph{i)}, Eq. \ref{isolem} holds. 
In the case
\emph{ii)}, to prove Eq. (\ref{isolem}) we need to change our choice of matrix representation for the qubit $\rB$.
Precisely,  we make the following change: 
\begin{equation}\label{eq:changeminus}
\begin{split}
S^\rB_{\sigma_x} & \mapsto \widetilde S^\rB_{\sigma_x}  =  - S^\rB_{\sigma_x} \\
S^\rB_{\sigma_y} & \mapsto \widetilde S^\rB_{\sigma_y}  =  - S^\rB_{\sigma_y} \\
S^\rB_{\sigma_z} & \mapsto \widetilde S^\rB_{\sigma_z}  =  - S^\rB_{\sigma_z} ,
\end{split}
\end{equation}
where $\sigma_z: = \varphi_1 - \varphi_2$.   Note that the inversion of the axes, sending $\sigma_k $ to $-\sigma_k$ for every $k=x,y,z$ is not an allowed physical transformation,
but this is not a problem here, because Eq. (\ref{eq:changeminus})  is just a new choice of matrix representation, in which the set of states of system $\rB$ is still represented by the Bloch sphere.

More concisely, the change of matrix representation $S^\rB  \mapsto \widetilde S^\rB$ can be expressed as
\begin{equation*}
S^{\rB}_\rho\mapsto
 \widetilde S^{\rB}_\rho:= Y \left[S^{\rB}_\rho \right]^T Y^\dag \qquad Y : =\begin{pmatrix}   0 & -1  \\  1 & 0 \end{pmatrix}.
\end{equation*}
Note that in the new representation $\widetilde S^{\rB}$ the physical transformation $\tU^*$ is still
represented as $\widetilde S^{\rB}_{\tU \rho} = U^* \widetilde S^{\rB}_\rho U^T$: indeed we have
 \begin{align*}
\widetilde S^{\rB}_{\tU^*\rho} &=  Y  \left[  S^{\rB}_{\tU^*  \rho} \right]^T Y^\dag \\
  &= Y  (U^*  S^{\rB}_\rho U^T)^T  Y^\dag\\
  &= Y  (U  \left[S^{\rB}_{\rho}\right]^T  U^\dag) Y^\dag\\
   &= (Y  U Y^\dag)  (Y  \left[S^{\rB}_{\rho}\right]^T Y^\dag)   (Y U^\dag Y^\dag)\\
  &=  U^* (Y\left[S^{\rB}_\rho\right]^T Y^\dag ) U^T \\
  &=U^* \widetilde S^{\rB}_\rho U^T,
\end{align*} 
having used the relations $Y^\dag Y = I$ and  $Y U Y^\dag = U^*$ for every $U \in \mathbb{SU} (2)$. 
Clearly, the change of standard representation $S \to \widetilde S$ for the qubit $\rB$ induces a change of
tensor representation $T \to \widetilde T$, where $\widetilde T $ is the tensor representation defined by $\widetilde T_{\rho
  \otimes \sigma}: = S^{\rA}_\rho \otimes \widetilde S^{\rB}_\sigma$.  With this change of representation, we have
\begin{equation*}
  \widetilde T_\Phi= \frac12
  \begin{pmatrix}
    1&0&0&1\\
    0&0&0&0\\
    0&0&0&0\\
    1&0&0&1
  \end{pmatrix}.
\end{equation*}
This concludes the proof of Eq. (\ref{isolem}).   

Let us now prove Eq. (\ref{phispan}).  Using the fact that
by definition $T_{\rho\otimes\tau}=(S^{\rA}_\rho\otimes S^{\rB}_\tau)$ one can directly verify the relation
\begin{equation*}
T_{\Phi}=S^{\rA}_\chi\otimes
S^{\rB}_\chi+\frac 1 4(S^{\rA}_{\sigma_x}\otimes S^{\rB}_{\sigma_x}-S^{\rA}_{\sigma_y}\otimes
S^{\rB}_{\sigma_y}+S^{\rA}_{\sigma_z}\otimes S^{\rB}_{\sigma_z}).
\end{equation*}   
This is precisely the matrix version of Eq. (\ref{phispan}).  \qed

Note that the choice of $S^\rB$ needed in Eq. (\ref{isolem})  is compatible with the choice of $S^\rB$ needed in lemma \ref{lem:almost}: indeed, to prove compatibility we only have to show that the representation $S_\rB$ used in Eq. (\ref{isolem}) has the property $[S^\rB_{\varphi_m} ]_{rs}= \delta_{mr} \delta_{ms}$, $m=1,2$.  This property is automatically guaranteed by the relation $\B {a_m}_\rA \K\Phi_{\rA\rB}  = 1/2 \K {\varphi_m}$, $m=1,2$  and by  Eq. (\ref{eq:pureblocks}) with $\alpha =0$ and $\beta = 1/2$.

\begin{corollary}
  In the standard representation the state $\Phi \in\Stset_1
  (\rA\rB)$ is represented by the matrix
  \begin{equation}
    S_{\Phi}=\frac12
    \begin{pmatrix}
      1&0&0&e^{i\theta}\\
      0&0&0&0\\
      0&0&0&0\\
      e^{-i\theta}&0&0&1
    \end{pmatrix}
  \end{equation}
  \label{cor:isostand}
\end{corollary}

\Proof  The thesis follows from theorem \ref{theo:sqrts} and lemma \ref{lem:almost}. \qed

We now define the reversible transformations $\tU_{x, \pi}$ and $\tU_{z, \frac{\pi} 2}$ as follows
\begin{equation}\label{eq:unitariegeneranti}
  \begin{array}{llll}
    S_{\tU_{x, \pi}\rho} & =  X S_\rho  X, \qquad   &X := & \begin{pmatrix}   0 & 1 \\  1 & 0\end{pmatrix}\\
    S_{\tU_{z, \frac {\pi} 2}\rho} & =  e^{-i \frac{\pi}4 Z} S_\rho e^{i \frac{\pi}4 Z}  , \qquad &  Z :=& \begin{pmatrix}   1 & 0 \\  0 & -1\end{pmatrix}.
  \end{array}
\end{equation}
Also, we define the states $\Psi, {\Phi_{z, \frac{\pi} 2}}$, and
$\Psi_{z,\frac {\pi } 2}$ as 
\begin{align*}
  \K\Psi  & :=  (\tU_{x,\pi}  \otimes \tI) \K {\Phi}  \\
  \K {\Phi_{z, \frac{\pi} 2} }&:=  (\tU_{z,\frac {\pi}  2}  \otimes \tI) \K {\Phi}  \\
  \K{\Psi_{z, \frac{\pi} 2}} &:= (\tU_{z,\frac {\pi} 2} \otimes \tI)
  \K \Psi
\end{align*}

\begin{lemma}
  The states $\Psi, \Phi_{z, \frac \pi 2} $, and $\Psi_{z, \frac \pi
    2} $ have the following tensor representation
  \begin{align}\label{phipsilem}
      T_{\Psi} =&\frac12
      \begin{pmatrix}
        0&0&0&0\\
        0&1&1&0\\
        0&1&1&0\\
        0&0&0&0\\
      \end{pmatrix},\quad
      T_{\Phi_{z, \frac{\pi} 2}}=&\frac12
      \begin{pmatrix}
        1&0&0&-i\\
        0&0&0&0\\
        0&0&0&0\\
        i&0&0&1\\
      \end{pmatrix}  \nonumber \\ 
      T_{\Psi_{z, \frac{\pi} 2}}=&\frac12
      \begin{pmatrix}
        0&0&0&0\\
        0&1&-i&0\\
        0&i&1&0\\
        0&0&0&0\\
      \end{pmatrix}.
  \end{align}
  Moreover, one has
  \begin{align}\label{phipsispan}
      \Psi  =&\chi_\rA\otimes\chi_\rB+\frac14(\sigma_x\otimes\sigma_x + \sigma_y\otimes\sigma_y-\sigma_z\otimes\sigma_z)\nonumber\\
      \Phi_{z, \frac {\pi} 2}  =&\chi_\rA\otimes\chi_\rB+\frac14(\sigma_y\otimes\sigma_x + \sigma_x\otimes\sigma_y+ \sigma_z\otimes\sigma_z)\nonumber\\
      \Psi_{z, \frac {\pi} 2}  =&\chi_\rA\otimes\chi_\rB+\frac14(\sigma_y\otimes\sigma_x - \sigma_x\otimes\sigma_y-\sigma_z\otimes\sigma_z)
  \end{align}
\end{lemma}

\Proof Eq. (\ref{phipsilem}) is obtained from Eq. (\ref{isolem}) by explicit calculation using lemma \ref{lem:tensoraction} and  Eq. (\ref{eq:unitariegeneranti}). Then, the
validity of Eq. (\ref{phipsispan}) is easily obtained from Eq. (\ref{phispan}) using the relations 
\begin{align*}
\tU_{x,\pi}  \K{\sigma_x}&= \K{\sigma_x}\\
\tU_{x,\pi} \K{\sigma_y}&= -\K{\sigma_y}\\
\tU_{x,\pi}  \K{\sigma_z}&= -\K{\sigma_z}
\end{align*}
and 
\begin{align*}
\tU_{z,\pi/2}  \K{\sigma_x}&= \K{\sigma_y}\\
\tU_{z,\pi/2} \K{\sigma_y}&= -\K{\sigma_x}\\
\tU_{z,\pi/2}  \K{\sigma_z}&= \K{\sigma_z}.
\end{align*}
  \qed

\begin{lemma}
  The states $\Psi, \Phi_{z, \frac{\pi} 2}$, and $\Psi_{z, \frac{\pi} 2}$ have a standard
  representation of the form
  \begin{equation}\label{phipsistand}
    \begin{array}{rl}
    S_{\Psi}=&\frac12
    \begin{pmatrix}
      0&0&0&0\\
      0&1&e^{i\gamma}&0\\
      0&e^{-i\gamma}&1&0\\
      0&0&0&0\\
    \end{pmatrix} \\ \\
   S_{\Phi_{z, \frac {\pi} 2}}=&\frac12
    \begin{pmatrix}
      1&0&0&\lambda ie^{i\theta}\\
      0&0&0&0\\
      0&0&0&0\\
      -\lambda ie^{-i\theta}&0&0&1
    \end{pmatrix} \\ \\
S_{\Psi_{z, \frac{\pi} 2}}=&\frac12
    \begin{pmatrix}
      0&0&0&0\\
      0&1&\mu ie^{i\gamma}&0\\
      0&-\mu ie^{-i\gamma}&1&0\\
      0&0&0&0\\
    \end{pmatrix}.
    \end{array}
  \end{equation}
  with $\theta$ as in corollary \ref{cor:isostand}, $\gamma \in [0, 2\pi)$ and $\lambda,\mu \in \{-1,1\}$.
\end{lemma}

\Proof Let us start from $\Psi$.  First, from Eq.(\ref{phipsispan}) it is immediate to obtain $\SC
{a_1 \otimes a_1} {\Psi} = \SC {a_2 \otimes a_2} { \Psi} =0$ and $\SC {a_1 \otimes a_2} {\Psi} =\SC
{a_2 \otimes a_1} \Psi =1/2$.  This gives the diagonal elements of $S_{\Psi}$.  Then, using theorem
\ref{theo:sqrts} we obtain that $S_\Psi$ must be as in Eq. (\ref{phipsistand}), for some value of
$\gamma$.  Let us now consider $\Phi_{z, \frac{\pi} 2}$. Again, the diagonal elements of the matrix
$S_{\Phi_{z, \frac {\pi } 2}}$ are obtained from Eq. (\ref{phipsispan}), which in this case yields
$\SC {a_1 \otimes a_1} {\Phi_{z, \frac{\pi} 2}} = \SC {a_2 \otimes a_2} {\Phi_{z, \frac{\pi} 2}}
=1/2$ and $\SC {a_1 \otimes a_2} {\Phi_{z, \frac {\pi} 2}} =\SC {a_2 \otimes a_1} {\Phi_{z,
    \frac{\pi} 2}} =0$. Hence, by theorem \ref{theo:sqrts} we must have
\begin{equation*}
   S_{\Phi_{z, \frac {\pi} 2}}=\frac12
    \begin{pmatrix}
      1&0&0& e^{i\lambda}\\
      0&0&0&0\\
      0&0&0&0\\
       e^{-i\lambda}&0&0&1
    \end{pmatrix}
\end{equation*}
for some value of $\lambda\in [0,2\pi)$.  Now, denote by $A$ the effect such that $\SC A \Phi =1$.  We then have 
\begin{align*} 
\SC A  {\Phi_{z, \frac {\pi} 2}} & = \Tr [ E_A  S_{\Phi_{z, \frac {\pi} 2}}]  =\Tr [ S_{\Phi}  S_{\Phi_{z, \frac {\pi} 2}}]\\
\SC A  {\Phi_{z, \frac {\pi} 2}} & = \Tr [ F_A  T_{\Phi_{z, \frac {\pi} 2}}]  =\Tr [ T_{\Phi}  T_{\Phi_{z, \frac {\pi} 2}}] = \frac 1 2,
\end{align*}  
having used theorem \ref{theo:E=S}, corollary \ref{cor:F=T}, and Eq. (\ref{phipsilem}).  Hence, we have $\Tr [ S_{\Phi}
S_{\Phi_{z, \frac {\pi} 2}}] = 1/2$, which implies $\lambda = \theta \pm \frac {\pi }2 $, as in Eq.
(\ref{phipsistand}).  Finally, the same arguments can be used for $\Psi_{z, \frac{\pi }2}$: The
diagonal elements of $S_{\Psi_{z, \frac {\pi} 2}}$ are obtained from the relations $\SC {a_1 \otimes
  a_1} {\Psi_{z, \frac{\pi} 2}} = \SC {a_2 \otimes a_2} {\Psi_{z, \frac{\pi} 2}} =0$ and $\SC {a_1
  \otimes a_2} {\Psi_{z, \frac {\pi} 2}} =\SC {a_2 \otimes a_1} {\Psi_{z, \frac{\pi} 2}} =1/2$,
which follow from Eq.  (\ref{phipsispan}).  This implies that the matrix $S_{\Psi_{z, \frac{\pi}
    2}}$ has the form
\begin{equation*}
  S_{\Psi_{z, \frac{\pi} 2}}=\frac12
  \begin{pmatrix}
    0&0&0&0\\
    0&1& e^{i\mu}&0\\
    0& e^{-i\mu}&1&0\\
    0&0&0&0\\
  \end{pmatrix},
\end{equation*}
for some $\mu \in [0,2\pi)$.  The relation $\Tr[S_{\Psi} S_{\Psi_{z, \frac {\pi} 2}} ] =\Tr[ T_{\Psi} T_{\Psi_{z,
    \frac {\pi} 2}}] =1/2$ then implies $\mu = \gamma \pm \frac {\pi} 2$. \qed

Let us now consider the four vectors $\Sigma_{x}^{(11,22)}, \Sigma_y^{(11,22)} ,\Sigma_{x}^{(12,21)},
\Sigma_y^{(12,21)}$ defined as follows
\begin{align}
  \Sigma_{x}^{(11,22)} & =  2\left( \Phi -  \chi_\rA \otimes \chi_\rB  - \frac 1 4  \sigma_z\otimes \sigma_z\right) \nonumber\\
  \Sigma_{y}^{(11,22)} & = 2 \left( \Phi_{z, \frac {\pi}  2} -  \chi_\rA \otimes \chi_\rB  - \frac 1 4  \sigma_z\otimes \sigma_z\right) \nonumber\\
  \Sigma_{x}^{(12,21)} & =  2\left(\Psi -  \chi_\rA \otimes \chi_\rB  + \frac 1 4  \sigma_z\otimes \sigma_z \right)\nonumber\\
  \Sigma_{x}^{(12,21)} & = 2 \left(\Psi_{z, \frac {\pi} 2} - \chi_\rA
    \otimes \chi_\rB + \frac 1 4 \sigma_z\otimes \sigma_z\right) .
\label{defSigma}
\end{align}
By the previous results, it is immediate to obtain the matrix representations of these vectors.  In
the tensor representation, using Eqs. (\ref{isolem}) and (\ref{phipsilem}) we obtain
\begin{align*}
&T_{\Sigma_{x}^{(11,22)}} =\begin{pmatrix}
      0&0&0&1\\
      0&0& 0&0\\
      0&  0& 0&0\\
      1&0&0&0\\
    \end{pmatrix},\quad
T_{\Sigma_{y}^{(11,22)}} =\begin{pmatrix}
      0&0&0&-i\\
      0&0& 0&0\\
      0&  0& 0&0\\
      i&0&0&0\\
    \end{pmatrix},\\
&T_{\Sigma_{x}^{(12,21)}} =\begin{pmatrix}
      0&0&0&0\\
      0&0& 1&0\\
      0& 1& 0&0\\
      0&0&0&0\\
    \end{pmatrix},\quad
T_{\Sigma_{y}^{(12,21)}} =\begin{pmatrix}
      0&0&0&0\\
      0&0& -i&0\\
      0& i& 0&0\\
      0&0&0&0\\
    \end{pmatrix},
\end{align*}
while in the standard representation, using Eqs. (\ref{cor:isostand})  and (\ref{phipsistand}),  we obtain  
\begin{align*}
S_{\Sigma_{x}^{(11,22)}} & =\begin{pmatrix}
      0&0&0&e^{i\theta}\\
      0&0& 0&0\\
      0&  0& 0&0\\
      e^{-i \theta}&0&0&0\\
    \end{pmatrix}\\
S_{\Sigma_{y}^{(11,22)}} & =\begin{pmatrix}
      0&0&0&-\lambda i e^{i \theta} \\
      0&0& 0&0\\
      0&  0& 0&0\\
      \lambda i e^{-i \theta}&0&0&0\\
    \end{pmatrix}\\
S_{\Sigma_{x}^{(12,21)}} &=\begin{pmatrix}
      0&0&0&0\\
      0&0& e^{i \gamma}&0\\
      0& e^{-i \gamma}& 0&0\\
      0&0&0&0\\
    \end{pmatrix}\\
S_{\Sigma_{x}^{(11,22)}} &=\begin{pmatrix}
      0&0&0&0\\
      0&0& -\mu i e^{i\gamma}&0\\
      0& \mu i e^{-\gamma}& 0&0\\
      0&0&0&0\\
    \end{pmatrix},
\end{align*}
Comparing the two matrix representations we are now in position to prove the desired result:
\begin{lemma}\label{lem:almostalmost} 
  With a suitable choice of axes, one has $S_{\sigma_k \otimes \sigma_l} = T_{\sigma_k \otimes
    \sigma_l}$ for every $k,l = x,y$.
\end{lemma}

\Proof For the face $(11,22)$, using the freedom coming from
Eqs.~\eqref{eq:itsimportant} and \eqref{eq:rotax}, we redefine the $x$
and $y$ axes so that $\sigma^{(11,22)}_x := \Sigma^{(11,22)}_x$ and
$\lambda \sigma^{(11,22)}_y := \Sigma^{(11,22)}_y$. In this way we have
\begin{equation*}
  S_{\Sigma_k^{(11,22)}} = T_{\Sigma_k^{(11,22)}} \qquad \forall k = x,y
\end{equation*} 
Likewise, for the face $(12,21)$ we redefine the $x$ and $y$ axes so that $\sigma^{(12,21)}_x :=
\Sigma^{(12,21)}_x$ and $\mu\sigma^{(12,21)}_y := \Sigma^{(12,21)}_y$, so that we have
\begin{equation*}
  S_{\Sigma_k^{(12,21)}} = T_{\Sigma_k^{(12,21)}} \qquad \forall k = x,y.
\end{equation*} 
Finally, using Eqs. (\ref{phispan}), (\ref{phipsispan}), and (\ref{defSigma}) we have the relations
\begin{align*}
  \sigma_x \otimes \sigma_x &=  \Sigma^{(11,22)}_x  +  \Sigma_x^{(12,21)}\\
  \sigma_y \otimes \sigma_y &=  \Sigma^{(11,22)}_x  - \Sigma_x^{(12,21)}\\
  \sigma_x \otimes \sigma_y &=  \Sigma^{(11,22)}_y  -  \Sigma_y^{(12,21)}\\
  \sigma_y \otimes \sigma_x &= \Sigma^{(11,22)}_y +
  \Sigma_y^{(12,21)}.
\end{align*}
Since $S$ and $T$ coincide on the right-hand side of each equality, they must also coincide on the
left-hand side.\qed

\begin{theorem}
  With a suitable choice of axes, the standard representation coincides with the
  tensor representation, that is, $S_\rho = T_\rho$ for every $\rho \in\Stset (\rA\rB)$.
  \label{theo:standeqtens}
\end{theorem}
\Proof Combining lemma \ref{lem:almost} with lemma \ref{lem:almostalmost} we obtain that $S$ and $T$
coincide on the tensor products basis $\mathcal B\times \mathcal B$, where $\mathcal B =
\{\varphi_1, \varphi_2, \sigma_x, \sigma_y\}$.  By linearity, $S$ and $T$ coincide on every state.
\qed

From now on, whenever we will consider a composite system $\rA\rB$ where $\rA$ and $\rB$ are
two-dimensional we will adopt the choice that guarantees that the standard
representation coincides with the tensor representation.

\subsection{Positivity of the matrices}
In this paragraph we show that the states in our theory can be represented by positive matrices.
This amounts to prove that for every system $\rA$, the set of states $\Stset_1(\rA)$ can be
represented as a subset of the set of density matrices in dimension $d_\rA$.  This result will be
completed in subsection \ref{sub:all}, where we will see that, in fact, every density matrix in
dimension $d_\rA$ corresponds to some state of $\Stset_1 (\rA)$.

The starting point to prove positivity is the following:
\begin{lemma}
  Let $\rA$ and $\rB$ be two-dimensional systems. Then, for every pure
  state $\Psi \in \Stset(\rA\rB)$ one has $S_\Psi\ge 0$.
  \label{lem:posfour}
\end{lemma}

\Proof Take an arbitrary vector $V \in{\mathbb C}^2 \otimes \Cmplx^2$, written in the Schmidt form
as $|V\> = \sum_{n=1}^2 \sqrt{\lambda_n} |v_n\>|w_n\>$.  Introducing the unitaries $U,V$ such that
$U|v_n\>=|n\>$ and $V|w_n\>=|n\>$ for every $n=1,2$ then we have $|V\> = (U^\dag \otimes V^\dag)
|W\>$, where $|W\> = \sum_{n=1}^2 \sqrt{\lambda_n} |n\>|n\>$.  Therefore, we have
\begin{align*}
\<  V|  S_{\Psi} |V\>    & = \<W |   S_{(\tU\otimes \tV) \Psi}  |W\>  \\
\end{align*}
where $\tU$ and $\tV$ are the reversible transformations defined by $S_{\tU \rho} = U S_\rho U^\dag
$ and $S_{\tV \rho} = V S_\rho V^\dag $, respectively ($\tU$ and $\tV$ are physical transformations by virtue of corollary \ref{cor:unitaryconjugation}).  Here we used the fact that the standard
two-qubit representation coincides with the tensor representation and, therefore, $S_{(\tU \otimes \tV)  \Psi}  =  (U \otimes V)  S_\Psi   (U \otimes V)^\dag$.  Denoting the pure state $(\tU
\otimes \tV) \K {\Psi}$ by $\K{\Psi'}$ we then have
\begin{align*}
\<  V|  S_{\Psi} |V\>  = &   \lambda_1   \left[  S_{\Psi'} \right]_{11,11}   +    \lambda_2   \left[  S_{\Psi'} \right]_{22,22}  \\
&+  2 \sqrt{\lambda_1 \lambda_2}  \mathsf{Re} \left([S_{\Psi'}]_{11,22}  \right)\\
\end{align*}
Since by theorem \ref{theo:sqrts} we have
$[S_{\Psi'}]_{11,22}=\sqrt{[S_{\Psi'}]_{11,11}[S_{\Psi'}]_{22,22}}e^{i\theta}$, we conclude 
\begin{equation*}
  \begin{split}
\<  V|  S_{\Psi} |V\>   = &   \lambda_1   \left[  S_{\Psi'} \right]_{11,11}   +    \lambda_2   \left[  S_{ \Psi'} \right]_{22,22}+  \\
&+  2 \cos \theta   \sqrt{\lambda_1 \lambda_2   [S_{\Psi'}]_{11,11}[S_{\Psi'}]_{22,22}  }  \\
  \ge &  \left (\sqrt{\lambda_1 [S_{\Psi'}]_{11,11}}  - \sqrt{\lambda_2 [S_{\Psi'}]_{22,22}} \right)^2  \ge 0.
  \end{split}
\end{equation*}
Since the vector $V\in \Cmplx^2 \otimes \Cmplx^2$ is arbitrary,  the matrix $S_\Psi$ is positive. \qed

\begin{corollary}\label{cor:posfour1}
Let $\rC$ be a system of dimension $d_\rC = 4$.  Then, with a suitable choice of matrix representation the pure states of $\rC$ are represented by positive matrices.
\end{corollary}
\Proof The system $\rC$ is operationally equivalent to the composite system $\rA\rB$, where $d_\rA = d_\rB =2$.  Let $\tU \in \Trnset(\rA\rB, \rC)$ be the reversible transformation implementing the equivalence. Now, we know that the states of $\rA\rB$ are represented by positive matrices.  If we define the basis vectors for $\rC$ by applying $\tU$ to the basis for $\rA\rB$, then we obtain that the states of $\rC$ are represented by the same matrices representing the states of $\rA\rB$. \qed

\begin{corollary}
  Let $\rA$ be a system with $d_\rA = 3$. With a suitable choice of matrix representation, the matrix
  $S_\varphi$ is positive for every pure state $\varphi\in\Stset(\rA)$.  \label{cor:posthree}
\end{corollary}

\Proof Let $\rC$ be a system with $d_\rC = 4$. By corollary
\ref{cor:posfour1} the states of $\rC$ are represented by positive
matrices.  Define the state $\omega: = \frac 1 3 (\varphi_{1} +
\varphi_{2} + \varphi_{3} )$, where $\{\varphi_m\}_{m=1}^4$ are four
perfectly distinguishable pure states.  By the compression axiom, the
face $F_\omega$ can be encoded in a three-dimensional system $\rD$ (corollary \ref{cor:equidim}). In
fact, since $\rD$ is operationally equivalent to $\rA$, the face
$F_\omega$ can be encoded in $\rA$.  Let $\tE \in \Trnset (\rD,\rA)$
and $\tD \in\Trnset( \rA,\rD)$ be the encoding and decoding operation,
respectively.  If we define the basis vectors for $\rA$ by applying
$\tE$ to the basis vectors for the face $F_{\omega}$,
then we obtain that the states of $\rA$ are represented by the same
matrices representing the states in the face $F_\omega$. Since these
matrices are positive, the thesis follows. \qed

From now on, for every three-dimensional system $\rA$ we will choose
the $x$ and $y$ axes so that $S_\rho$ is positive for every $\rho \in
\Stset (\rA)$.
\begin{corollary}\label{cor:rankone}  
  Let $\varphi \in\Stset_1 (\rA)$ be a pure state with $d_\rA=3$. Then, the corresponding matrix
  $S_\varphi$, given by
\begin{equation}\label{rankone}
S_\varphi = \begin{pmatrix} 
p_1 & \sqrt{p_1 p_2}  e^{i\theta_{12}}  & \sqrt{p_1 p_3}  e^{i \theta_{13}}\\
\sqrt{p_1 p_2}  e^{-i\theta_{12}} & p_2 & \sqrt{p_2 p_3} e^{i\theta_{23}} \\
\sqrt{p_1 p_3}  e^{-i\theta_{13}}  &  \sqrt{p_2 p_3} e^{-i\theta_{23}} & p_3
\end{pmatrix}.
\end{equation}
satisfies the property 
\begin{align*}
e^{i \theta_{13}} =e^{i( \theta_{12} + \theta_{23})}.
\end{align*} 
Equivalently, $S_\varphi = |v\>\<v|$,
where $v \in \Cmplx^3$ is the vector given by $|v\> : = (\sqrt{p_1} , \sqrt{p_2} e^{-i \theta_{12}}
, \sqrt{p_3} e^{-i\theta_{13}} )^T$.
\end{corollary}
\Proof The relation can be trivially satisfied when $p_i = 0$ for some $i \in\{1,2,3\}$.  Hence, let
us assume $p_1, p_2,p_3>0$.  Computing the determinant of $S_{\varphi}$ one obtains $\det
(S_\varphi) = 2 p_1 p_2 p_3 [\cos (\theta_{12} + \theta_{23} - \theta_{13}) -1]$.  Since $S_\varphi$
is positive, we must have $\det (S_\varphi) \ge 0$.  If $p_1, p_2, p_3 > 0$ the only possibility is
$\theta_{13} = \theta_{12} + \theta_{23} \mod 2 \pi$. \qed

Corollary \ref{cor:rankone} can be easily extended to systems of arbitrary dimension. To this
purpose, we choose the $x-$ and $y-$axes in such a way that the projection of every state $\rho
\in\Stset_1 (\rA)$ on a three-dimensional face is represented by a positive matrix.
\begin{lemma}
  If $\varphi \in\Stset_1 (\rA)$ is a pure state and $d_\rA =N$, then
  $S_\varphi = |v\>\<v|$, where $v \in\Cmplx^N$ is the vector given by
  $v:= (\sqrt{p_1}, \sqrt{p_2} e^{-i \alpha_{2}} , \dots,
  \sqrt{p_{N}} e^{-i\alpha_{N}})^T$ with $\alpha_i \in [0,2\pi) \quad \forall i =2, \dots,N$.
  \label{lem:rankoneN}
\end{lemma}  
 
\Proof Consider a triple $V = \{p,q,r\} \subseteq\{1,\dots,N\}$. Then the state $\Pi_V|\varphi)$ is
proportional to a pure state of a three dimensional system, whose representation $S_{\Pi_V\varphi}$
is the $3\times3$ square sub-matrix of $S_\varphi$ with elements $[S_{\varphi}]_{kl}=\sqrt{p_k
  p_k}e^{i\theta_{kl}}$, $(k,l)\in V\times V$.  Now, corollary \ref{cor:rankone} forces the relation
$e^{i\theta_{pr}}=e^{i(\theta_{pq} + \theta_{qr})}$.  Since this relation must hold for every choice
of the triple $V=\{p,q,r\}$, if we define $\alpha_p:=\theta_{p1}$, then we have
$e^{i\theta_{pq}}=e^{i(\theta_{p1}+\theta_{1q})}=e^{i(\theta_{p1}-\theta_{q1})}=e^{i(\alpha_p-\alpha_q)}$.
It is then immediate to verify that $S_{\varphi}=|v\>\<v|$, where $v=(\sqrt{p_1}, \sqrt{p_2} e^{-i
  \alpha_{2}} , \dots, \sqrt{p_{N}} e^{-i\alpha_{N}})^T$.  \qed

In conclusion, we proved the following
\begin{corollary}
  For every system $\rA$, the state space $\Stset_1 (\rA)$ can be represented as a subset of the set
  of density matrices in dimension $d_\rA$.
\end{corollary}  
\Proof For every state $\rho \in\Stset_1 (\rA)$ the matrix $S_\rho$ is Hermitian by construction,
with unit trace by corollary \ref{cor:unittrace}, and positive since it is a convex mixture of
positive matrices. \qed
  
\subsection{Quantum theory in finite dimensions}\label{sub:all}
Here we conclude our derivation of quantum theory by showing that every density matrix in dimension
$d_\rA$ corresponds to some state $\rho \in\Stset_1 (\rA)$.

We already know from the superposition principle (lemma
\ref{lem:allprobN}) that for every choice probabilities
$\{p_i\}_{i=1}^{d_\rA}$ there is a pure state $\varphi\in\Stset_1
(\rA)$ such that $\{p_i\}_{i=1}^{d_\rA}$ are the diagonal elements of
$S_\varphi$. Thus, the set of density matrices corresponding to pure
states contains at least one matrix of the form $S_\varphi =
|v\>\<v|$, with $|v\> = (\sqrt {p_1}, \sqrt{p_2} e^{-i \beta_{2}} ,
\dots, \sqrt{p_{d_\rA}} e^{-i \beta_{d_\rA}})$. It only remains to prove
that every possible choice of phases $\beta_{i}\in[0,2\pi)$
corresponds to some pure state.

Recall that for a face $F \subseteq \Stset_1 (\rA)$ we defined the group $\grp G_{F, F^\perp}$ to be
the group of reversible transformations $\tU \in\grp G_\rA$ such that $\tU = _{\omega_F} \tI_\rA$
and $\tU = _{\omega_F^\perp} \tI_\rA$.  We then have the following
\begin{lemma}\label{lem:blockfase}
  Consider a system $\rA$ with $d_\rA = N$. Let
  $\{\varphi_i\}_{i=1}^{N} \subset \Stset_1 (\rA)$ be a maximal set of
  perfectly distinguishable pure states, $F$ be the face identified by
  $\omega_F = 1/(N-1) \sum_{i=1}^{N-1} \varphi_i$ and $F^\perp$ its
  orthogonal face, identified by the state $\varphi_N$.  If $\tU$ is a
  reversible transformation in $\grp G_{F, F^\perp}$, then the action
  of $\tU$ is given by
  \begin{equation}
    S_{\tU \rho}  = U S_\rho U^\dag \qquad U = 
    \left( \begin{array}{ccc|c} 
      &&  & 0\\
       & I_{N-1} & & \vdots  \\
       && & 0\\
        \hline 
        0 &\dots& 0  &  e^{-i\beta}
      \end{array} \right)
    \label{blockfase}
  \end{equation}
  where $I_{N-1}$ is the $(N-1) \times (N-1)$ identity
  matrix and $\beta \in [0,2\pi)$.
\end{lemma}

\Proof Consider an arbitrary state $\rho\in\Stset_1(\rA)$ and its matrix representation
\begin{equation*}\label{effemat}
  S_\rho=\left(
    \begin{array}{c|c}
      S_{\Pi_F\rho}&{\bf f}\\
      \hline {\bf f}^\dag&S_{\Pi_F^\perp\rho}
    \end{array}\right),
\end{equation*}
where ${\bf f} \in \Cmplx^{N-1}$ is a suitable vector.  Since $\tU=_{\omega_F}\tI_\rA$ and
$\tU=_{\omega_F^\perp}\tI_\rA$, we have that
\begin{equation*}\label{gmat}
  S_{\tU\rho}=\left(
    \begin{array}{c|c}
      S_{\Pi_F\rho}&{\bf g}\\
      \hline {\bf g}^\dag&S_{\Pi_F^\perp\rho}
    \end{array}\right),
\end{equation*}
where ${\bf g} \in\Cmplx^{N-1}$ is a suitable vector.  To prove Eq.
(\ref{blockfase}), we will now prove that ${\bf g} = e^{i\beta} {\bf
  f}$ for some suitable $\beta\in [0,2\pi)$.

Let us start from the case $N =3$.  Since $\tU \K {\varphi_i }= \K{\varphi_i} \quad \forall
i=1,2,3$, we have $\B {a_i} \tU = \B {a_i} \quad \forall i= 1,2,3$ (lemma  \ref{lem:aA=a}).  This implies that $\tU$ sends
states in the face $F_{13}$ to states in the face $F_{13}$: indeed,  for every $\rho \in F_{13}$  one has $\B{a_{13}} \tU \K\rho =
\SC{a_{13}} \rho = 1$, which implies $\tU\rho  \in F_{13}$  (lemma \ref{lem:identifytest}).  In other words, the restriction of $\tU$ to the
face $F_{13}$ is a reversible qubit transformation.  Therefore, the action of $\tU$ on a state $\rho
\in F_{13}$ must be given by
\begin{equation*}
S_{\tU \rho}   = \begin{pmatrix} 
\rho_{11}  & 0  & \rho_{13}  e^{i\beta} \\
0 &               0 &  0 \\
\rho_{31} e^{-i \beta}  & 0  & \rho_{33},  
\end{pmatrix} 
\end{equation*}
for some $\beta \in [0,2\pi)$.  Similarly, we can see that $\tU$ sends states in the face $F_{23}$
to states in the face $F_{23}$. Hence, for every $\sigma \in F_{23}$ we have
\begin{align*}
S_{\tU \sigma}   = \begin{pmatrix} 
0  &  0  & 0  \\
0 &   \sigma_{22}             &  \sigma_{23}  e^{i\beta'} \\
0  & \sigma_{32}  e^{-i\beta'}    & \rho_{33} 
\end{pmatrix} 
\end{align*}
for some $\beta'\in[0,2 \pi )$.  We now show that $e^{i\beta'} =e^{i \beta}$.  To see that, consider
a generic state $\varphi \in \Stset_1 (\rA)$, with the property that $p_i =\SC {a_i} \varphi >0$ for
every $i = 1,2,3$ (such state exists due to the superposition principle of theorem \ref{theo:superpositiongeneral}).  Writing $S_{\varphi}$ as in Eq. (\ref{rankone}) we then have
\begin{align*}
 & S_{\tU \varphi}  =\\
  & =    \begin{pmatrix} 
    p_1 & \sqrt{p_1 p_2}  e^{i\theta_{12}}  & \sqrt{p_1 p_3}  e^{i  (\theta_{13} + \beta)}\\
    \sqrt{p_1 p_2}  e^{-i\theta_{12}} & p_2 & \sqrt{p_2 p_3} e^{i (\theta_{23} + \beta')} \\
    \sqrt{p_1 p_3} e^{-i (\theta_{13} + \beta)} & \sqrt{p_2 p_3}
    e^{-i (\theta_{23} + \beta')} & p_3
  \end{pmatrix}
\end{align*}
Now, since $\varphi$ and $\tU \varphi$ are pure states, by corollary \ref{cor:rankone} we must have
\begin{align*}
e^{i \theta_{13}} &= e^{i (\theta_{12} + \theta_{23})}\\
e^{i (\theta_{13} +\beta)} &= e^{i
  (\theta_{12} +\theta_{23} + \beta')}.
\end{align*} 
By comparison we obtain $e^{i\beta} = e^{i \beta'}$. This proves Eq.  (\ref{blockfase}) for $N=3$.
The proof for $N >3$ is then immediate: for every three-dimensional face $F_{pq N }$ the action of
$\tU$ is given Eq. (\ref{blockfase}) for some $\beta_{pq}$.  However, since the two faces $F_{pqN}$
and $F_{pq'N}$ overlap on $\varphi_p$ we must have $\beta_{pq} = \beta_{pq'}$. Similarly $\beta_{pq}
= \beta_{p'q}$. We conclude that $\beta_{pq} =\beta$ for every $p,q$.  This proves Eq.
(\ref{blockfase}) in the general case. \qed

We now show that every possible phase shift in Eq. (\ref{blockfase}) corresponds to a physical transformation:

\begin{lemma} A transformation $\tU$ of the form of Eq.  (\ref{blockfase}) is a reversible
  transformation for every $\beta\in [0,2\pi)$.
\end{lemma}
\Proof By lemma \ref{lem:blockfase}, the group $\grp G_{F, F^\perp}$ is a subgroup of $U(1)$.  Now,
there are two possibilities: either $\grp G_{F, F\perp}$ is a (finite) cyclic group or $\grp
G_{F,F^\perp}$ coincides with $U(1) $. However, we know from theorem \ref{theo:circ} that $\grp
G_{F, F^\perp}$ has a continuum of elements. Hence, $\grp G_{F, F^\perp} \simeq U(1)$ and $\beta$
can take every value in $[0,2\pi)$.  \qed
 
An obvious corollary of the previous lemmas is the following
    
\begin{corollary}
  The transformation $\tU_{{\boldsymbol\beta}}$ defined by
  \begin{equation}\label{tuttelefasi3}
    S_{\tU_{\boldsymbol\beta} \rho} =  U S_\rho  U^\dag ,
  \end{equation}
  where $U$ is the diagonal matrix with diagonal elements $(1,e^{i\beta_1},\dots,e^{i\beta_{N-1}})$
  is a reversible transformation for every vector $\boldsymbol{\beta}:= (\beta_2, \dots, \beta_N)\in
  [0,2\pi) \times \dots \times [0,2\pi)$.
\end{corollary}

This leads directly to the conclusion of our derivation:
\begin{theorem}\label{theo:Iseethelight}
 For every system $\rA$, the state space $\Stset_1 (\rA)$ is the set of all density matrices on the Hilbert space  $\Cmplx^{d_\rA}$. 
 \end{theorem}

\Proof Let $N = d_\rA$.  For every choice of probabilities ${\mathbf p} = (p_1, \dots, p_N)$ there
exists at least one pure state $\varphi_{\mathbf p}$ such that $p_k = \SC {a_k}{\varphi_{\mathbf
    p}}$ for every $k=1, \dots, N$(lemma \ref{lem:allprobN} ).  This state is represented by the
matrix $S_{\varphi_{\mathbf p}} = |v_{\mathbf p}\>\<v_{\mathbf p}|$ with $|v_{\mathbf p}\> =
(\sqrt{p_1}, \sqrt{p_2} e^{-i \alpha_2},\dots, \sqrt{p_{N} }e^{-i \alpha_N} )^T$ (lemma
\ref{lem:rankoneN}).  Finally, we can transform $\varphi_{\mathbf p}$ with every reversible
transformation $\tU_{\boldsymbol{\beta}}$ defined in Eq.  (\ref{tuttelefasi3}), thus obtaining
$S_{\tU_{\boldsymbol\beta} \varphi_{\mathbf p}} = U_{\boldsymbol{\beta}} |v_{\mathbf
  p}\>\<v_{\mathbf p}| U_{\boldsymbol{\beta}}^\dag$ where $U_{\boldsymbol{\beta}} | v_{\mathbf p} \>
=(\sqrt{p_1}, \sqrt{p_2} e^{-i (\alpha_2+\beta_2)}, \dots \sqrt{p_N} e^{-i (\alpha_N +
  \beta_N)})^T$.  Since $\mathbf p$ and $\boldsymbol{\beta}$ are arbitrary, this means that every
rank-one density matrix corresponds to some pure state.  Taking the possible convex mixtures we obtain that
every $N\times N$ density matrix corresponds to some state of system $\rA$. \qed

Choosing a suitable representation $\rho \mapsto S_\rho$, we proved that for every system $\rA$ the
set of normalized states $\Stset_1 (\rA)$ is the whole set of density matrices in dimension $d_\rA$.
Thanks to the purification postulate, this is enough to prove that all the effects $\Cntset (\rA)$
and all the transformations $\Trnset(\rA,\rB)$ allowed in our theory are exactly the effects and the
transformations allowed in quantum theory. Precisely we have the following

\begin{corollary}
  For every couple of systems $\rA$ and $\rB$ the set of physical transformations $\Trnset (\rA,
  \rB)$ coincides with the set of all completely positive trace-non increasing maps from $M_{d_\rA}(\Cmplx)$ to $M_{d_\rB}(\Cmplx)$.
 \end{corollary}
 \Proof We proved that our theory has the same normalized states of quantum theory.  On the other
 hand, quantum theory is a theory with purification and in quantum theory the possible physical transformations are quantum operations, i.e. completely positive trace-preserving maps.   The thesis then follows from the fact that two
 theories with purification that have the same set of normalized states are necessarily the same
 (theorem \ref{theo:statesspecify}). \qed

\section{Conclusion}\label{sec:conclusion}
  
Quantum theory can be derived from purely informational principles.  In particular, it belongs to a
broad class of theories of information-processing that includes classical and quantum information
theory as  special cases.  Within this class, quantum theory is identified uniquely by the
purification postulate, stating that the ignorance about a part is always compatible with the
maximal knowledge of the whole in an essentially unique way.  This postulate appears as the origin
of the key features of quantum information
processing, such as no-cloning, teleportation, and error correction (see also Ref.
\cite{purification}). 
The general vision underlying the present work is that the main
primitives of quantum information processing should be derived
directly from the principles, without the abstract mathematics of
Hilbert spaces, in order to make the revolutionary aspects of quantum
information immediately accessible and to place them in the broader
context of the fundamental laws of physics.


Finally, we would like to comment on possible generalizations of our work.  As in any axiomatic
construction, one can ask how the results change when the principles are modified.  For example, one
may be interested in relaxing the local distinguishability axiom and in considering theories, like
quantum theory on real Hilbert spaces, where global measurements are essential to characterize the
state of a composite system.  In this direction, the results of Ref. \cite{purification} suggest
that also quantum theory on real Hilbert spaces can be derived from the purification principle,
after that the local distinguishability requirement has been suitably relaxed. A possible way to
weaken the local distinguishability requirement is to assume only the property of \emph{local
  distinguishability from pure states} proposed in Ref. \cite{purification}: this property states
that the probability of distinguishing two states by local measurements is larger than $1/2$
whenever one of the two states is pure.  A different way to relax local distinguishability would be
to assume the property of 2-\emph{local tomography} proposed in Ref. \cite{hw}, which requires that the
state of a multipartite system can be completely characterized using only measurements on bipartite
subsystems. This property is equivalent to 2-\emph{local distinguishability}, defined as the requirement that two different states of a multipartite system can be distinguished
with probability of success larger than $1/2$ using only local measurements or measurements on bipartite
subsystems.

A more radical generalization of our work would be to relax the assumption of causality.  This would
be particularly important for the discussion of quantum gravity scenarios, where the causal
structure is not given a priori but is part of the dynamical variables of the theory.  In this
respect, the contribution of our work is twofold.  First, it makes evident how fundamental is the
assumption of causality in the ordinary formulation of quantum theory: the whole formalism of
quantum states as density matrices with unit trace, quantum measurements as resolutions of the
identity, and quantum channels as trace-preserving maps is crucially based on it.  Technically
speaking, the fact that the normalization of a state is given by a single linear functional (the
trace, in quantum theory) is the signature of causality.  This partly explains the troubles and
paradoxes encountered when trying to combine the formalism of density matrices with non-causal
evolutions, as in Deutsch's model for close timelike curves \cite{deutsch,bennett}.  Moreover, given
that the usual notion of normalization has to be abandoned in the non-causal scenario, and that the
ordinary quantum formalism becomes inadequate, one may ask in what sense a theory of quantum gravity
would be ``quantum".  The suggestion coming from our work is that a ``quantum" theory is a theory
satisfying the purification principle, which can be suitably formulated even in the absence of
causality \cite{inprep}. The discussion of theories with purification in the non-causal scenario is
an exciting avenue of future research.

\acknowledgments 
The authors are grateful to S Flammia for adding to the package qcircuit the commands needed for the diagrammatic representation of states and effects. 

Research at Perimeter Institute for Theoretical Physics is supported in part by the Government of Canada through NSERC and by the Province of Ontario through MRI.  
Research at QUIT is supported by the University of Pavia and by INFN.  Perimeter Institute and INFN supported the mutual exchange of scientific visits that made this work possible.  


\end{document}

%% file: myQcircuit.tex
\usepackage[matrix,frame,arrow]{xy}
\usepackage{amsmath}

\newcommand{\qw}[1][-1]{\ar @{-} [0,#1]}
\newcommand{\qwx}[1][-1]{\ar @{-} [#1,0]}

\newcommand{\gate}[1]{*{\xy *+<.6em>{#1};p\save+LU;+RU **\dir{-}\restore\save+RU;+RD **\dir{-}\restore\save+RD;+LD **\dir{-}\restore\POS+LD;+LU **\dir{-}\endxy} \qw}

\newcommand{\measureD}[1]{*{\xy*+=+<.5em>{\vphantom{\rule{0em}{.1em}#1}}*\cir{r_l};p\save*!R{#1} \restore\save+UC;+UC-<.5em,0em>*!R{\hphantom{#1}}+L **\dir{-} \restore\save+DC;+DC-<.5em,0em>*!R{\hphantom{#1}}+L **\dir{-} \restore\POS+UC-<.5em,0em>*!R{\hphantom{#1}}+L;+DC-<.5em,0em>*!R{\hphantom{#1}}+L **\dir{-} \endxy} \qw}

\newcommand{\multimeasureD}[2]{*+<1em,.9em>{\hphantom{#2}}\save[0,0].[#1,0];p\save !C *{#2},p+LU+<0em,0em>;+RU+<-.8em,0em> **\dir{-}\restore\save +LD;+LU **\dir{-}\restore\save +LD;+RD-<.8em,0em> **\dir{-} \restore\save +RD+<0em,.8em>;+RU-<0em,.8em> **\dir{-} \restore \POS !UR*!UR{\cir<.9em>{r_d}};!DR*!DR{\cir<.9em>{d_l}}\restore \qw}
\newcommand{\control}{*!<0em,.025em>-=-{\bullet}}
\newcommand{\controlo}{*-<.21em,.21em>{\xy *=<.59em>!<0em,-.02em>[o][F]{}\POS!C\endxy}}

\newcommand{\multigate}[2]{*+<1em,.9em>{\hphantom{#2}} \qw \POS[0,0].[#1,0];p !C *{#2},p \save+LU;+RU **\dir{-}\restore\save+RU;+RD **\dir{-}\restore\save+RD;+LD **\dir{-}\restore\save+LD;+LU **\dir{-}\restore}
\newcommand{\ghost}[1]{*+<1em,.9em>{\hphantom{#1}} \qw}
\newcommand{\Qcircuit}[1][0em]{\xymatrix @*[o] @*=<#1>}

\newcommand{\pureghost}[1]{*+<1em,.9em>{\hphantom{#1}}}
\newcommand{\multiprepareC}[2]{*+<1em,.9em>{\hphantom{#2}}\save[0,0].[#1,0];p\save !C
  *{#2},p+RU+<0em,0em>;+LU+<+.8em,0em> **\dir{-}\restore\save +RD;+RU **\dir{-}\restore\save
  +RD;+LD+<.8em,0em> **\dir{-} \restore\save +LD+<0em,.8em>;+LU-<0em,.8em> **\dir{-} \restore \POS
  !UL*!UL{\cir<.9em>{u_r}};!DL*!DL{\cir<.9em>{l_u}}\restore}
\newcommand{\prepareC}[1]{*{\xy*+=+<.5em>{\vphantom{#1\rule{0em}{.1em}}}*\cir{l^r};p\save*!L{#1} \restore\save+UC;+UC+<.5em,0em>*!L{\hphantom{#1}}+R **\dir{-} \restore\save+DC;+DC+<.5em,0em>*!L{\hphantom{#1}}+R **\dir{-} \restore\POS+UC+<.5em,0em>*!L{\hphantom{#1}}+R;+DC+<.5em,0em>*!L{\hphantom{#1}}+R **\dir{-} \endxy}}
\newcommand{\poloFantasmaCn}[1]{{{}^{#1}_{\phantom{#1}}}}

%% file: QMfromPuriPRA-resub.bbl
\begin{thebibliography}{99}
\bibitem{anchese...}  In fact, the equivalence of all purifications under local reversible
  transformations is tightly related to what seems the main point of Schr\"odinger's paper, namely
  that all possible ensembles of a mixed state can be ``steered" by measurements on the purifying
  system.  This property is an immediate consequence of the
  equivalence of all purifications.
\bibitem{nothingquantum}  In quantum theory the trivial system $\rI$ is represented by the one-dimensional Hilbert space $\spc H \simeq \Cmplx$.
\bibitem{singlepurifying} In principle, the purifying system for the state $R_{i}$ might depend on
  the value of $i$, let us call it $C_i$.   However,   since  there is a finite number of outcomes one can always choose a
  single purifying system by defining the composite system $\rC = \rC_1 \dots \rC_{N}$ (here we labelled the outcomes with the natural numbers  $\{1, \dots, N\}$).  If
  $\Psi_{i} \in \Stset(\rA\rC_i)$ is a purification of $R_{i}$ and $\psi_{i} $ is any pure state on
  the composite system including all system but the $i$-th, then also $\Psi_{i}' = \Psi_{i} \otimes
  \psi_{i} \in \Stset(\rA\rC)$ is a purification of $R_i$.
\end{thebibliography}
